\documentclass[aps,prx,twocolumn,showpacs]{revtex4-2}
\pdfoutput=1
\usepackage{hyperref}
\hypersetup{
colorlinks = true, linkcolor = [rgb]{0,0,1}, citecolor = [rgb]{0,0,1},
                            urlcolor = [rgb]{0,0,1}}
\usepackage{graphicx,epsf}
\usepackage{amsmath}
\usepackage{amssymb}
\usepackage{color}
\usepackage{esint}
\usepackage{wasysym}

\newcommand{\bB}{{\bf B}}

\newcommand{\bD}{{\bf D}}
\newcommand{\bh}{{\bf h}}

\newcommand{\bS}{{\bf S}}
\newcommand{\bR}{{\bf R}}
\newcommand{\bq}{{\bf q}}
\newcommand{\bQ}{{\bf Q}}
\newcommand{\bp}{{\bf p}}
\newcommand{\bk}{{\bf k}}

\newcommand{\taub}{\mbox{\boldmath $\tau $}}

\begin{document}

\title{
  Thermal Hall conductivity of a valence bond solid phase in the
  square lattice $J_1$-$J_2$ antiferromagnet Heisenberg model
  with a Dzyaloshinskii-Moriya interaction
}

\author{Lucas S. Buzo and R. L. Doretto}
\affiliation{Instituto de F\'isica Gleb Wataghin,
                  Universidade Estadual de Campinas,
                  13083-859 Campinas, SP, Brazil}

\date{\today}

\begin{abstract}
We calculate the thermal Hall conductivity $\kappa_{xy}$ for the
columnar valence-bond solid phase of a two-dimensional frustrated
antiferromagnet. 
In particular, we consider the square lattice spin-$1/2$ $J_1$-$J_2$ 
antiferromagnetic Heisenberg model with an additional 
Dzyaloshinskii-Moriya interaction between the spins and in the
presence of an external magnetic field. 
We concentrate on the intermediate parameter region
of the $J_1$-$J_2$ model, where a quantum paramagnetic phase is stable,
and consider a Dzyaloshinskii-Moriya vector pattern associated with the 
couplings between the spins in the CuO$_2$ planes of the YBCO compound.
We describe the columnar valence-bond solid phase within the
bond-operator formalism, which allows us to map the Heisenberg model
into an effective interacting boson model written in terms of triplet operators.
The effective boson model is studied within the harmonic approximation
and the triplon excitation bands of the columnar valence-bond solid
phase is determined.
We then calculate the Berry curvature and the Chern numbers of the triplon
excitation bands and, finally, determine the thermal Hall conductivity
due to triplons as a function of the temperature.
We find that the Dzyaloshinskii-Moriya interaction yields a finite
Berry curvature for the triplon bands, but the corresponding Chern
numbers vanish. Although the triplon excitations are topologically
trivial, the thermal Hall conductivity of the columnar valence-bond
solid phase in the square lattice antiferromagnet is finite at low temperatures.
Our results complement a previous study [Phys. Rev. {\bf B} 99, 165126 (2019)] 
concerning the thermal Hall effect due to spinons of a spin-liquid
phase on a square-lattice.
We also comment on the relations of our results with a {\sl no-go} condition
for a thermal Hall effect [Phys. Rev. Lett. {\bf 104}, 066403 (2010)] 
previously derived for ordered magnets.
\end{abstract}

\maketitle

\section{Introduction}
\label{sec:intro}

The topological properties of the elementary excitations of insulating
quantum magnets \cite{review-sachdev,review-balents}
have been receiving some attention in recent years
\cite{murakami17,ar-thermal-review}.
Long-range ordered phases in ferromagnets
\cite{owerre16-2,owerre16, malki19,wu21,kim22,li13} 
and in antiferromagnets (AFMs) \cite{kovalev16,fiete18,mertig22}
with topologically nontrivial magnon excitations,
in addition to valence bond solid (VBS) phases in AFMs
\cite{penc15,malki17,coldea17,joshi17}
with topologically nontrivial triplon excitations have been reported.
Such systems, whose excitations bands have nonzero Chern numbers,
are the bosonic analogues of the Chern band insulators:
noninteracting fermionic systems whose electronic bands are
characterized by a finite Chern number
\cite{haldane88,kane13,rachel18}.

Experimentally, the nontrivial topological properties of insulating quantum
magnets can be probed, for instance, via measurements of the thermal
Hall conductivity $\kappa_{xy}$ \cite{lee10,prl-ryo11,prb-ryo11}.
Differently from Chern band insulators, whose nontrivial topological
properties yield an anomalous quantum Hall effect (a quantized Hall
conductance in the absence of an external magnetic field) \cite{rmp-mc23},
magnons and triplons are charge-neutral excitations, and therefore,
do not respond to an applied electric field \cite{ar-thermal-review}. 
However, in the presence of a temperature gradient, magnons and
triplons may induce a transverse (Hall) heat current,
in addition to the (usual) longitudinal one.
Indeed, a magnon thermal Hall effect has been described
in ferromagnets on honeycomb \cite{owerre16,wu21,kim22},
Shastry-Sutherland \cite{malki19},
pyrochlore \cite{prb-ryo11,tokura12},
and perovskite \cite{tokura12} lattices, and
in AFMs on kagome \cite{fiete18},
honeycomb \cite{mertig22},
and square \cite{hotta19} lattices,
while a triplon thermal Hall effect has been predicted for the
Shastry-Sutherland compound SrCu$_2$(BO$_3$)$_2$ \cite{penc15,malki17}.
One should also mention the spin Nernst effect of magnons
(the analogue of the spin Hall effect for electrons)
in AFMs on a honeycomb lattice \cite{xiao16,kovalev16},
a thermal Hall effect due to bosonic spinons \cite{doki18,rhine19}
in spin liquid phases,
and the recent proposal of a triplon thermal Hall effect induced
by an electric field \cite{esaki23}.
Interestingly, although a serie of studies has been devoted to
the magnon thermal Hall effect in insulating quantum magnets, 
the triplon thermal Hall effect has received few attention.

\begin{figure*}[t]
\centerline{\includegraphics[width=4.4cm]{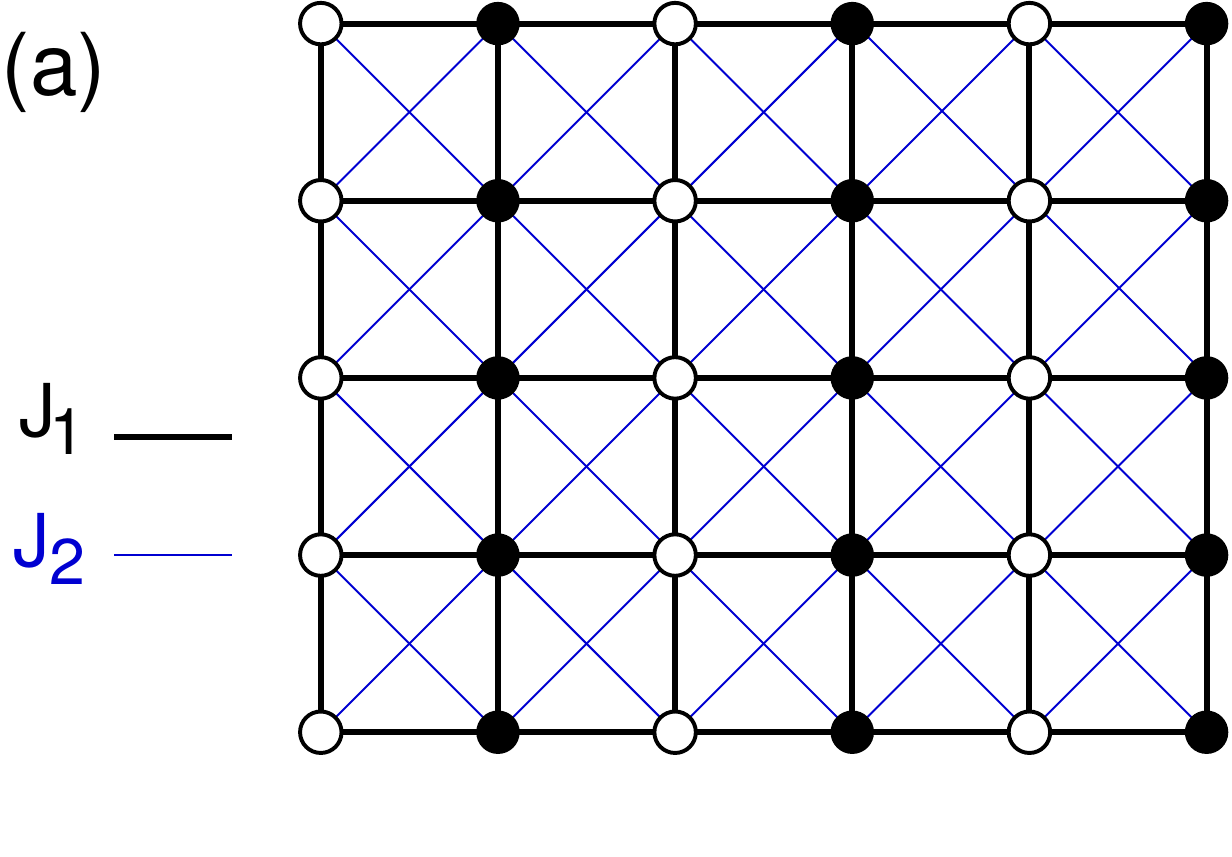}
            \hskip1.0cm
            \includegraphics[width=3.0cm]{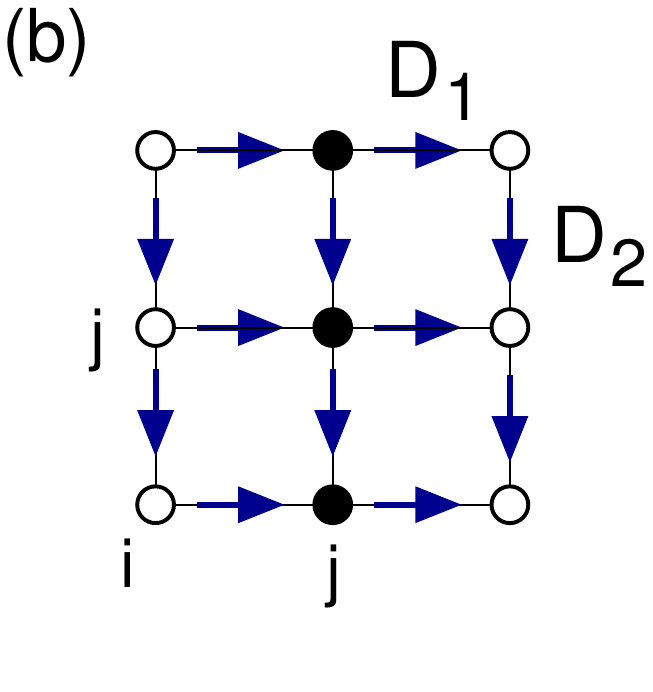}
            \hskip1.0cm
            \includegraphics[width=4.0cm]{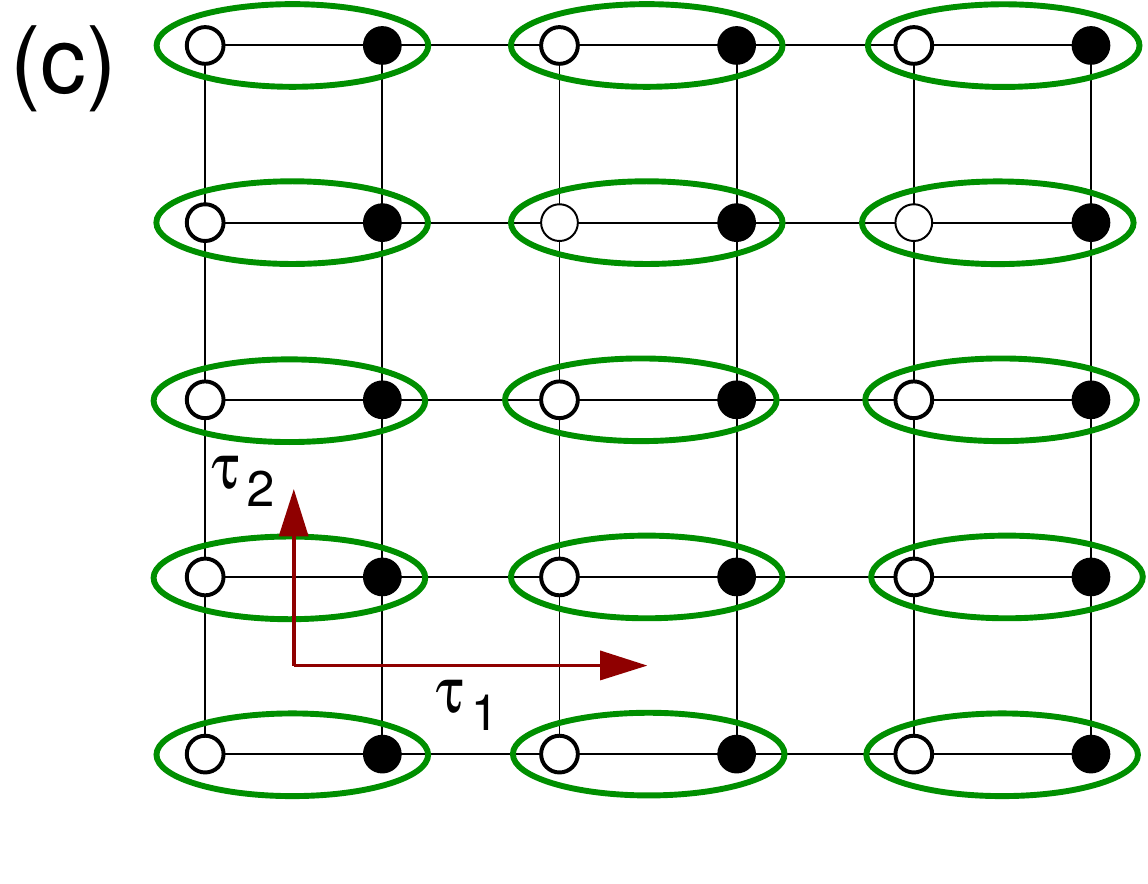}
            \hskip1.0cm
            \includegraphics[width=2.7cm]{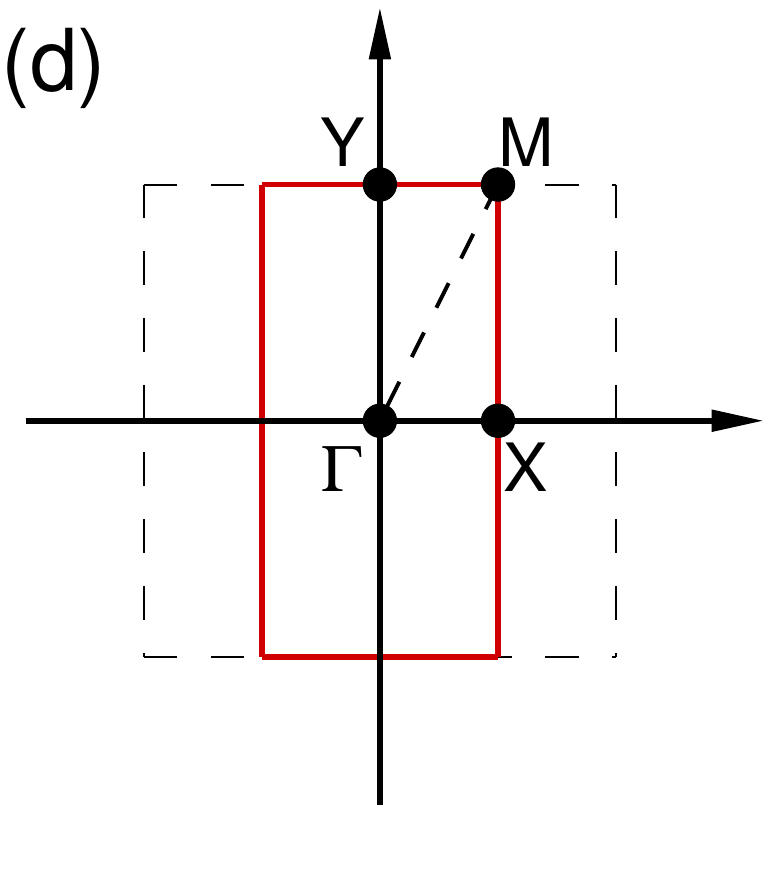}
          }
\caption{Schematic representations: 
 (a) The square lattice spin-$1/2$ $J_1$-$J_2$ AFM Heisenberg model \eqref{ham-j1j2}.
 (b) The DM vectors $\bD_{ij}$ [Eq.~\eqref{ham-dm}] for the spins in the CuO$_2$ planes
 of the YBCO compound; $\bD_1 = (D,0,0)$ and $\bD_2 = (0,-D,0)$. 
 (c) The columnar VBS ground state; the green ellipses indicate that the spins $\bS^1$ (open circle) and 
 $\bS^2$ (filled circle) form a singlet state;
 $\taub_1$ and $\taub_2$ [Eq.~\eqref{tau-col}] are the primitive
 vectors of the dimerized lattice $\mathcal{D}$ defined by the (green) singlets.
 (d) Brillouin zones of the dimerized (solid red line) and the original square (dashed black line) lattices;
 $\mathbf{X} = (\pi/2,0)$, $\mathbf{M} = (\pi/2,\pi)$,  and $\mathbf{Y} = (0,\pi)$
 with the lattice spacing $a$ of the original square lattice being set to 1.
}
\label{fig:model}
\end{figure*}

The magnon thermal Hall effect in insulating quantum magnets was
theoretically studied in Refs.~\cite{lee10, prl-ryo11,prb-ryo11}.
Based on linear spin-wave theory results, Katsura {\sl et al.} \cite{lee10} 
show that the lattice geometry may constraint the presence or absence
of a thermal Hall effect, namely, the thermal Hall conductivity should
vanish (a {\sl no-go theorem}) in quantum magnets whose unit cells
share edges, such as the ones realized in triangular and square lattices.
Moreover, starting from a Kubo formula, it was shown that the thermal Hall
conductivity is finite for a ferromagnet on a (corner sharing) kagome
lattice \cite{lee10}. 
Performing a semiclassical analysis and using liner response theory,
Matsumoto and Murakami \cite{prl-ryo11,prb-ryo11} determine the thermal
transport coefficients for a long-range magnetic ordered phase in a ferromagnet
and show that the thermal Hall conductivity $\kappa_{xy}$ can be written in
terms of the Berry curvatures of the magnon excitation bands. 
It was verify that 
\cite{owerre16-2,owerre16,chen18,wu21,malki19,tokura12,li13,xiao16,kovalev16,
  hotta19,penc15,malki17,coldea17,joshi17,joshi19} 
one important ingredient that may yield a finite Berry curvature for
the bosonic (magnon and triplon) excitation bands is the 
Dzyaloshinskii-Moriya (DM) interaction \cite{dm01,dm03,dm02},
an interaction between localized spins which is associated with
the spin-orbit coupling, and therefore, acts as a kind of
pseudomagnetic field for magnons and triplons.
It should be mentioned that 
the formalism \cite{prl-ryo11,prb-ryo11} is indeed quite general, such that the
expression derived for the thermal Hall conductivity $\kappa_{xy}$ can be applied
to any quantum magnet with bosonic elementary excitations.

Concerning the {\sl no-go} condition \cite{lee10} for the thermal
Hall effect in insulating quantum magnets, it is interesting that some
exceptions to this rule were reported \cite{hotta19, rhine19}.
Kawano and Hotta \cite{hotta19} find a finite
thermal Hall conductivity for a canted AFM ordered phase realized in a
noncentrosymmetric square-lattice AFM;
interestingly, the presence/absence of magnon edge modes is here
characterized by a Z$_2$ topological invariant instead of a Chern number.  
Samajdar {\sl et al.} \cite{rhine19} study a spin-liquid phase on a
square-lattice AFM with a DM interaction term, whose DM vectors,
a set of vectors that characterizes a DM interaction
(see Sec~\ref{sec:model} below for details),  
were chosen as the ones associated with the couplings between the
spins in the CuO$_2$ planes of the La$_2$CuO$_4$ and
YBa$_2$Cu$_3$O$_6$ (YBCO) compounds 
\cite{dm-cuprates01,dm-cuprates02,dm-cuprates03};
the spin-liquid phase is described within the Schwinger-boson formalism; 
interestingly, they find that the thermal Hall conductivity is finite only when
DM vectors related to the YBCO compound are considered, i.e., the
symmetry-allowed DM interaction in the CuO$_2$ planes of the 
La$_2$CuO$_4$ compound does not yield a finite thermal Hall response;
such results are in agreement with a symmetry analysis also presented;
moreover, although $\kappa_{xy}$ is finite, the (bosonic) spinon
excitations of the spin-liquid phase are topologically trivial.
It is important to mention that such choices for the DM vector patterns
were motivated by previous measurements of the thermal Hall 
conductivity in the pseudogap phase of cuprate superconductors 
\cite{kappa-LSCO}, which found a large negative thermal Hall
conductivity at low temperatures;
later, it was found some evidence that such a large thermal Hall
conductivity in La$_2$CuO$_4$ is due to phonons
\cite{kappa-LSCO2,kappa-LSCO3}.

In this paper, we study the triplon thermal Hall effect, in particular,
we calculate the thermal Hall conductivity $\kappa_{xy}$ 
for a VBS phase realized in a square-lattice frustrated AFM.
We consider the spin-$1/2$ $J_1$-$J_2$ AFM Heisenberg model on a square
lattice [Fig.~\ref{fig:model}(a)], with an additional DM interaction
between the nearest-neighbor  spins [Fig.~\ref{fig:model}(b)] and in the presence of an
external magnetic field.  
We focus on the intermediate parameter region 
$0.4\, J_1 \lesssim J_2 \lesssim 0.6\, J_1$ of the $J_1$-$J_2$ model,
where a quantum paramagnetic phase sets in, and, in particular,
consider the columnar VBS phase [Fig.~\ref{fig:model}(c)]. 
The main purpose of this paper is to verify whether a 
DM interaction in the square-lattice AFM yields a columnar VBS phase with
topologically nontrivial triplon excitations and   
whether the {\sl no-go} theorem for the thermal Hall effect derived in
Ref.~\cite{lee10} for an ordered magnet also applies for a VBS phase;
moreover, motivated by the results derived in Ref.~\cite{rhine19}, we consider the 
DM vector pattern corresponding to the YBCO compound;
we would like to verify whether the finite thermal Hall conductivity found
for the spin-liquid phase on a square lattice is a generic feature of quantum
paramagnetic phases, i.e., it is a feature displayed by both 
spin-liquid and VBS phases on a square lattice;
the idea is that our results should complement the previous analysis
reported in Ref.~\cite{rhine19}.

\subsection{Overview of the results}
\label{sec:overview}

We study a spin-$1/2$ frustrated AFM on a square lattice with the
aid of the bond-operator formalism \cite{sachdev90}.
We show that the columnar VBS phase can be described by an effective
interacting boson model expressed in terms of triplet operators. 
We consider the effective boson model in the (lowest-order) harmonic
approximation and determine the triplon excitation bands.
Moreover, we calculate the Berry curvatures and Chern numbers of the
triplon bands.
Following the lines of the formalism \cite{prl-ryo11,prb-ryo11}, which
was previously applied in the study of the magnon thermal Hall effect,
we finally determine the thermal Hall conductivity $\kappa_{xy}$
due to the triplons as a function of the temperature.
We find that the Berry curvature of the triplon bands are finite in
some regions of the first (dimerized) Brillouin zone, but the triplons
are topologically trivial, since the Chern numbers of the triplon
bands vanish.
Our main find is indeed the determination of the dependence of the
thermal Hall conductivity $\kappa_{xy}$  with the temperature $T$:
In spite of the fact that the triplon excitations are topologically trivial,
$\kappa_{xy}$ is finite at low temperatures;
such a result agrees with Ref.~\cite{rhine19}, where
$\kappa_{xy}$ is determined for a spin liquid phase realized in a 
two-dimensional AFM whose Heisenberg model includes the same DM
interaction considered in our paper,
but it is in disagreement with the {\sl no-go} condition \cite{lee10}
for a thermal Hall effect due to magnons, previously determined for long-range ordered
magnets.

\subsection{Outline}
\label{sec:outline}

Our paper is organized as follows.
In Sec.~\ref{sec:model}, we introduce the Hamiltonian of the
spin-$1/2$ frustrated AFM Heisenberg model on a square lattice
considered in our study.  
In Sec.~\ref{sec:bond}, we briefly review the bond-operator
representation \cite{sachdev90} for spin operators, a formalism which
allows us to describe the columnar VBS phase, and derive an
effective interacting boson model in terms of triplet operators.  
The effective boson model is analysed within the harmonic
approximation in Sec.~\ref{sec:harm} and the region of stability of
the columnar VBS phase and the energy of the elementary triplon
excitations are determined.
Sec.~\ref{sec:berry} is devoted to the calculation of the Berry
curvatures of the triplon excitation bands and the determination of
the corresponding Chern numbers.
The thermal Hall conductivity $\kappa_{xy}$ for the columnar VBS
phase is discussed in Sec.~\ref{sec:kappa}.
Finally, a brief summary of our main findings are provided in the 
Sec.~\ref{sec:summary}.
Some further details about the effective boson model and the
analytical procedure employed in the diagonalization of the effective
boson model within the harmonic approximation can be found 
in the three Appendices.

\section{Frustrated square-lattice antiferromagnet}
\label{sec:model}

Let us consider a frustrated spin-$1/2$ Heisenberg
AFM on a square lattice described by the Hamiltonian
\begin{equation}
  \mathcal{H} = \mathcal{H}_J + \mathcal{H}_{\rm DM} + \mathcal{H}_B,
\label{ham}
\end{equation}
where $\mathcal{H}_J$ is the Hamiltonian of the $J_1$-$J_2$
square-lattice AFM Heisenberg model, 
\begin{equation}
  \mathcal{H}_J = J_1\sum_{\langle ij \rangle} \bS_i\cdot\bS_j 
                            + J_2\sum_{\langle\langle ij \rangle\rangle} \bS_i\cdot\bS_j,
\label{ham-j1j2}
\end{equation}
$\mathcal{H}_{\rm DM}$ is the DM interaction term \cite{dm01,dm03,dm02},
\begin{equation}
  \mathcal{H}_{\rm DM} = \sum_{\langle ij \rangle} \bD_{ij} \cdot \left( \bS_i \times \bS_j \right),
\label{ham-dm}
\end{equation}
and $\mathcal{H}_B$ is the Zeeman term that describes the coupling of
the spins with an external magnetic field $\bB$,
\begin{equation}
  \mathcal{H}_B = -g\mu_B \bB \cdot \sum_i \bS_i
                         \equiv -\bh \cdot \sum_i \bS_i.
\label{ham-B}
\end{equation}
Here $\bS_i$ is a spin-$1/2$ operator at site $i$ and $J_1 > 0$ and
$J_2 > 0$ are, respectively, the nearest-neighbor and
next-nearest-neighbor exchange couplings as illustrated in Fig.~\ref{fig:model}(a).
$\bD_{ij}$ is a DM vector that couples the spins $\bS_i$ and
$\bS_j$; we consider, in particular, the DM vector pattern shown in
Fig.~\ref{fig:model}(b), which corresponds to the symmetry-allowed
couplings between the spins in the CuO$_2$ planes of the YBCO compound
\cite{dm-cuprates01,dm-cuprates02,dm-cuprates03}.   
Finally, $g$ is the electron g factor and $\mu_B$ the Bohr magneton. 
It should be mentioned that the AFM Heisenberg model \eqref{ham}, without
the next-nearest-neighbor exchange coupling $J_2$ and an additional
symmetric pseudodipolar interaction between the spins ($\Gamma$ term),
was previously considered in the study of the cuprate superconductors,
as discussed, e.g., in Refs.~\cite{gooding05, marcelo06, lara06}.

The $J_1$-$J_2$ model \eqref{ham-j1j2} has been receiving a lot of
attention in the last few years and now its phase diagram at temperature $T=0$
is well established
\cite{docout88,huse89,moreo89,sondhi90,sachdev90,oliveira91,chubukov91,read91,
  schulz92,dagotto91,igarashi93,schulz95,oitmaa96,zhito96,singh99,kotov99,
  sorella00, beca01,takano03,zhang03,mila06,sirker06,
  darradi08,arlego08,cirac09,beach09, ralko09,ortiz09,
  arlego11,richter12, yu12,jiang12,mezzacapo12,li12,wang13,hu13,
  gu14,chou14, eggert14, doretto14,gong14,
  richter15,imada15,wang16, yang16,sheng18, becca18,dong18,wang18,
  didier17,capponi19,yu19,doretto20,nomura20,ferrari20,
  hasik21,liu22,liu23,luo23}:
A semiclassical N\'eel long-range ordered phase with ordering
wave vector $\bQ = (\pi,\pi)$ sets in for $J_2 \lesssim 0.4\, J_1$,
a quantum paramagnetic phase is stable in the intermediate parameter
region $0.4\, J_1 \lesssim J_2 \lesssim 0.6\, J_1$, and 
a stripe long-range ordered  phase  with $\bQ = (\pi,0)$ or $(0,\pi)$ is
the ground state for $J_2 \gtrsim 0.6\, J_1$.  
Interestingly, the nature of the quantum paramagnetic phase is still an open issue. 
Among the several proposals that have been made for the ground state
of the model \eqref{ham-j1j2} within this intermediate parameter
region, one should mention:     
The (dimerized) columnar  
\cite{singh99,kotov99,sheng18}
and staggered 
\cite{eggert14}
VBSs, where both translational and rotational lattice symmetries are broken;
the (tetramerized) plaquette VBS, where only the translational lattice
symmetry is broken
\cite{zhito96,sorella00,takano03,mila06,ortiz09,yu12,doretto14,gong14}; 
a mixed columnar-plaquette VBS \cite{ralko09};
gapless \cite{beca01,zhang03,wang13,hu13,richter15,becca18,dong18}
and 
gapped \cite{jiang12,mezzacapo12,li12,yang16}
spin-liquid ground states.
More recently, it was found that, within the intermediate parameter region, 
a gapless spin-liquid phase sets in for $J_2 \lesssim 0.53  J_1$ while  
a VBS is stable for $J_2 \gtrsim 0.53 J_1$ \cite{imada15,wang18,nomura20,ferrari20,liu22}; 
such results qualitatively agree with previous density matrix renormalization
group (DMRG) calculations \cite{gong14}, which indicate 
a gapless phase for $0.44\, J_1 < J_2 < 0.50\, J_1$ and a  
VBS one for $0.50\, J_1 < J_2 < 0.61\, J_1$, although with a plaquette
singlet pattern.

Although the $J_1$-$J_2$ model \eqref{ham-j1j2}  has been
extensively studied, few attention has been devoted
to the description of the effects of an additional (anisotropic) DM
interaction between the spins.
Exact diagonalization results \cite{voigt96} for the Heisenberg model \eqref{ham},
without an external magnetic field, found that the extension of the
quantum paramagnetic region of the $J_1$-$J_2$ model
\eqref{ham-j1j2} is affected by the presence of a finite DM interaction.
More recently, a Majorana fermion representation for the spin
operators was employed in order to describe a chiral spin-liquid phase
of the model \eqref{ham} 
\cite{merino22}:
based on a mean-field approach, the stability of such a spin-liquid phase 
was studied;   
exact diagonalization results were also reported.
In both studies \cite{voigt96} and \cite{merino22}, DM vectors
associated with cuprate superconductor compounds were considered.

In the following, we consider the AFM Heisenberg model \eqref{ham} 
within the intermediate parameter region 
$0.4\, J_1 \lesssim J_2 \lesssim 0.6\, J_1$ for the next-nearest
neighbor exchange coupling $J_2$
and DM interaction $D \le 0.50\, J_1$, where
\begin{equation}
   D = |\bD_1| = |\bD_2|,
\end{equation}  
see Fig.~\ref{fig:model}(b).
In particular, we concentrate on the columnar VBS phase 
illustrated in Fig.~\ref{fig:model}(c).
Indeed, the region of stability of the
columnar VBS phase for the $J_1$-$J_2$ model \eqref{ham-j1j2}
and the spectrum of the elementary (triplon) excitations were
determined by one of us within the bond-operator formalism in
Ref.~\cite{doretto20}.

A few remarks here concerning the choice of the model \eqref{ham}
to study a triplon thermal Hall effect are in order:
As mentioned in the Introduction, one of the motivations for our study
are the results \cite{rhine19} concerning the thermal Hall effect due
to spinons of a spin-liquid phase on a square lattice;
recall that our idea is to complement such previous study, but now
focus on a VBS phase on a square lattice.
Although an AFM Heisenberg model with only nearest-neighbor coupling $J_1$
was considered in Ref.~\cite{rhine19}, one needs to consider the
$J_1$-$J_2$ model, since only in the presence of a next-nearest-neighbor 
exchange coupling $J_2$ a (columnar)  VBS phase could be
obtained within a mean-field approximation (see Sec.~\ref{sec:omega} below).
The choice of DM vectors associated with the YBCO
compound is based on the fact that only in this case the spin-liquid phase
discussed in Ref.~\cite{rhine19} displays a finite thermal Hall
conductivity; recall that the thermal Hall conductivity vanishes for
the DM interaction corresponding to the La$_2$CuO$_4$ compound.  
Finally, one needs to introduce an external magnetic field in
order to obtain three triplon bands well separated
(see Sec.~\ref{sec:omega} below), and therefore, 
properly define a Chern number for each triplon band.

\section{Bond operator Formalism}
\label{sec:bond}

The bond-operator representation for spin operators introduced by Sachdev and
Bhatt \cite{sachdev90} is a formalism that allows us to describe a
dimerized VBS phase. In this section, this formalism is briefly summarized,     
following the lines of Refs.~\cite{leite19,doretto20}.

Let us consider the Hilbert space of two $S=1/2$ spins, $\bS^1$
and $\bS^2$, which is made out of a singlet $| s \rangle$ and
three triplet $| t_\alpha \rangle$ states: 
\begin{eqnarray}
| s \rangle   &=& \frac{1}{\sqrt{2}}\left(|\uparrow \downarrow \rangle
                   -|\downarrow \uparrow \rangle \right), 
\;\;\;\;\;\;\;
| t_x \rangle = \frac{1}{\sqrt{2}}\left(|\downarrow \downarrow \rangle
                   -|\uparrow \uparrow \rangle \right), 
\nonumber \\
| t_y \rangle &=& \frac{i}{\sqrt{2}}\left(|\uparrow \uparrow \rangle
                   +|\downarrow \downarrow \rangle \right), 
\;\;\;\;\;\;\;
| t_z \rangle = \frac{1}{\sqrt{2}}\left(|\uparrow \downarrow \rangle
                   +|\downarrow \uparrow \rangle \right).
\nonumber \\
\end{eqnarray}
Let us define a set of boson operators,
$s^\dagger$ and $t^\dagger_\alpha$, with $\alpha = x$, $y$, $z$,
which respectively creates the singlet and the three triplet states
out of a fictitious vacuum $|0\rangle$,
\begin{equation}
  | s \rangle = s^\dagger |0\rangle \;\;\;\;\;\; {\rm and} 
  \;\;\;\;\;\;
  | t_\alpha \rangle = t_\alpha^\dagger |0\rangle.
\end{equation}
In order to remove unphysical states from the enlarged Hilbert space, 
one should introduce the constraint
\begin{equation}
  s^\dagger s + \sum_\alpha t^\dagger_\alpha t_\alpha = 1.
\label{constraint}
\end{equation}
One then calculates the matrix elements of each component of the 
spin operators $\bS^1$ and $\bS^2$
within the basis $|s\rangle$ and $|t_\alpha\rangle$,
i.e., one determines $\langle s | S^\mu_\alpha | s \rangle$, 
$\langle s | S^\mu_\alpha | t_\beta \rangle$, and
$\langle t_\gamma | S^\mu_\alpha | t_\beta \rangle$,
with $\mu = 1$, $2$ and $\alpha$, $\beta$, $\gamma = $ $x$, $y$, $z$. 
Based on the set of obtained results, one easily concludes that 
the components of the spin operators $\bS^1$ and $\bS^2$
can be expanded in terms of the boson operators $s^\dagger$ and
$t^\dagger_\alpha$ as 
\begin{equation}
   S^{1,2}_\alpha = \pm\frac{1}{2}\left(s^\dagger t_\alpha + t^\dagger_\alpha s
                                 \mp i\epsilon_{\alpha\beta\gamma}t^\dagger_\beta t_\gamma \right),
\label{spin-bondop}
\end{equation}
where $\epsilon_{\alpha\beta\gamma}$ is the completely antisymmetric
tensor with $\epsilon_{xyz} = 1$ and the summation convention over
repeated indices is assumed. 
Adding a site index $i$ to the singlet and triplet operators, i.e.,
defining the boson operators
$s^\dagger_i$ and $t^\dagger_{i\,\alpha}$,
one generalizes the bond-operator representation \eqref{spin-bondop} for
spin operators $\bS_i$ on a given lattice.

One should mention that a generalization of the bond-operator
representation \eqref{spin-bondop} for a tetramerized (plaquette) VBS,
which includes two singlet, nine triplet, and five quintet boson operators,
was introduced by one of us in Ref.~\cite{doretto14}.

\subsection{Effective boson model}
\label{sec:boson-model}

With the aid of the generalized bond-operator representation \eqref{spin-bondop}, 
we now map the AFM Heisenberg model \eqref{ham} into an effective
boson model that is written in terms of the triplet
operators $t_{i\,\alpha}$. Such an effective model allows us
to describe the corresponding columnar VBS phase.
In order to perform the mapping, one needs to express the
Hamiltonian \eqref{ham} in terms of the underline dimerized lattice
$\mathcal{D}$, which is defined by the singlet (dimer) arrangement of the
columnar VBS state, see Fig.~\ref{fig:model}(c).

Let us first consider the $J_1$-$J_2$ model \eqref{ham-j1j2}.
In terms of the underline dimerized lattice $\mathcal{D}$,
the Hamiltonian \eqref{ham-j1j2} assumes the form
\begin{align}
 \mathcal{H}_J   =  & \sum_{i \in \mathcal{D}} 
      J_1 \left(   \bS^1_i\cdot\bS^2_i  +  \bS^1_i\cdot\bS^1_{i+2}  +
                      \bS^2_i\cdot\bS^2_{i+2}  + \bS^2_i\cdot\bS^1_{i+1}  \right)
\nonumber \\
   &+  J_2\left(  \bS^1_i\cdot\bS^2_{i+2}  + \bS^2_i\cdot\bS^1_{i+2} \right)
\nonumber \\
   &+ J_2 \left( \bS^2_i\cdot\bS^1_{i+1+2}  + \bS^2_i\cdot\bS^1_{i+1-2} \right),
\label{ham12-dimer}
\end{align}
where $i$ is a site of the dimerized lattice $\mathcal{D}$,
$\bS^1_i$ and $\bS^2_i$ are the two spins within each unit cell,  
and the (lower) index $n = 1,2$ indicates the dimer 
nearest-neighbor vectors $\taub_n$,
\begin{equation}
   \taub_1 = 2 a \hat{x},  \;\;\;\;\;\;\;\;\;\;\;
   \taub_2 = a \hat{y}, 			
\label{tau-col}
\end{equation} 	
with $a$ being the lattice spacing of the original square
lattice, see Fig.~\ref{fig:model}(c). In the following, we set $a=1$.
Substituting the generalized bond-operator representation
\eqref{spin-bondop} into the Hamiltonian \eqref{ham12-dimer}, one finds  
that \cite{doretto20}
\begin{equation}
 \mathcal{H}_{J} = \mathcal{H}_{J,0} + \mathcal{H}_{J,2} + \mathcal{H}_{J,3} + \mathcal{H}_{J,4},
\label{h-boson1}
\end{equation}
where each $\mathcal{H}_{J,n}$ term has $n$ triplet operators as shown
in Eq.~\eqref{h-boson1ap}.   
Finally, the constraint \eqref{constraint} is treated on average via
a Lagrange multiplier $\mu$: one then adds the following term 
to the Hamiltonian \eqref{h-boson1}
\[
  -\mu \sum_i \left( s_i^\dagger s_i + t_{i \alpha}^\dagger t_{i \alpha}  - 1 \right).
\]

In the bond-operator formalism, a dimerized VBS ground state, such as the
columnar VBS one,  
can be viewed as a condensate of the singlets $s_i$. One then sets 
\begin{equation}  
  s_i^\dagger = s_i  = \langle s_i^\dagger \rangle
                              =  \langle s_i \rangle \rightarrow  \sqrt{N_0}
\label{condensate}
\end{equation}
in the Hamiltonian \eqref{h-boson1} and, therefore, arrives at an
effective boson model expressed only in terms of the triplet boson
operators $t_{i \alpha}$. As discussed in Sec.~\ref{sec:harm} below, the
constants $N_0$ and $\mu$ are self-consistently calculated for fixed
values of the next-nearest-neighbor exchange coupling  $J_2$, the DM
interaction $D$, and the external magnetic field $h$.

It is useful to write the Hamiltonian \eqref{h-boson1} in momentum
space. Considering the Fourier transform of the triplet operators
$t_{i\alpha}$,
\begin{equation}
   t_{i \alpha} = \frac{1}{ \sqrt{N'} } \sum_{\bk \in {\rm BZ}} 
                  e^{i\bk \cdot \bR_i } \: t_{\bk \alpha},
\label{fourier}
\end{equation}
with $\bR_i$ being a vector of the dimerized lattice $\mathcal{D}$, 
$N' = N/2$ the number of dimers ($N$ is the number of sites of the
original square lattice), and the momentum sum running over the
(dimerized) first Brillouin zone [see Fig.~\ref{fig:model}(d)],
one shows that the four terms $\mathcal{H}_{J,n}$ of the Hamiltonian
\eqref{h-boson1} assume the form
\begin{align}
 \mathcal{H}_{J,0}  &= -\frac{3}{8}J_1NN_0 - \frac{1}{2}\mu N(N_0 - 1),
\label{h0} \\
\nonumber \\
 \mathcal{H}_{J,2} &= \sum_\bk \left[ A_\bk t^\dagger_{\bk\alpha}t_{\bk\alpha}
                  + \frac{1}{2}B_\bk \left( t^\dagger_{\bk\alpha}t^\dagger_{-\bk\alpha}
                  + {\rm H.c.}\right) \right],
\label{h2}  \\
\nonumber \\
 \mathcal{H}_{J,3} &= \frac{1}{2\sqrt{N'}}\epsilon_{\alpha\beta\lambda}\sum_{\bp,\bk}\xi_{\bk-\bp}
                  \; t^\dagger_{\bk-\bp\alpha}t^\dagger_{\bp\beta}t_{\bk\lambda} + {\rm H.c.},
\label{h3} \\
\nonumber \\
 \mathcal{H}_{J,4} &= \frac{1}{2N'}\epsilon_{\alpha\beta\lambda}\epsilon_{\alpha\mu\nu}
                  \sum_{\bq,\bp,\bk} \gamma_\bk \;
                   t^\dagger_{\bp+\bk\beta}t^\dagger_{\bq-\bk\mu}t_{\bq\nu}t_{\bp\lambda},
\label{h4}
\end{align}
with the coefficients $A_\bk$, $B_\bk$, $\xi_\bk$, and $\gamma_\bk$ given by
\begin{align}
  A_\bk =& \frac{1}{4}J_1 - \mu + B_\bk, 
\nonumber \\
\nonumber \\
  B_\bk =& -\frac{1}{2}N_0\left[ J_1\cos(2k_x) - 2 (J_1 - J_2)\cos(k_y)  \right.
\nonumber \\
            &+ \left.  J_2\cos(2k_x + k_y) + J_2\cos(2k_x - k_y) \right],        
  \nonumber \\
  \nonumber \\
 \xi_\bk  =& -\sqrt{N_0}\left[J_1\sin(2k_x) + J_2\sin(2k_x + k_y)  \right.
\nonumber \\
                &+ \left.  J_2\sin(2k_x - k_y) \right],
\nonumber \\
\label{coefs01} \\
 \gamma_\bk =& -\frac{1}{2}\left[  J_1\cos(2k_x) + 2(J_1 + J_2)\cos k_y \right.
\nonumber \\
                    &+ \left. J_2\cos(2k_x + k_y)  +  J_2\cos(2k_x - k_y) \right].
\nonumber
\end{align}
We should mention that the effective boson model \eqref{h-boson1} was
previously derived by one of us in Ref.~\cite{doretto20}. For
completeness, we provide here all the details of the procedure that,
in the following, will be applied to the DM term \eqref{ham-dm} and
the Zeeman coupling \eqref{ham-B}.

In terms of the sites $i$ of the underline dimerized lattice
$\mathcal{D}$, the DM interaction \eqref{ham-dm} reads
\begin{align}
  \mathcal{H}_{\rm DM} = \sum_{i \in \mathcal{D}} &\Big[
      \bD_{i,i}\cdot(\bS_i^1 \times \bS_i^2) + \bD_{i,i+1} \cdot (\bS_{i}^2 \times \bS_{i+1}^1) \Big. 
\nonumber \\
 & + \Big. \bD_{i,i+2} \cdot (\bS_i^2 \times \bS_{i+2}^2)   \Big.
\nonumber \\
 & + \Big. \bD_{i,i+2}\cdot(\bS_i^1 \times \bS_{i+2}^1) \Big],
\label{ham-dm-dimer}
\end{align}
where the DM vectors $\bD_{i,j}$ are given by [see Fig.~\ref{fig:model}(b)]
\begin{align}
      &\bD_{i,i} = \bD_{i,i+1} = \bD_1 = (D,0,0),
\nonumber \\
      &\bD_{i,i+2} = \bD_2 = (0,-D,0).
\label{dm-vectors}
\end{align}
Again, substituting the bond-operator representation \eqref{spin-bondop}
generalized to the lattice case into the Hamiltonian
\eqref{ham-dm-dimer}, we show that
\begin{equation}
  \mathcal{H}_{\rm DM} = \mathcal{H}_{\rm DM,1} + \mathcal{H}_{\rm DM,2}
                              + \mathcal{H}_{\rm DM,3} + \mathcal{H}_{\rm DM,4},
\label{h-boson2}
\end{equation}
where the $\mathcal{H}_{\rm DM,n}$ term contains $n$ triplet operators,
see Eq.~\eqref{h-boson2ap} for details.
After replacing the singlet operators by its average value
\eqref{condensate}, we performe a Fourier transform with the aid of
Eq.~\eqref{fourier} and find the expressions of the $\mathcal{H}_{\rm DM,n}$ terms
in momentum space, namely, 
\begin{align}
  \mathcal{H}_{\rm DM,2}  =& -i \sum_\bk \Big[
                       \Big(  C_\bk  \, \epsilon_{y\beta\gamma}  
                              + D_\bk  \, \epsilon_{x\beta\gamma}  \Big)  t^\dagger_{\bk\beta}t_{\bk\gamma}  \Big.
\nonumber \\                        
    &+  \Big. \frac{1}{2}C_\bk  \, \epsilon_{y\beta\gamma} \left(  t_{-\bk\beta}t_{\bk\gamma} - t^\dagger_{\bk\gamma}t^\dagger_{-\bk\beta}\right)  \Big.
\nonumber \\
    &+ \Big. \frac{1}{2} D_\bk  \, \epsilon_{x\beta\gamma} \left(  t_{-\bk\beta}t_{\bk\gamma} - t^\dagger_{\bk\gamma}t^\dagger_{-\bk\beta}\right)    \Big],   
\label{hdm2} \\
\nonumber \\ 
  \mathcal{H}_{\rm DM,3}  = & +i \frac{1}{2\sqrt{N'}}  \epsilon_{x\beta\gamma} \epsilon_{\gamma\mu\nu}
                            \sum_{\bk,\bp}  \xi^{\rm DM}_{\bk-\bp}  \, \times
 \nonumber \\
                      &  \times \left( t^\dagger_{\bk-\bp \beta}t^\dagger_{\bp \mu} t_{\bk \nu}
                                             - t^\dagger_{\bk \nu}t_{\bk-\bp \beta} t_{\bp \mu}  \right),
\label{hdm3} \\ 
\nonumber \\
  \mathcal{H}_{\rm DM,4}  =& -i \frac{1}{2N'} \epsilon_{\beta\mu\nu} \epsilon_{\lambda\mu'\nu'} \sum_\bk
                                 \Big( \epsilon_{x\beta\lambda} \, \gamma^{\rm DM}_{x,\bk}  \Big. 
\nonumber \\
           & - \Big.  \epsilon_{y\beta\lambda} \, \gamma^{\rm DM}_{y,\bk}  \Big) 
                    t^\dagger_{\bp+\bk\mu}  t^\dagger_{\bq-\bk\mu'}  t_{\bp\nu}  t_{\bq\nu'},
\label{hdm4} 
\end{align}
where the coefficients $C_\bk$, $D_\bk$, $\xi^{\rm DM}_\bk$, and
$\gamma^{\rm DM}_{\alpha,\bk}$ are defined as
\begin{align}
  C_\bk =& D N_0 \sin(k_y),           
\quad\quad\quad  
  D_\bk = \frac{1}{2} D N_0 \sin(2 k_x),
\nonumber \\
\nonumber \\
  \xi^{\rm DM}_\bk =& D \sqrt{N_0} \cos\left( 2k_x \right),    
\label{coefs02} \\
\nonumber \\
 \gamma^{\rm DM}_{\alpha,\bk} =& D\left( \frac{1}{2} \delta_{\alpha,x} \sin\left( 2k_x \right) 
                                              + \delta_{\alpha,y} \sin k_y \right).
\nonumber 
\end{align}

A few remaks here about the linear term $\mathcal{H}_{\rm DM,1}$
are in order: 
The linear term couples the singlet $s_i$ and the
triplet $t_{i x}$ operators, see Eq.~\eqref{h-boson2ap};
it can be removed via an unitary transformation performed in each site of the
dimerized lattice $\mathcal{D}$;
we follow the lines of Ref.~\cite{coldea17} and consider an unitary
transformation up to first order in the parameter $\alpha = D/(2J_1)$;
as described in details in Appendix~\ref{ap:linear-term}, one of the
effects of such an unitary transformation is to modify the coefficient
$D_\bk$ of the quadratic Hamiltonian $\mathcal{H}_{\rm DM,2}$,
namely,  $D_\bk \rightarrow D_\bk + D'_\bk$,    
with $D'_\bk$ given by Eq.~\eqref{coefs04};
the new coefficient $D_\bk$ (the one considered in Sec.~\ref{sec:harm} below)
then reads
\begin{align}
  D_\bk  = & D N_0 \sin(2 k_x) + \frac{1}{2} D N_0 \frac{J_2}{J_1} \left[ 
                   \sin\left( 2 k_x + k_y \right) \right.
\nonumber \\
                  & \left. + \sin\left( 2 k_x - k_y \right) \right].
\label{coefs03}
\end{align}

Finally, one easily rewrites the Zeeman coupling \eqref{ham-B} in terms of
the dimerized lattice $\mathcal{D}$:
\begin{equation}
  \mathcal{H}_B = -\sum_{i \in \mathcal{D}} h_\alpha
    \left( \bS^1_{i\alpha}  +   \bS^2_{i\alpha} \right).
\label{ham-B-dimer}
\end{equation}
With the aid of Eq.~\eqref{spin-bondop}, one finds the expansion of the
Hamiltonian \eqref{ham-B-dimer} in terms of the triplet operators
$t_{i \alpha}$ [see Eq.~\eqref{h-boson3ap}] and, after the Fourier
transform \eqref{fourier},  shows that
\begin{equation}
 \mathcal{H}_{B} = i\epsilon_{\alpha\beta\gamma} \sum_\bk
    h_{\alpha} \, t_{\bk \beta}^\dagger t_{\bk \gamma},
\label{h-boson3}
\end{equation}
where $h_\alpha$, with $\alpha = x,y,z$, is the $\alpha$-component of
the external magnetic field $\bh = g\mu_B\bB$.


\section{Harmonic approximation}
\label{sec:harm}

In this section, the effective boson model defined by the Hamiltonians
\eqref{h-boson1}, \eqref{h-boson2}, and \eqref{h-boson3} is studied
in the lowest-order (harmonic) approximation. In this case, only terms
up to quadratic order in the triplet operators $t_{\bk \alpha}$ are
retained, i.e., one considers 
\begin{equation}
 \mathcal{H}  \approx \mathcal{H}_{J,0} + \mathcal{H}_{J,2}
                     + \mathcal{H}_{\rm DM,2} + \mathcal{H}_B.
\label{ham-harm}
\end{equation}

\begin{figure*}[t]
\centerline{\includegraphics[width=6.0cm]{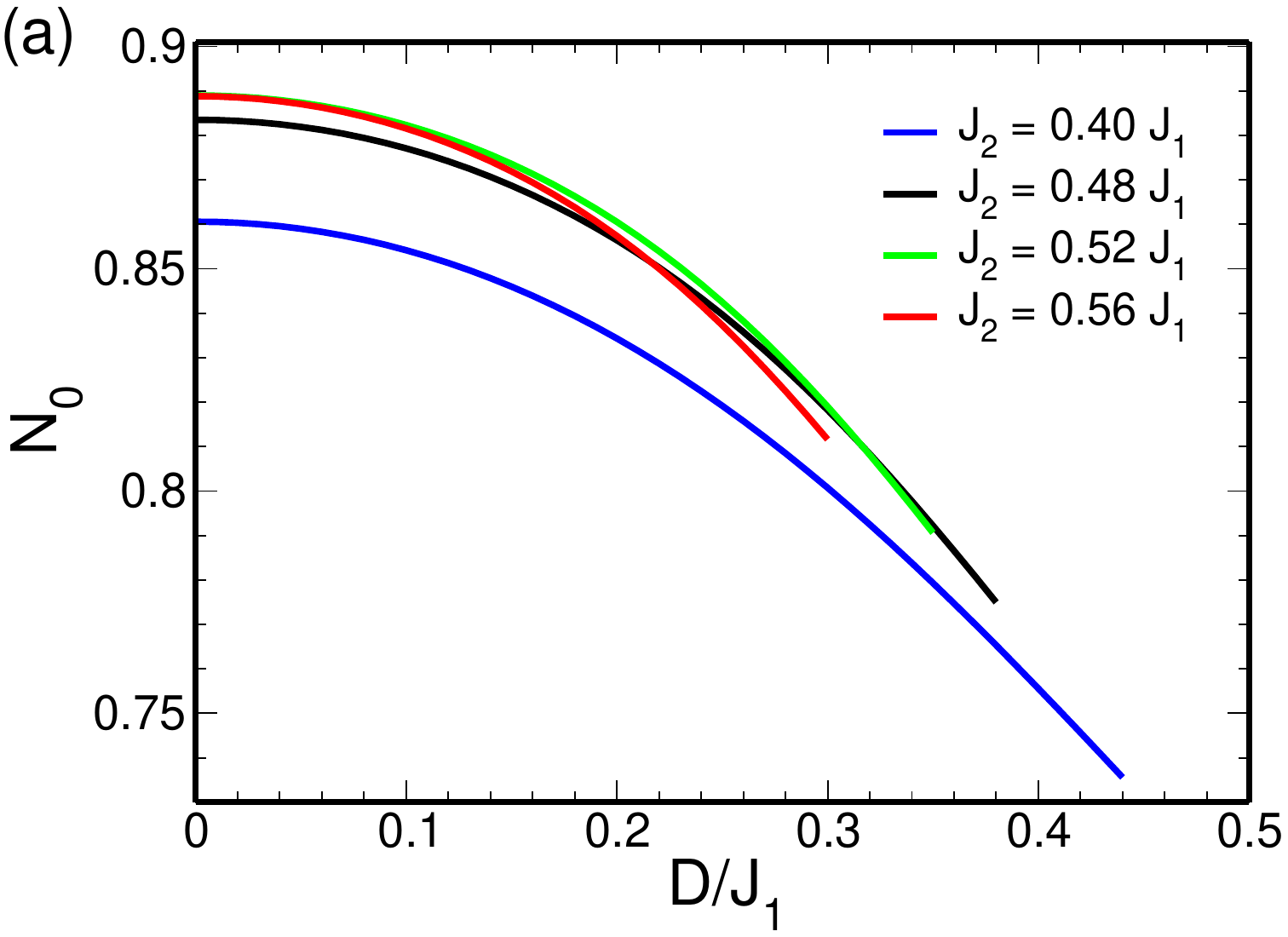}
                   \hskip-0.2cm
                   \includegraphics[width=6.0cm]{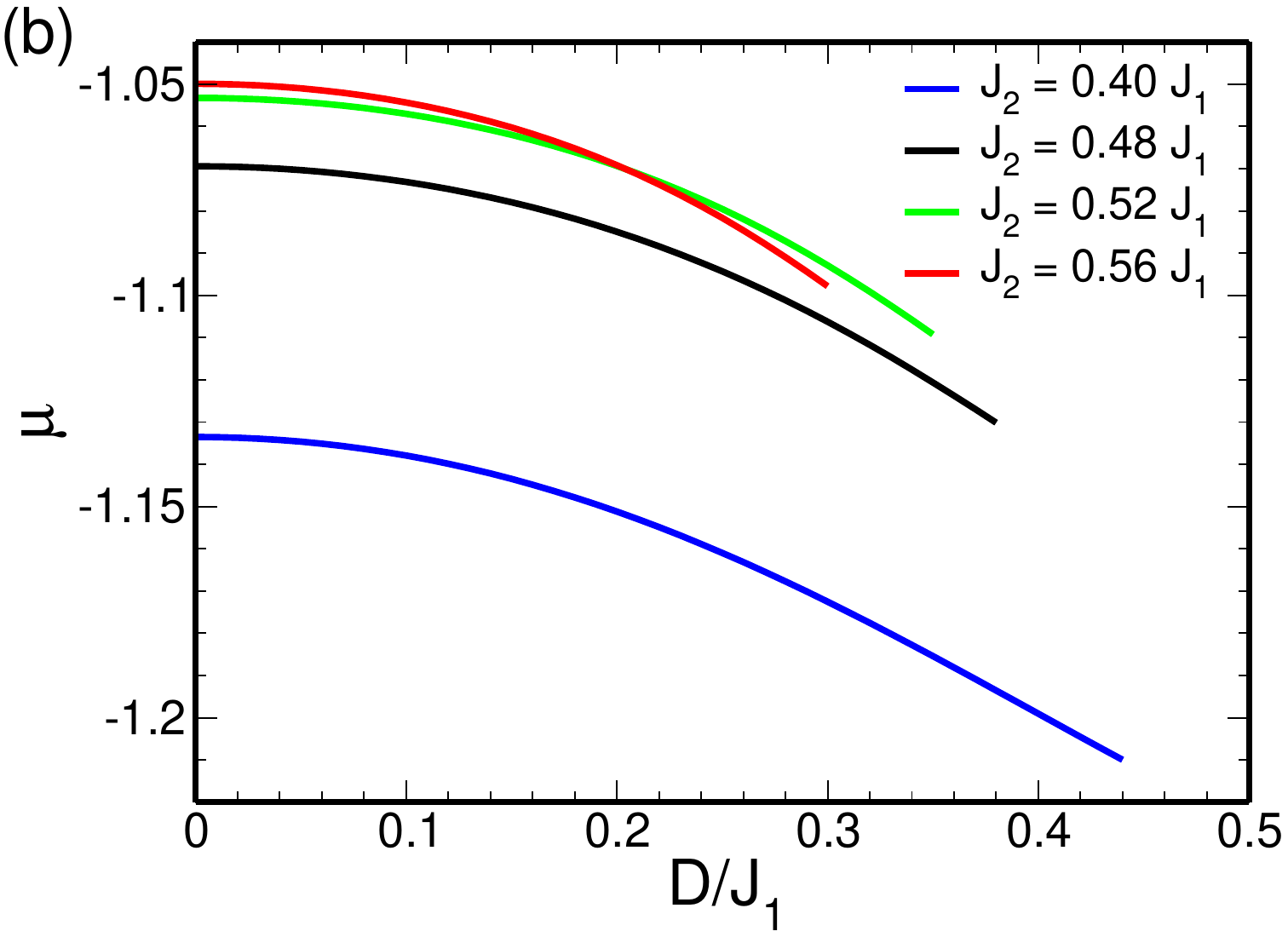}
                   \hskip-0.2cm 
                   \includegraphics[width=6.0cm]{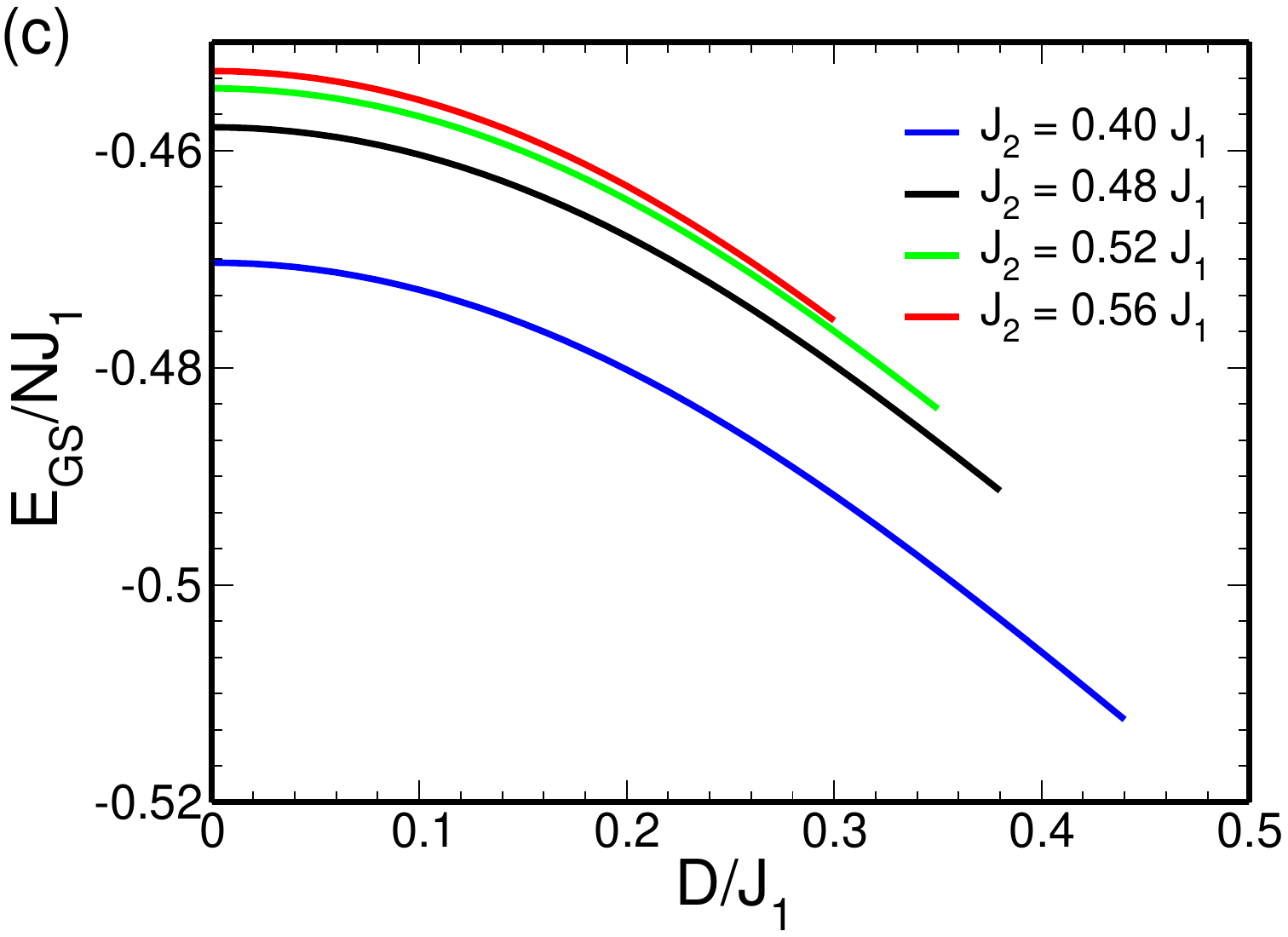}
}
\centerline{\includegraphics[width=6.0cm]{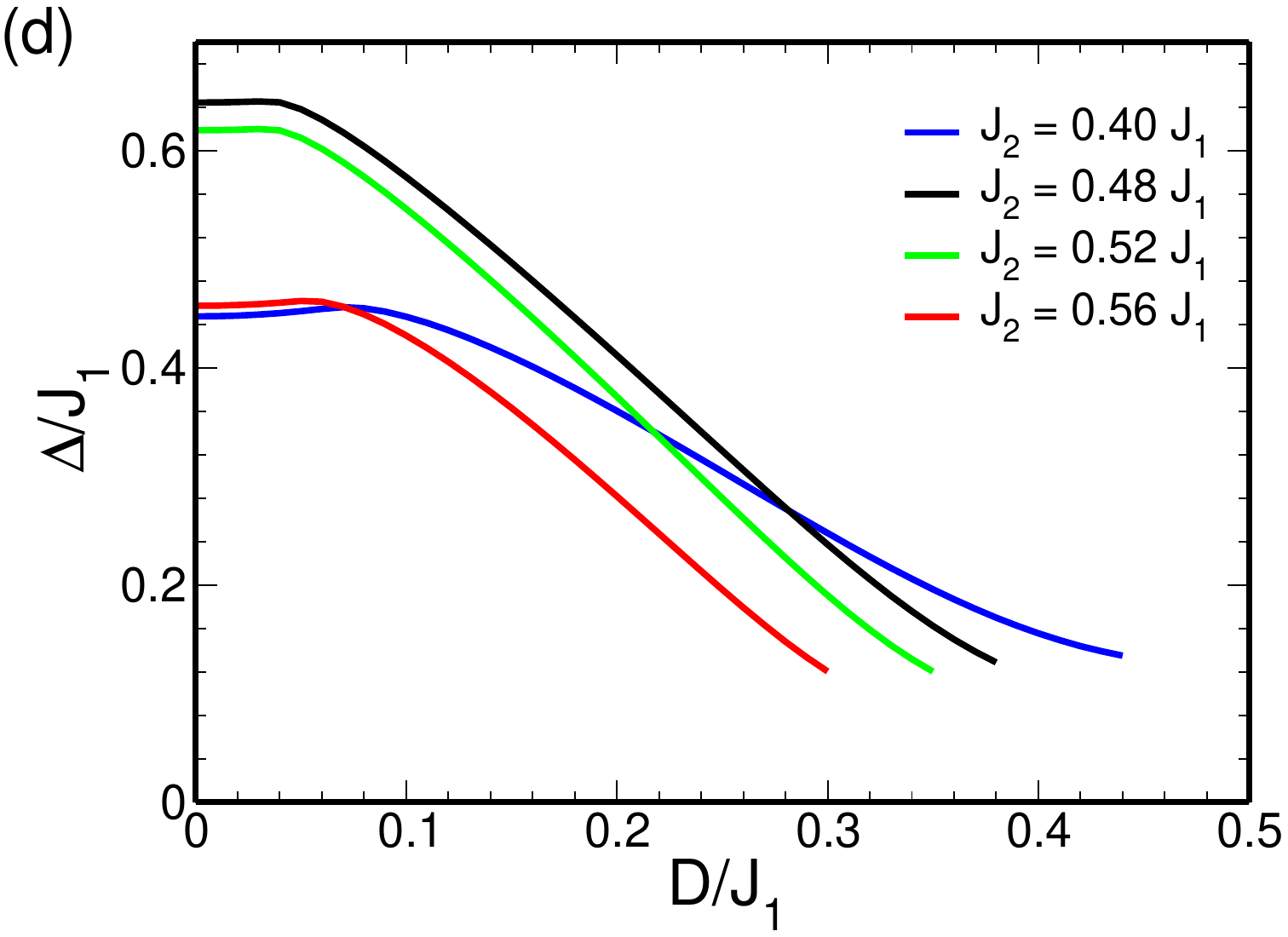}
                   \hskip1.0cm
                   \includegraphics[width=6.0cm]{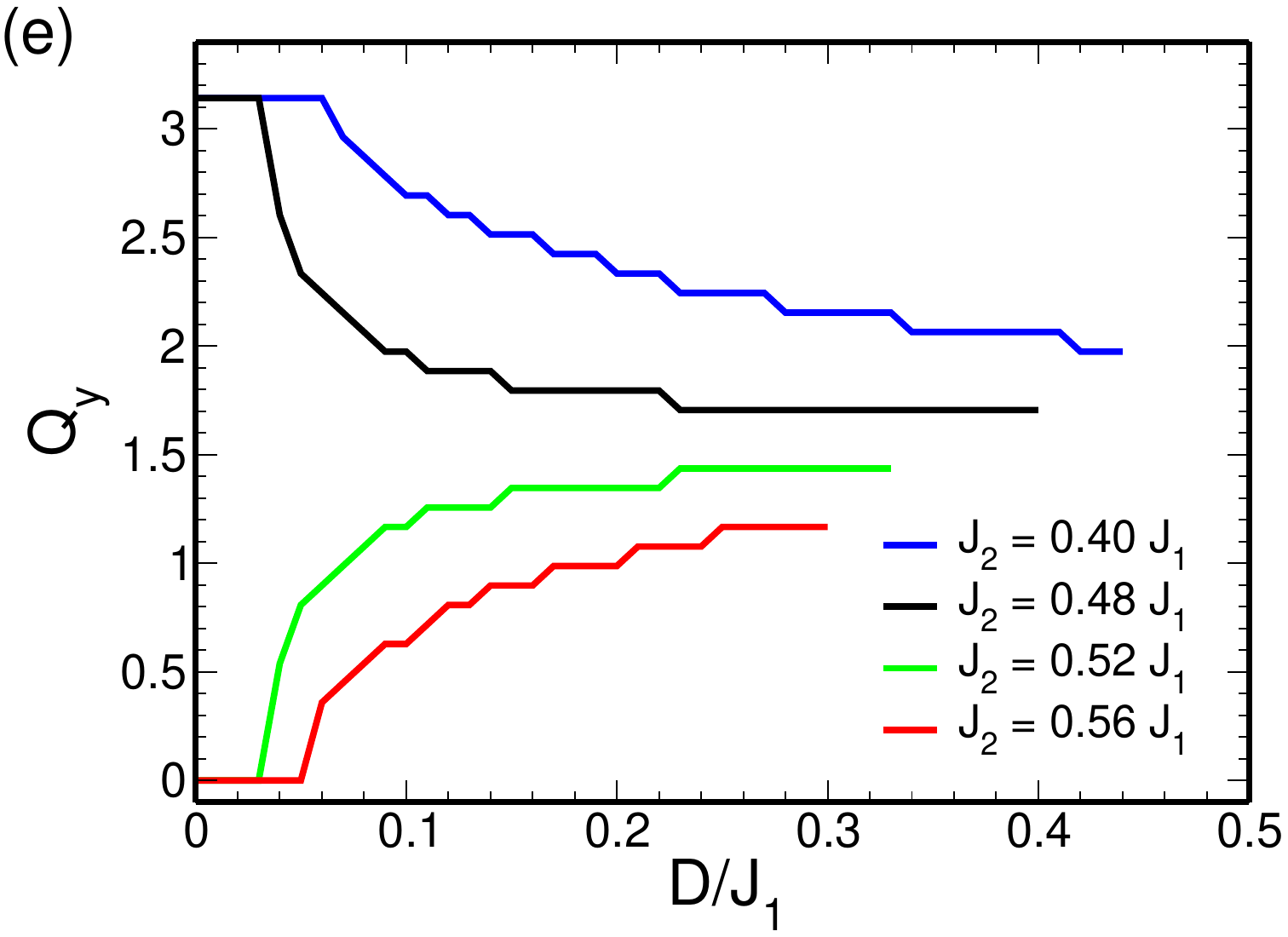}
}
\caption{Parameters (a) $N_0$ and (b) $\mu$,
  (c) the ground-state energy $E_{GS}/(N J_1)$,
  (d) the triplon gap $\Delta/J_1$, and
  (e) the $y$-component $Q_y$ of the momentum $\bQ = (0,Q_y)$ associated
  with the triplon gap
  as functions of the DM
  interaction  $D/J_1$ for the columnar VBS ground state (harmonic approximation).
  Next-nearest-neighbor exchange couplings 
  $J_2 = 0.40\, J_1$ (blue line),
  $J_2 = 0.48\, J_1$ (black line),
  $J_2 = 0.52\, J_1$ (green line), and 
  $J_2 = 0.56\, J_1$ (blue line)
  and the external magnetic field $h_z = 0.05\, J_1$.
}
\label{fig:nzero}
\end{figure*}

In order to diagonalize the quadratic Hamiltonian \eqref{ham-harm}, it is useful
to rewrite it in matrix form,  
\begin{equation}
   \mathcal{H} = E_0 + \frac{1}{2} \sum_\bk \Psi_\bk^\dagger \hat{H}_\bk \Psi_\bk,
\label{ham-harm-mat}
\end{equation}
where 
\begin{equation}
  E_0 = \mathcal{H}_{J,0} - \frac{3}{2}\sum_\bk A_\bk
\end{equation}
is a constant,
\begin{equation}
 \Psi_\bk^\dagger = \begin{pmatrix}
       t_{\bk,x}^\dagger & t_{\bk,y}^\dagger & t_{\bk,z}^\dagger & t_{-\bk,x} & t_{-\bk,y} & t_{-\bk,z} \end{pmatrix}
\label{psi-spinor}
\end{equation}
is a six-component vector, and
\begin{equation}
 \hat{H}_\bk = \begin{pmatrix}
                        \hat{A}_\bk + \hat{H}_B & \hat{B}_\bk \\
                        \hat{B}_\bk      &    \hat{A}_\bk - \hat{H}_B
                       \end{pmatrix}
\label{hk-mat}
\end{equation}
is a $6 \times 6$ matrix with the $3 \times 3$ matrices 
$\hat{A}_\bk$, $\hat{B}_\bk$, and  $\hat{H}_B$ defined as
\begin{align}
  \hat{A}_\bk &= \begin{pmatrix}
                         A_\bk & 0 & i C_\bk  \\
                         0 & A_\bk & -i D_\bk \\
                        -i C_\bk & i D_\bk & A_\bk
                         \end{pmatrix},
\nonumber \\
\nonumber \\  
   \hat{B}_\bk &= \begin{pmatrix}
                          B_\bk & 0 & i C_\bk  \\
                          0 & B_\bk & -i D_\bk  \\
                         -i C_\bk & i D_\bk & B_\bk
                         \end{pmatrix},
\label{AB-matrices} \\
\nonumber \\ 
   \hat{H}_B &= \begin{pmatrix}
                         0 & ih_z & -ih_y  \\
                         -ih_z & 0 & ih_x \\
                         ih_y & -ih_x & 0
                         \end{pmatrix}.
\nonumber
\end{align}  
Here, the coefficients $A_\bk$ and $B_\bk$ are given by Eq.~\eqref{coefs01}
while the coefficients $C_\bk$ and $D_\bk$ are, respectively, given by
Eqs.~\eqref{coefs02} and \eqref{coefs03}; as mentioned in the previous
section and discussed in Appendix~\ref{ap:linear-term},
the expression \eqref{coefs03} for the
coefficient $D_\bk$ includes the effects of the linear term
$\mathcal{H}_{\rm DM,1}$ [Eq.~\eqref{h-boson2ap}] associated with the DM
interaction \eqref{ham-dm}.  Finally, $h_\alpha = g\mu_B B_\alpha$,
with $\alpha = x,y,z$, are the components of the external magnetic
field; in the following, we assume that $\bB = B\hat{z}$.

The diagonalization of the $6 \times 6$ problem defined by
Eqs.~\eqref{ham-harm-mat}--\eqref{hk-mat} is
rather involved. It is then interesting to employ the procedure
described in Refs.~\cite{colpa78,blaizot}: Since we are dealing with a
bosonic Hamiltonian, instead of the matrix \eqref{hk-mat}, we 
should diagonalize 
\begin{equation}
  \hat{I}_B\hat{H}_\bk
   \quad\quad {\rm with} \quad\quad
  \hat{I}_B = \begin{pmatrix}
                      \hat{I} & 0 \\
                       0 & -\hat{I}
                     \end{pmatrix}
\label{ib-hk-mat}                     
\end{equation}
and $\hat{I}$ being the $3 \times 3$ identity matrix. 
The positive eigenvalues $\omega^\alpha_\bk$ of the matrix \eqref{ib-hk-mat},
with $\alpha = x,y,z$, are indeed the roots of a cubic polynomial and are shown
in Appendix~\ref{ap:diag}, see Eq.~\eqref{omega}.
After the diagonalization, the Hamiltonian \eqref{ham-harm-mat}
assumes the form
\begin{equation}
 \mathcal{H} = E_{GS} + \frac{1}{2} \sum_\bk \Phi_\bk^\dagger \hat{H}'_\bk \Phi_\bk,
\label{ham-mat-diag}
\end{equation}
where
\begin{equation}
  E_{GS} = - \frac{3}{8} J_1 N_0 N - \frac{1}{2}\mu N(N_0 - 1)
            + \frac{1}{2} \sum_{\bk,\alpha} \left(\omega_\bk^\alpha -  A_\bk \right)
\label{egs}
\end{equation}
is the ground-state energy,
\begin{equation}
  \Phi_\bk^\dagger = \begin{pmatrix}
                                 b_{\bk,x}^\dagger & b_{\bk,y}^\dagger & b_{\bk,z}^\dagger & b_{-\bk,x} & b_{-\bk,y} & b_{-\bk,x}
                                \end{pmatrix}
\end{equation}
is a six-component vector whose components are the
new boson (triplon) operators $b_{\bk,\alpha}$, and
\begin{equation}
 \hat{H}'_\bk = {\rm diag}\left( \omega^x_\bk, \, \omega^y_\bk, \, \omega^z_\bk,
                                             \, \omega^x_\bk, \, \omega^y_\bk, \, \omega^z_\bk \right)
\label{ham-harm-mat-diag}
\end{equation}
is a $6 \times 6$ diagonal matrix.
The eigenvalues $\omega^\alpha_\bk$ are indeed the energies of the triplon
excitations above the VBS ground state. Moreover, the relation between
the triplet $t$ and triplon $b$ boson operators reads
\begin{equation}
   \Psi_\bk = \hat{T}_\bk \Phi_\bk,
\label{psi-phi}
 \end{equation}
where the $6 \times 6$ matrix $\hat{T}_\bk$ assumes the form 
\begin{equation}
\hat{T}_\bk = \begin{pmatrix}
                      \hat{U}_\bk  & \hat{Y}_\bk \\
                      \hat{V}_\bk  & \hat{X}_\bk	
\end{pmatrix}
\label{Tmatrix1}
\end{equation}
and obeys the condition \cite{colpa78,blaizot}
\begin{equation}
  \hat{T}_\bk \hat{I}_B \hat{T}^\dagger_\bk = \hat{I}_B,
\label{conditionT}
\end{equation}
with the matrix $\hat{I}_B$ defined in Eq.~\eqref{ib-hk-mat}.
Here $\hat{U}_\bk$, $\hat{V}_\bk$,  $\hat{Y}_\bk$, and $\hat{X}_\bk$
are $3 \times 3$ matrices whose expressions of the corresponding
matrix elements in terms of the coefficients $A_\bk$, $B_\bk$,
$C_\bk$, and $D_\bk$, the magnetic field $h_z$, and the triplon
energies $\omega^\alpha_\bk$ can be found in Appendix~\ref{ap:diag}.

In order to determine the ground-state energy \eqref{egs} and the triplon
dispersion relations $\omega^\alpha_\bk$, we still need to determine
the parameters $\mu$ and $N_0$.
These two parameters follow from the saddle-point conditions
$\partial E_{GS} / \partial N_0 = 0$ and $\partial E_{GS} / \partial \mu = 0$,
that yield the two self-consistent equations
\begin{align}
  \mu &=  -\frac{3}{4} J_1 + \frac{1}{2 N'} \sum_{\bk,\alpha}
                   \left( \frac{\partial \omega_\bk^\alpha}{\partial N_0} - \frac{B_\bk}{N_0} \right),
\nonumber \\
\label{eqs-self} \\
  N_0 &= 1 + \frac{1}{2 N'} \sum_{\bk, \alpha}
               \left( \frac{\partial \omega_\bk^\alpha}{\partial \mu} + 1 \right).
\nonumber
\end{align}
The above set of self-consistent equations is numerically solved, and
therefore, the parameters $\mu$ and $N_0$ are determined
for fixed values of the next-nearest-neighbor exchange coupling $J_2$,
the DM interaction $D$, and the external magnetic field $h_z$.

\subsection{Region of stability of the columnar VBS phase and the triplon excitations}
\label{sec:omega}

\begin{figure}[t] 
\centerline{\includegraphics[width=7.5cm]{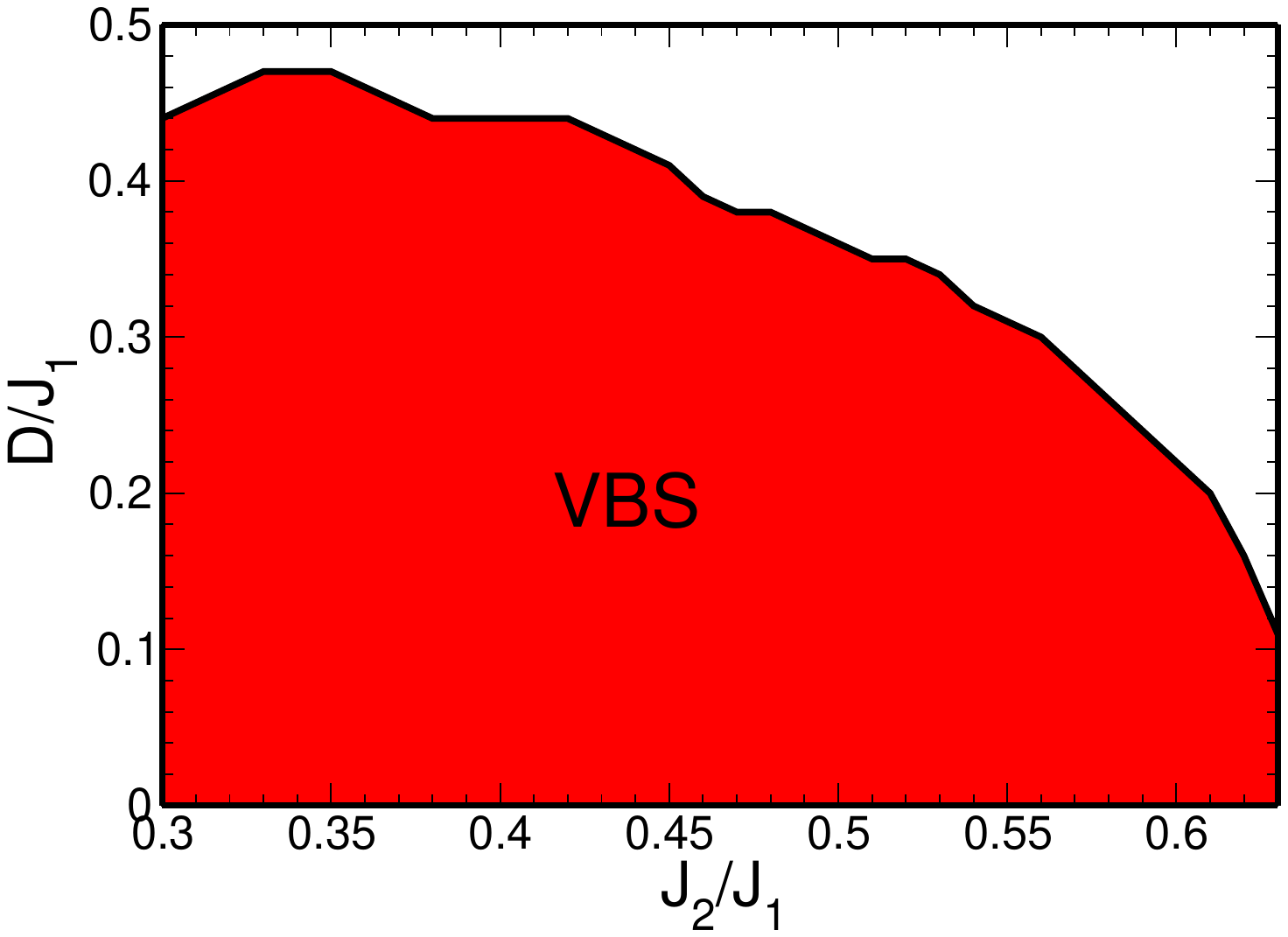}
}
\caption{Region of stability of the columnar VBS phase
  [Fig.~\ref{fig:model}(c)] of the Heisenberg model \eqref{ham}
  determined within the bond-operator formalism at the harmonic
  approximation. The external magnetic field $h_z = 0.05\, J_1$.
}
\label{fig:phasediag}
\end{figure}

Figures~\ref{fig:nzero}(a) and (b) show, respectively, the parameters
$N_0$ and $\mu$ as functions of the DM interaction $D$, for fixed
values of the next-nearest-neigbhor exchange coupling $J_2$ and 
external magnetic field $h_z = 0.05\, J_1$, both determined via
a numerical solution of the set of self-consistent equations
\eqref{eqs-self}. One sees that the parameter $N_0$ decreases with
the increasing of the DM interaction $D$, which indicates that the
stability of the columnar VBS phase decreases as the DM interaction
increases. Indeed, the region of stability of the columnar VBS phase, i.e., the
region of the parameter space where (numerical) solutions for the
self-consistent equations \eqref{eqs-self} can be found, is indicated in
Fig.~\ref{fig:phasediag}.
Notice that:
For $D=0$, the columnar VBS phase is stable within the intermediate
parameter region $0.30\, J_1 \le J_2 \le 0.63\, J_1$, which is larger
than the one ($0.4\, J_1 \lesssim  J_2 \lesssim 0.6\, J_1$) expected 
for the quantum paramagnetic phase of the $J_1$-$J_2$ model, 
a feature of the harmonic approximation found in our previous studies
\cite{doretto14,leite19,doretto20};
as the exchange coupling $J_2$ increases, the $D_{\rm max}$ below
which the VBS phase is stable decreases, i.e., the effect of the DM
interaction on the VBS seems to be distinct for the
small and large $J_2$ parameter regions;
interestingly, such a distinct behaviour for $J_2$ below and above
$J_2 \sim 0.5 \, J_1$ qualitatively agrees with
Refs.~\cite{gong14,imada15,wang18,nomura20,ferrari20,liu22},
whose results indicate that two different phases may set in within the
quantum paramagnetic region of the $J_1$-$J_2$ model (see also
Sec.~\ref{sec:model} above). 
Finally, it should be mentioned that the region of
stability of the columnar VBS phase in the absence of the external
magnetic field (not shown here) is almost equal to the one found for 
$h_z = 0.05\, J_1$.

The ground-state energy \eqref{egs} in terms of the DM interaction $D$
and for fixed values of the exchange coupling $J_2$ is shown in
Fig.~\ref{fig:nzero}(c). For all values of the coupling $J_2$, we find
that the ground-state energy $E_{GS}$ decreases with the DM
interaction $D$. Moreover, for a fixed DM interaction, we
notice that  $E_{GS}$ increases with the increasing of the exchange
coupling $J_2$ up to $J_2 \sim 0.57\, J_1$ and then it decreases.
Indeed, for $D = h = 0$,  a quite similar behaviour was observed for
the ground-state energy, namely, a monotonical increasing of $E_{GS}$
with $J_2$ up to  $J_2 = 0.57\, J_1$ (see Fig.~3(a) from Ref.~\cite{doretto20}).

\begin{figure*}[t]
\centerline{\includegraphics[width=7.7cm]{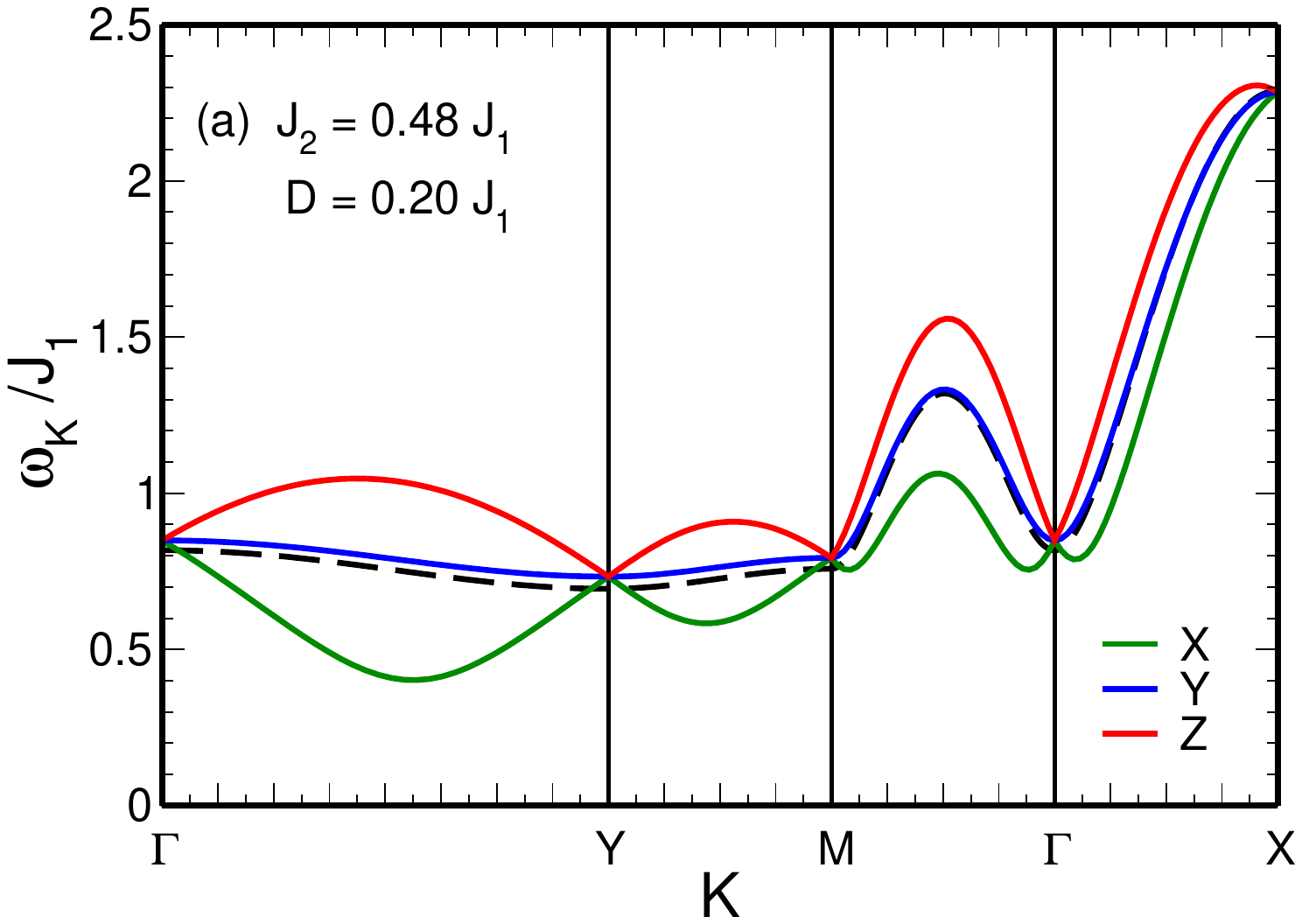}
                  \hskip1.0cm
                  \includegraphics[width=7.7cm]{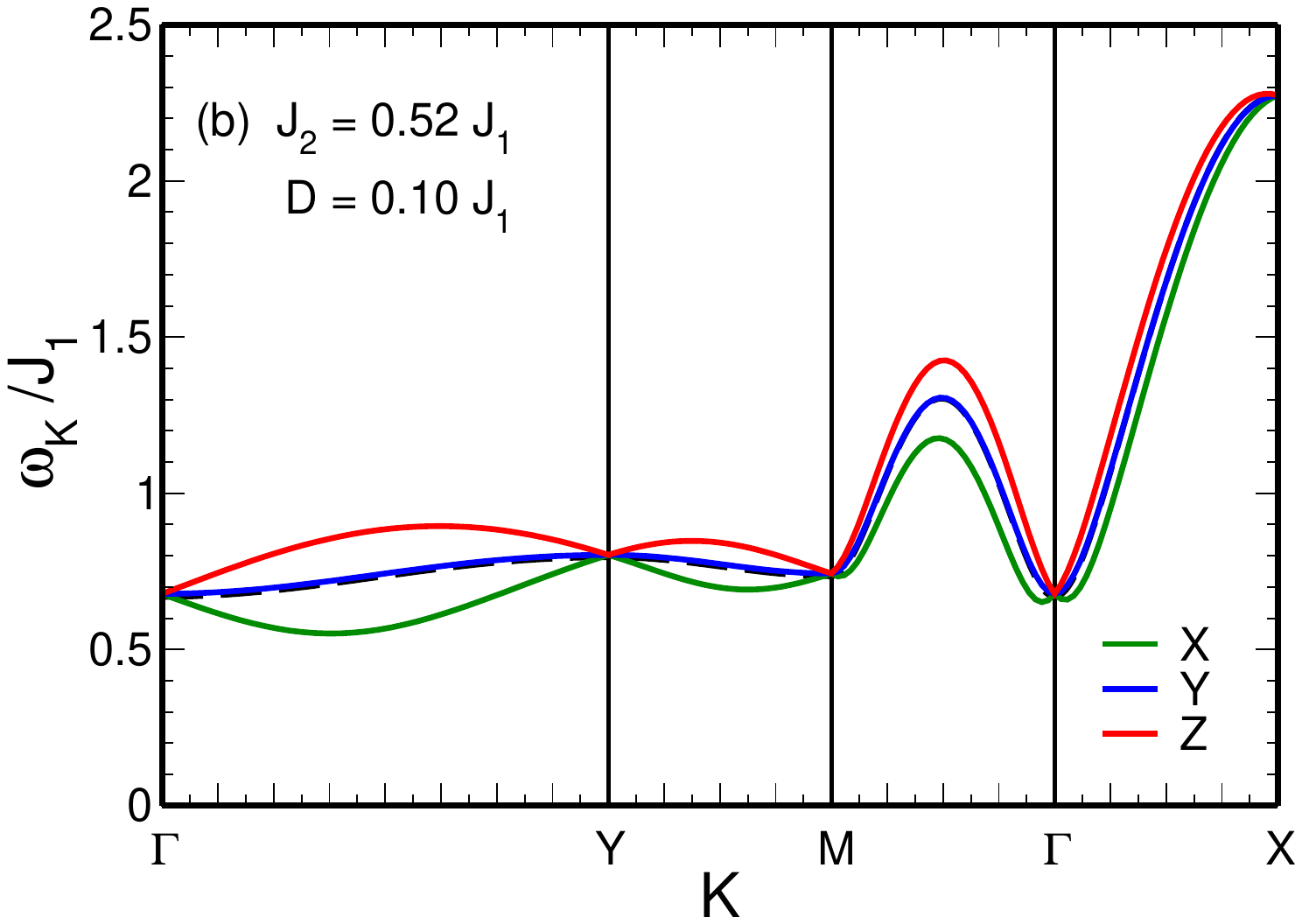}
}
\centerline{\includegraphics[width=7.7cm]{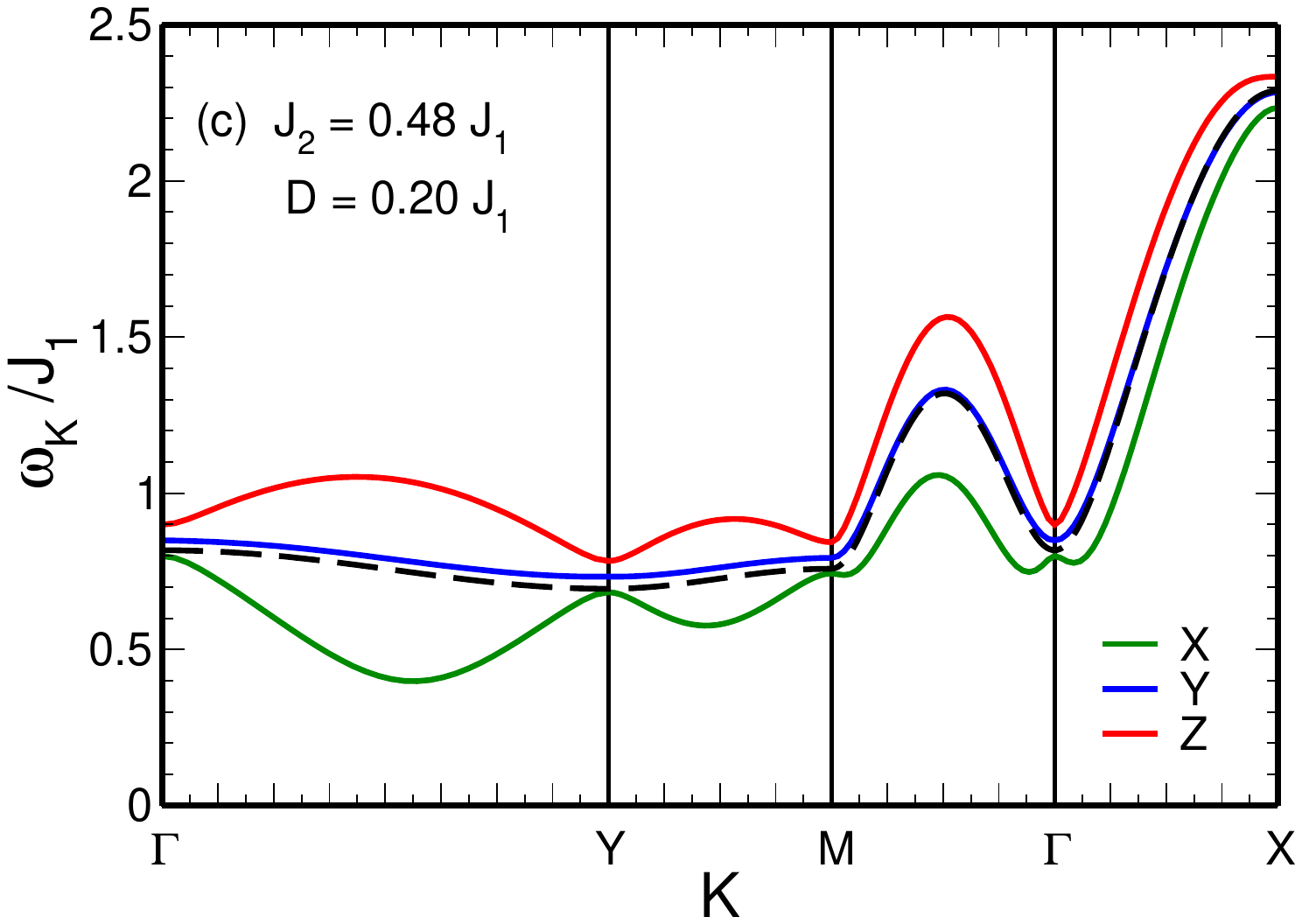}
                  \hskip1.0cm
                  \includegraphics[width=7.7cm]{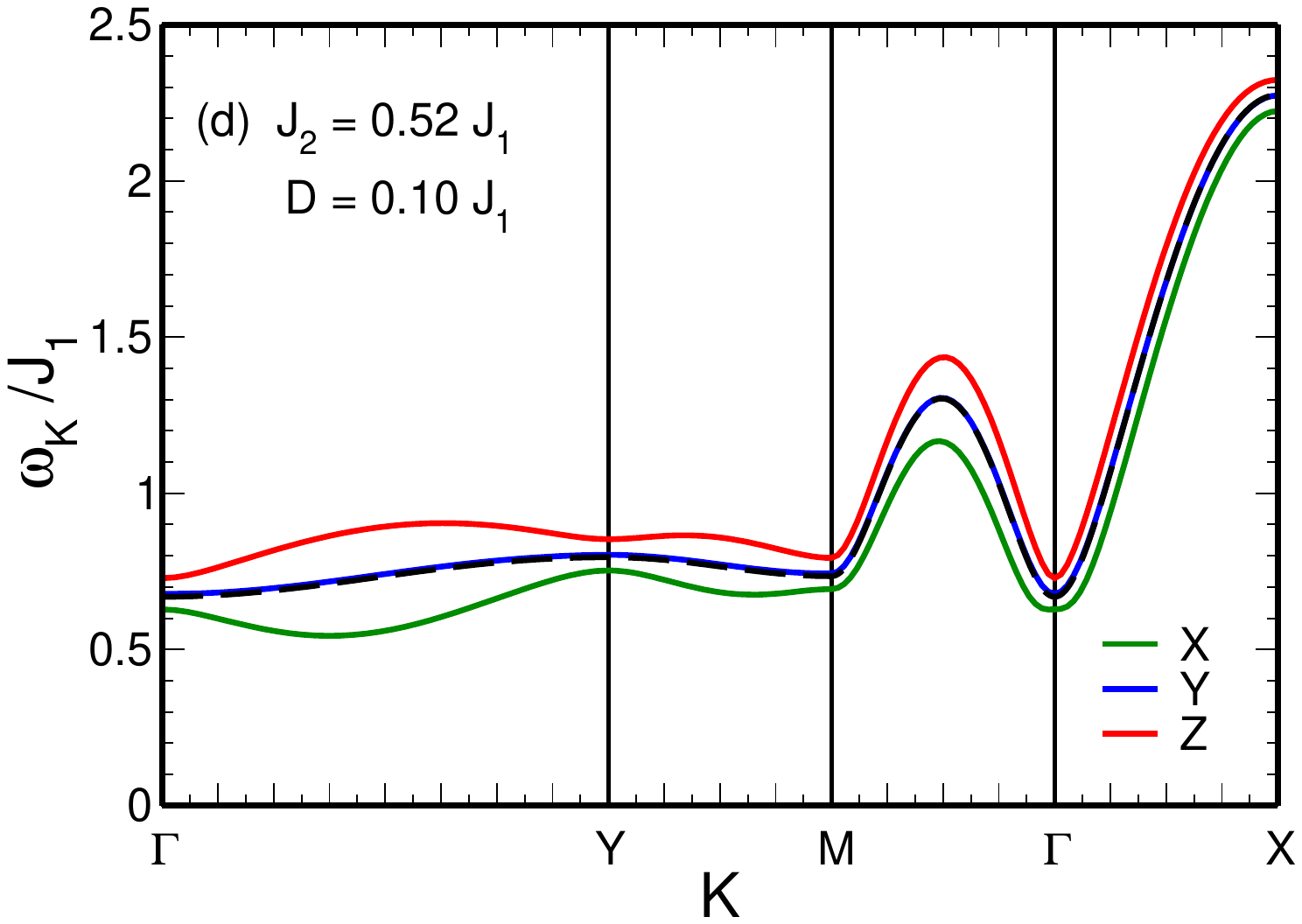}
}
\caption{Triplon dispersion relations
  $\omega^x_\bk$ (solid green line),
  $\omega^y_\bk$ (solid blue line), and
  $\omega^z_\bk$ (solid red line)
  [see Eq.~\eqref{omega}] of the columnar VBS phase
  at the harmonic approximation along paths in the dimerized Brillouin zone
  [Fig.~\ref{fig:model}(d)]:
  (a) and (c) $J_2 = 0.48\, J_1$ and $D = 0.20\, J_1$ and
  (b) and (d) $J_2 = 0.52\, J_1$ and $D = 0.10\, J_1$.
  The external magnetic field
  $h_z = 0 $ (upper panels) and 
  $h_z = 0.05\, J_1$ (lower panels). 
  Dashed black line: triplon dispersion relation for the model
  \eqref{ham} in the absence of the DM and Zeeman terms (see Fig.~4
  from Ref.~\cite{doretto20}). 
}
\label{fig:omega}
\end{figure*}

Figures~\ref{fig:omega}(a) and (c) show the energy $\omega^\alpha_\bk$ of the
triplon excitations for $J_2 = 0.48\, J_1$ and $D = 0.20\, J_1$, with
$h_z = 0$ [Fig.~\ref{fig:omega}(a)] and 
$h_z = 0.05\, J_1$ [Fig.~\ref{fig:omega}(c)] (solid lines),
in addition to the triplon spectrum for $D = h = 0$
(dashed black line), which is included here for comparison 
(see Fig.~4(a) from Ref.~\cite{doretto20}). 
As expected for a VBS phase, the triplon excitation spectrum is gapped.
One sees that the triple degeneracy of the triplon bands, a feature displayed by
the $J_1$-$J_2$ model \eqref{ham-j1j2}, is partially lifted by the DM interaction, since
the triplon bands touch at the 
$\Gamma$, $\mathbf{X}$, $\mathbf{M}$, and $\mathbf{Y}$ points of the
dimerized first Brillouin zone [see Fig.~\ref{fig:model}(d)] for
$D \not=0$ and $h = 0$ [Fig.~\ref{fig:omega}(a)]. 
The three triplon bands are completely separated only in the
presence of both DM interaction and external magnetic field
[Fig.~\ref{fig:omega}(c)]:
Indeed, the DM interaction alone may not completely lift the
degeneracy of an excitation spectrum as found, e.g., 
for magnons in magnets with
long-range order \cite{owerre16,malki19,hotta19}
and triplons on a Shastry-Sutherland lattice \cite{penc15,malki17}.
Moreover, the triplon gap $\Delta$ (the minimum of the $\omega^x_\bk$ triplon band)
is associated with an incommensurate momentum $\bQ = (0, 1.7952)$,
for $h_z = 0.05\, J_1$;
such a behaviour should be contrasted with the triplon spectrum for $D = h = 0$, 
whose triplon gap is located at the (commensurate) $\mathbf{Y}$ point  
of the dimerized  Brillouin zone.
Similar considerations hold for the triplon excitation spectrum for 
$J_2 = 0.52\, J_1$ and $D = 0.10\, J_1$,  with
$h_z = 0$ [Fig.~\ref{fig:omega}(b)] and 
$h_z = 0.05\, J_1$ [Fig.~\ref{fig:omega}(d)]:
Here, the triplon gap is also located at an incommensurate momentum
[$\bQ = (0, 1.1669)$ for $h_z = 0.05\, J_1$] while,
for $D=h=0$, the triplon gap is at the centre of the dimerized
Brillouin zone, the $\Gamma$ point.
In the following, one considers a finite external magnetic field 
($h_z = 0.05\, J_1$), since one needs well separated triplon bands in
order to properly define a Chern number and determine the topological
aspect of the triplon excitations.

\begin{figure*}[t]
  \centerline{\includegraphics[width=6.2cm]{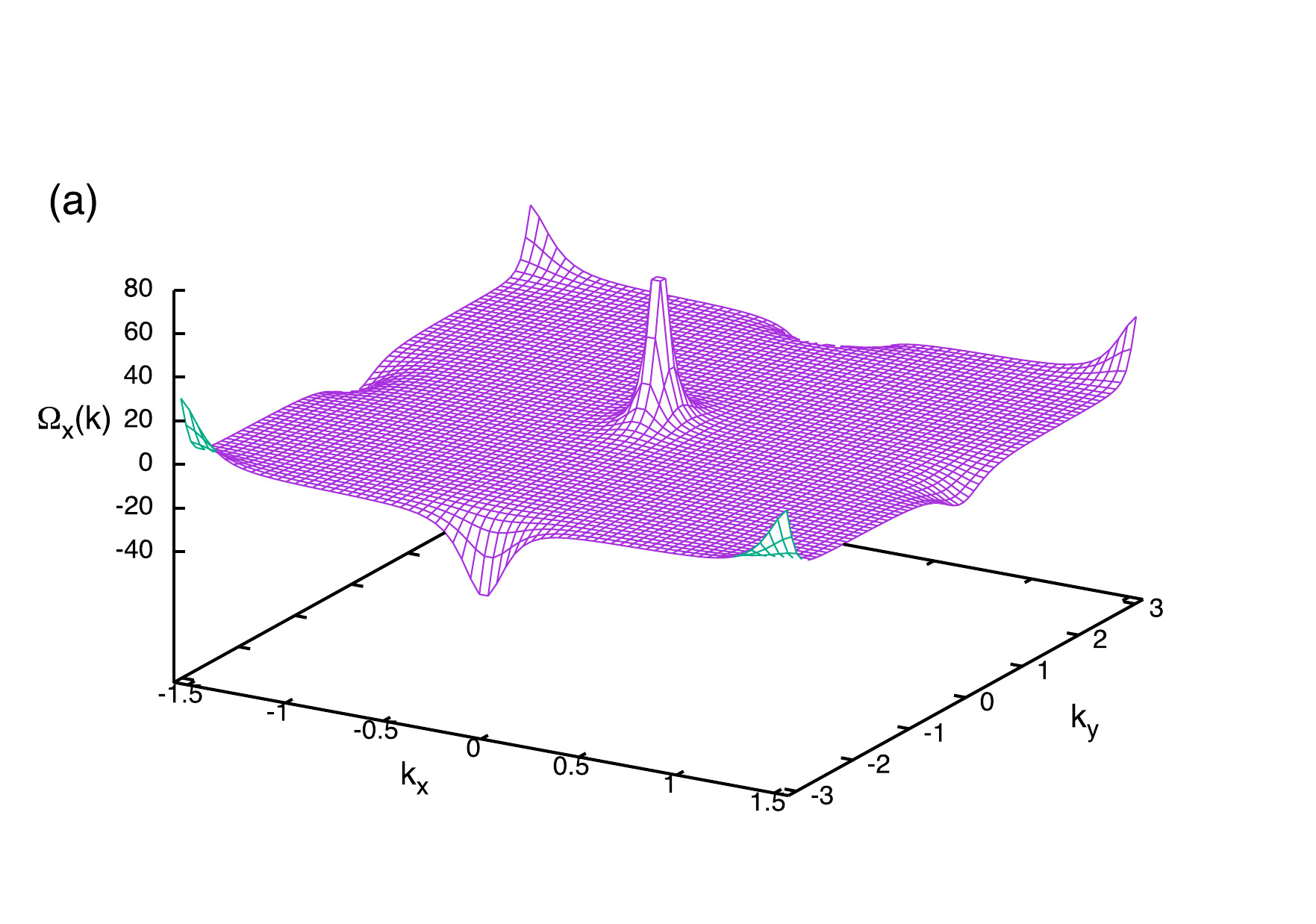}
                    \includegraphics[width=6.2cm]{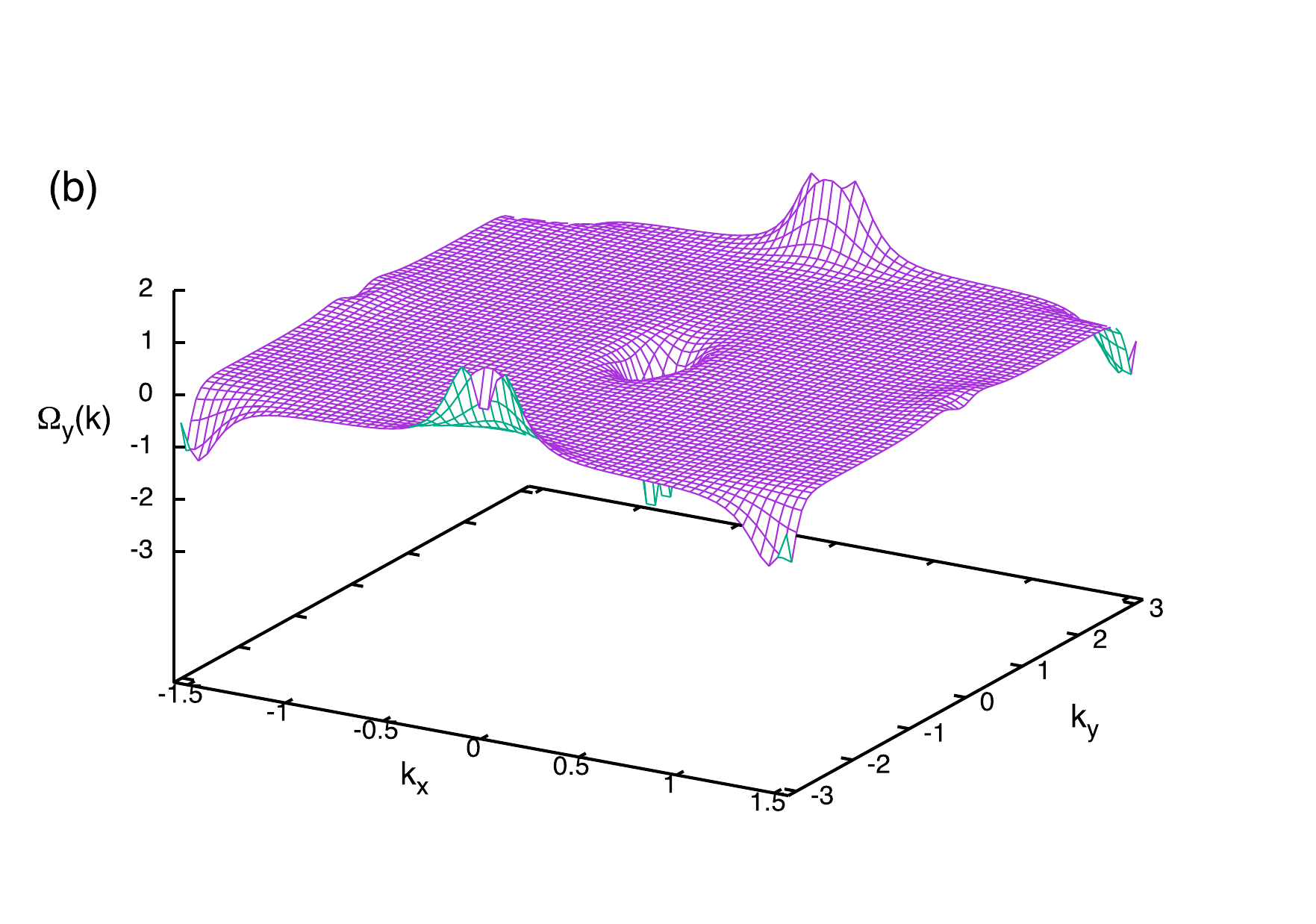}
                    \includegraphics[width=6.2cm]{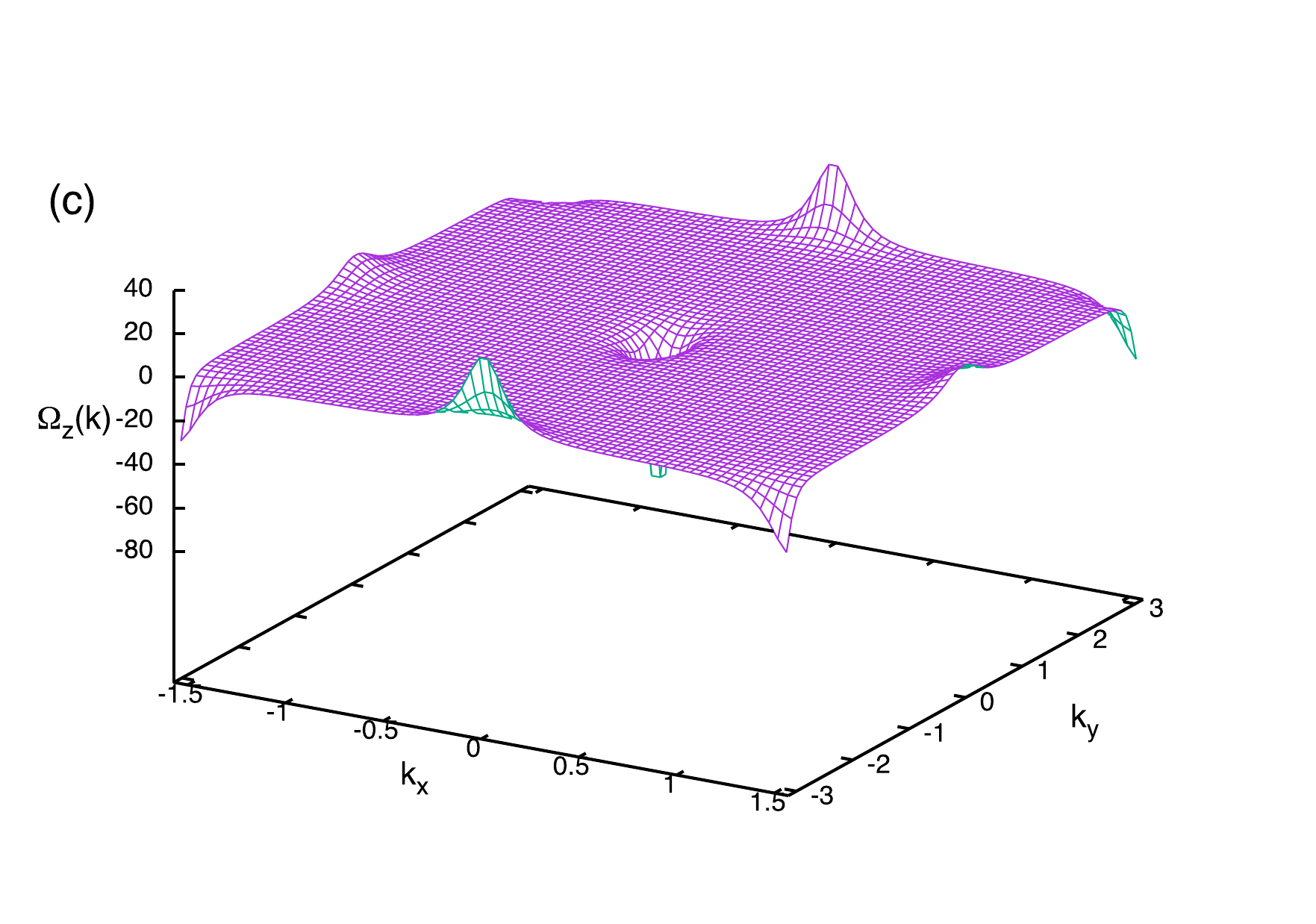}    
}
  \centerline{\includegraphics[width=6.2cm]{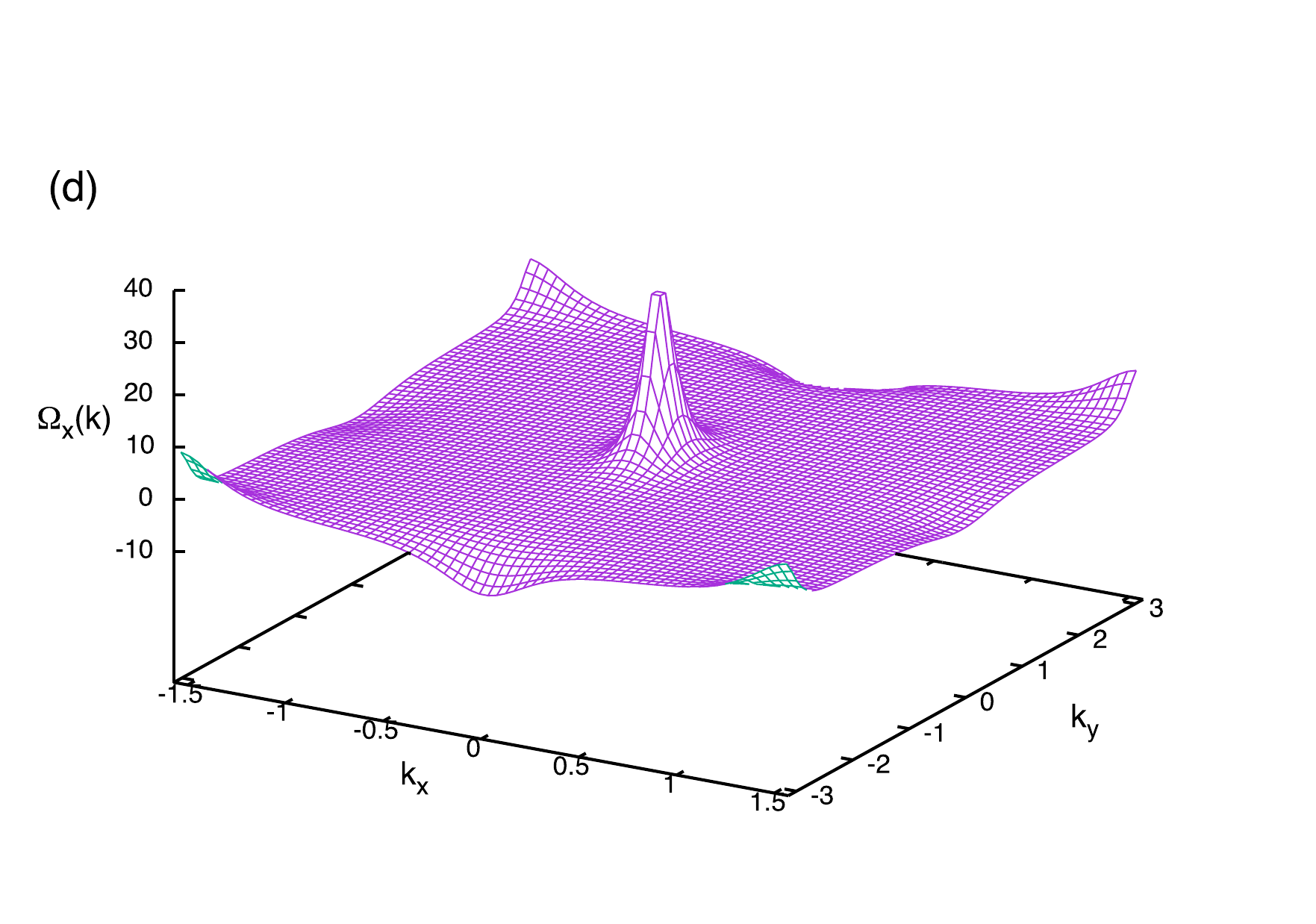}
                    \includegraphics[width=6.2cm]{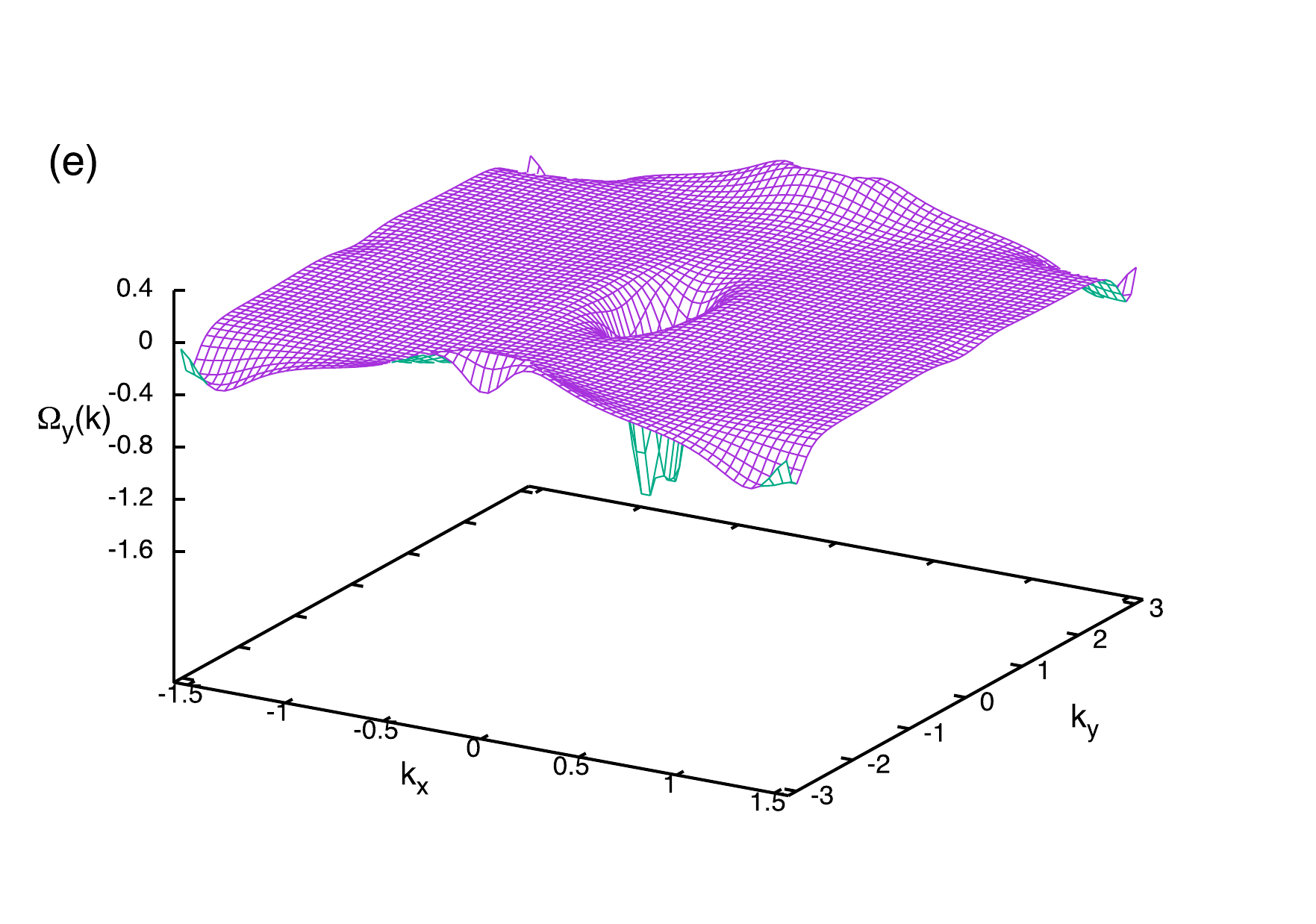}
                    \includegraphics[width=6.2cm]{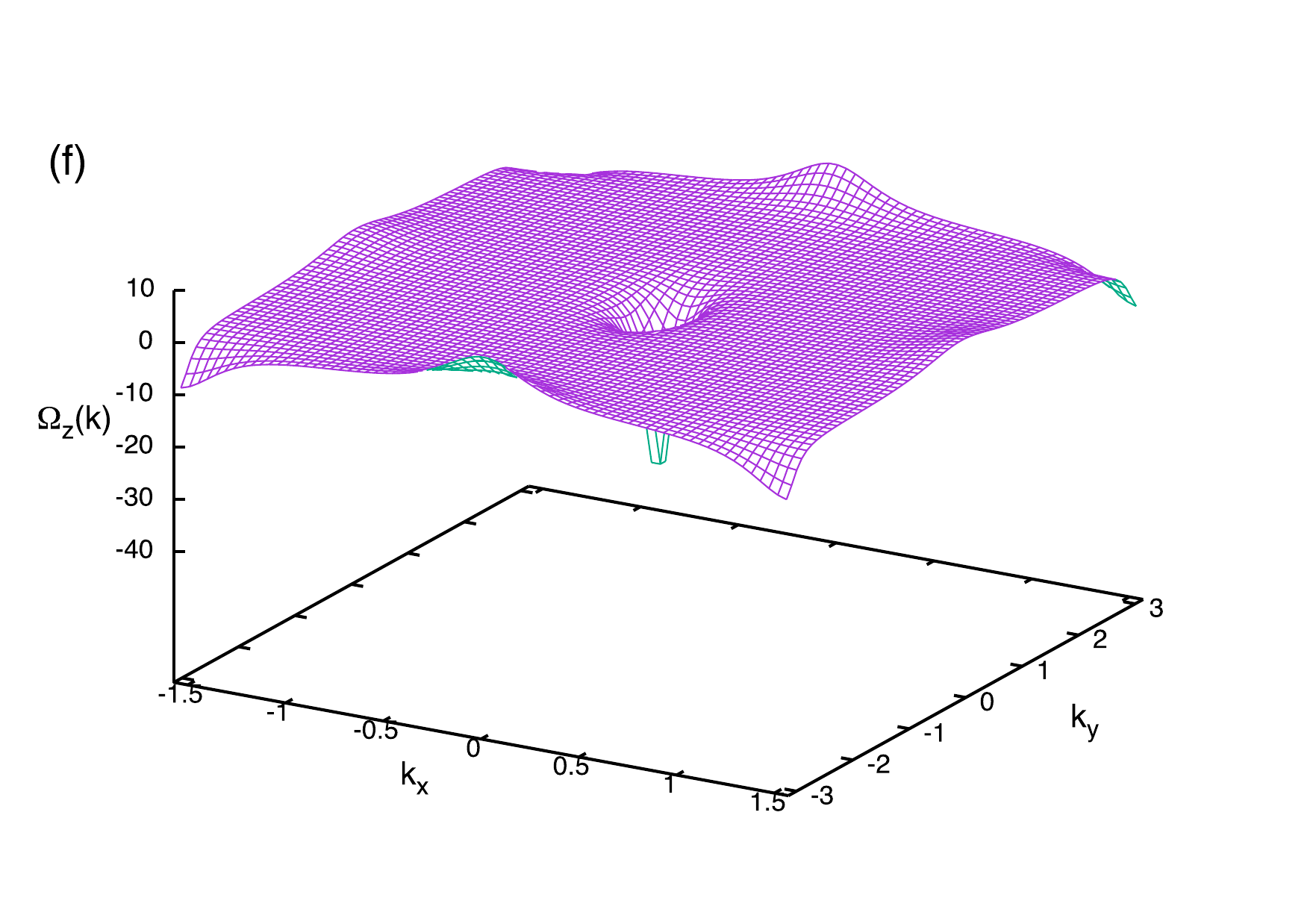} 
}                
\caption{Berry curvatures $\Omega_\alpha(\bk)$ [Eq.~\eqref{Berry}]
  of the triplon dispersion relations $\omega^\alpha_\bk$ shown in Fig.~\ref{fig:omega}: 
  (a) $\Omega_x(\bk)$, (b) $\Omega_y(\bk)$, and  (c) $\Omega_z(\bk)$ 
  for $J_2 = 0.48\, J_1$ and $D = 0.20\, J_1$;
  (d) $\Omega_x(\bk)$, (e) $\Omega_y(\bk)$, and  (f) $\Omega_z(\bk)$   
  for $J_2 = 0.52\, J_1$ and $D = 0.10\, J_1$.
  The external magnetic field $h_z = 0.05\, J_1$.  
}
\label{fig:berry}
\end{figure*}

The behaviour of the triplon gap $\Delta$ in terms of the DM
interaction $D$ for selected values of the exchange coupling $J_2$ is
shown in Fig.~\ref{fig:nzero}(d). One sees that the triplon gap
decreases with the DM interaction, regardless the value of the
exchange coupling $J_2$. The smallest value of the gap, 
$\Delta \sim 0.12 \, J_1$, corresponds to the parameter $D_{\rm max}$ above
which numerical solutions for the self-consistent equations
\eqref{eqs-self} are no longer found. 
Moreover, as illustrated in Fig.~\ref{fig:nzero}(e), the momentum $\bQ$ associated
with the triplon gap moves along the $\Gamma$-$\mathbf{Y}$ line of the
dimerized Brillouin zone: 
For $J_2 < 0.50 \, J_1$, the momentum $\bQ$
moves from the commensurate $\mathbf{Y}$ point to an incommensurate
value $\bQ = (0,Q_y)$ as the DM interaction increases; 
for $J_2 > 0.50 \, J_1$, the momentum $\bQ$ moves from the
commensurate $\Gamma$ point to an incommensurate 
momentum $\bQ = (0,Q_y)$ with the increasing of the parameter $D$.
Such set of results indicates that the black line in Fig.~\ref{fig:phasediag} 
may signal a quantum phase transition from a VBS phase to a noncollinear
long-range ordered magnetic phase with incommensurate ordering wave
vector $\bQ$.     
Due to the behaviour of the triplon gap $\Delta$ at the border of the
VBS region shown in Fig.~\ref{fig:phasediag}, at the moment, it is not
clear whether such a quantum phase transition is a first-order
transition or a continuous one; such an issue, which needs some
further studies, is beyond the scope of our paper.

Finally, it is interesting to compare our findings with the ones
determined by Merino and Ralko \cite{merino22} for the Heisenberg
model \eqref{ham} in the absence of an external magnetic field and
considering DM vectors associated with cuprate superconductors
(see also Sec.~\ref{sec:model} above). 
Based on a Majorana fermion representation for the spin operators, the
mean-field results indicate some evidence for a VBS phase within the
intermediate parameter region of the $J_1$-$J_2$ model, even in the
presence of a finite DM interaction, in agreement with our findings summarized
in Fig.~\ref{fig:phasediag}. 
Moreover, a continuous quantum phase transition from the VBS phase to
an magnetic ordered phase is observed as the DM coupling increases for
a fixed value of the next-nearest-neighbor exchange coupling $J_2$
(see Fig.~3 from Ref.~\cite{merino22}).

\section{Berry curvatures and Chern numbers}
\label{sec:berry}

In this section, we study the topological properties of the triplon
bands $\omega^\alpha_\bk$. The main idea is to check whether the DM
interaction \eqref{ham-dm}, with the DM vectors \eqref{dm-vectors},
yields a columnar VBS phase for the square lattice $J_1$-$J_2$ model \eqref{ham-j1j2}
with topologically nontrivial triplons, as previously found for a Heisenberg AFM
on a Shastry-Sutherland lattice \cite{penc15,malki17,coldea17}.

For quadratic bosonic Hamiltonians such as Eq.~\eqref{ham-harm-mat},
which includes the anomalous terms $t_{-\bk,\alpha} t_{\bk,\beta}$
and $t^\dagger_{-\bk,\alpha} t^\dagger_{\bk,\beta}$,
the Berry curvature $\Omega_\alpha(\bk)$ of the (triplon)
excitation band $\omega^\alpha_\bk$ is given by \cite{ryo13,ryo14}         
\begin{equation}
  \Omega_\alpha(\bk) = i \epsilon_{\mu \nu} \left[
      \hat{I}_B \frac{\partial \hat{T}_\bk^\dagger}{\partial k_{\mu}}
      \hat{I}_B \frac{\partial \hat{T}_\bk}{\partial k_{\nu}} \right]_{\alpha\alpha},
\label{Berry}
\end{equation}
with $\alpha = 1, 2, \ldots, 6$ and $\mu,\nu = x,y$.
Here $\hat{I}_B$ and $\hat{T}_\bk$ are, respectively, the $6 \times 6$
matrices \eqref{ib-hk-mat} and \eqref{Tmatrix1}, 
$\epsilon_{\mu \nu}$ is the completely antisymmetric
tensor with $\epsilon_{xy} = 1$, and
$[ \quad ]_{\alpha\alpha}$ indicates the diagonal element
of the corresponding square matrix. 
Due to the form of the diagonal Hamiltonian \eqref{ham-harm-mat-diag}, the Berry
curvatures of the $\omega^x_\bk$, $\omega^y_\bk$, and $\omega^z_\bk$
triplon bands are given by
$\Omega_1(\bk) \equiv \Omega_x(\bk)$,
$\Omega_2(\bk) \equiv \Omega_y(\bk)$, and
$\Omega_3(\bk) \equiv \Omega_z(\bk)$, respectively.
We calculate the Berry curvatures \eqref{Berry} of the triplon bands
$\omega^\alpha_\bk$ using the analytical expressions for the matrix elements of the
$\hat{T}_\bk$ matrix shown in Appendix~\ref{ap:diag}.

Figures~\ref{fig:berry}(a), (b), and (c) show the Berry curvatures
\eqref{Berry} respectively for the triplon bands 
$\omega^x_\bk$, $\omega^y_\bk$, and $\omega^z_\bk$ displayed in
Fig.~\ref{fig:omega}(c), which correspond to the parameters
$J_2 = 0.48\, J_1$, $D = 0.20\, J_1$, and $h_z = 0.05\, J_1$ of the model \eqref{ham}.
For the three triplon bands, one sees that the Berry curvatures vanish
for almost all points of the dimerized Brillouin zone [Fig.~\ref{fig:model}(d)],
except in the vicinity of the $\Gamma$, $\mathbf{X}$, $\mathbf{M}$, and $\mathbf{Y}$ points.
One notices that $\Omega_x(\bk) \simeq -\Omega_z(\bk)$, in addition to
the fact that the peak intensities for $\Omega_x(\bk)$ and $\Omega_z(\bk)$
are larger than the corresponding ones for $\Omega_y(\bk)$.
Importantly, one verifies that the peak intensities of the Berry curvatures
decrease as the DM interaction $D$ decreases and almost vanish
for $D = 10^{-3}\, J_1$
(the analytical expressions derived for the matrix elements of the
$\hat{T}_\bk$ matrix are not suitable for the case $D = 0$, see
Appendix~\ref{ap:diag} for details);
such a feature indicates that the DM interaction \eqref{ham-dm}, with the DM
vectors \eqref{dm-vectors}, indeed yields a finite Berry curvature for the triplons.  
Similar qualitatively features are found for the three triplon bands
shown in Fig.~\ref{fig:omega}(d), which are associated with the parameters 
$J_2 = 0.52\, J_1$, $D = 0.10\, J_1$, and $h_z = 0.05\, J_1$,
see Figs.~\ref{fig:berry}(d), (e), and (f).

\begin{table}[b]
\centering
\begin{small}
\caption{Chern numbers of the $\omega^x_\bk$, $\omega^y_\bk$, and $\omega^z_\bk$ 
  triplon bands of the columnar VBS phase of the model \eqref{ham} for
  $J_2 = 0.48$ and $J_2 = 0.52\, J_1$ and the DM interaction
  $D = 0.10$ and $0.20\, J_1$.
  The external magnetic field $h_z = 0.05\, J_1$.}
\label{table}
\begin{tabular}{c c c c c c c}
\hline
 &&&& $J_2 = 0.48\, J_1$ && $J_2 = 0.52\, J_1$ \\  
\hline
$D$ &$\quad$ &  $\alpha$ & $\quad$ & $C_\alpha$  &$\quad$ &  $C_\alpha$  \\
\hline
  $0.10$ && $x$ && $0$ &&  $0$\\
  $0.10$ && $y$ && $0$ &&  $0$\\
  $0.10$ && $z$ && $0$ &&  $0$\\
  $0.20$ && $x$ && $+2.08 \times 10^{-4}$ &&  $+1.07 \times 10^{-3}$\\
  $0.20$ && $y$ && $+2.09 \times 10^{-4}$ &&  $+0.88 \times 10^{-3}$\\
  $0.20$ && $z$ && $-4.18 \times 10^{-4}$ & & $-1.95 \times 10^{-3}$\\
\hline
\end{tabular}
\end{small}
\end{table}

Once the Berry curvatures $\Omega_\alpha(\bk)$ of the triplon bands
$\omega^\alpha_\bk$ are determined, we can calculate the
Chern number of each triplon band, which is defined as the integral of the
Berry curvature over the dimerized Brillouin zone \cite{ryo13,ryo14}, namely,           
\begin{equation}
  C_\alpha = \frac{1}{2\pi} \int_{BZ} d^2k \, \Omega_\alpha(\bk).
\label{chern}
\end{equation}
We determine the Chern numbers $C_\alpha$  
via a numerical integration of Eq.~\eqref{chern}.

The Chern numbers of the triplon bands for
exchange couplings $J_2 = 0.48$ and $0.52\, J_1$ and
DM interactions $D = 0.10$ and $0.20\, J_1$ are displayed in Table~\ref{table}.
One finds that, regardless the value of the exchange coupling $J_2$ and
the DM interaction $D$, the Chern numbers of the three triplon bands
vanish: such results are related to the symmetries of the Berry
curvatures, as exemplified in Fig.~\ref{fig:berry}.
Therefore, although the Berry curvatures are finite in some regions of the
dimerized Brillouin zone, the triplon excitations of the columnar VBS
phase of the Heisenberg model \eqref{ham} are topologically trivial.

It should be mention that our findings are similar to the ones
obtained in Ref.~\cite{rhine19} for a spin-liquid phase on a square lattice: 
The Berry curvatures of the (bosonic) spinon bands display finite values in some
regions of the Brillouin zone (see Figs.~7(e) and (f) from Ref.~\cite{rhine19}),
but the corresponding Chern numbers vanish.

\begin{figure*}[t]
\centerline{\includegraphics[width=8.0cm]{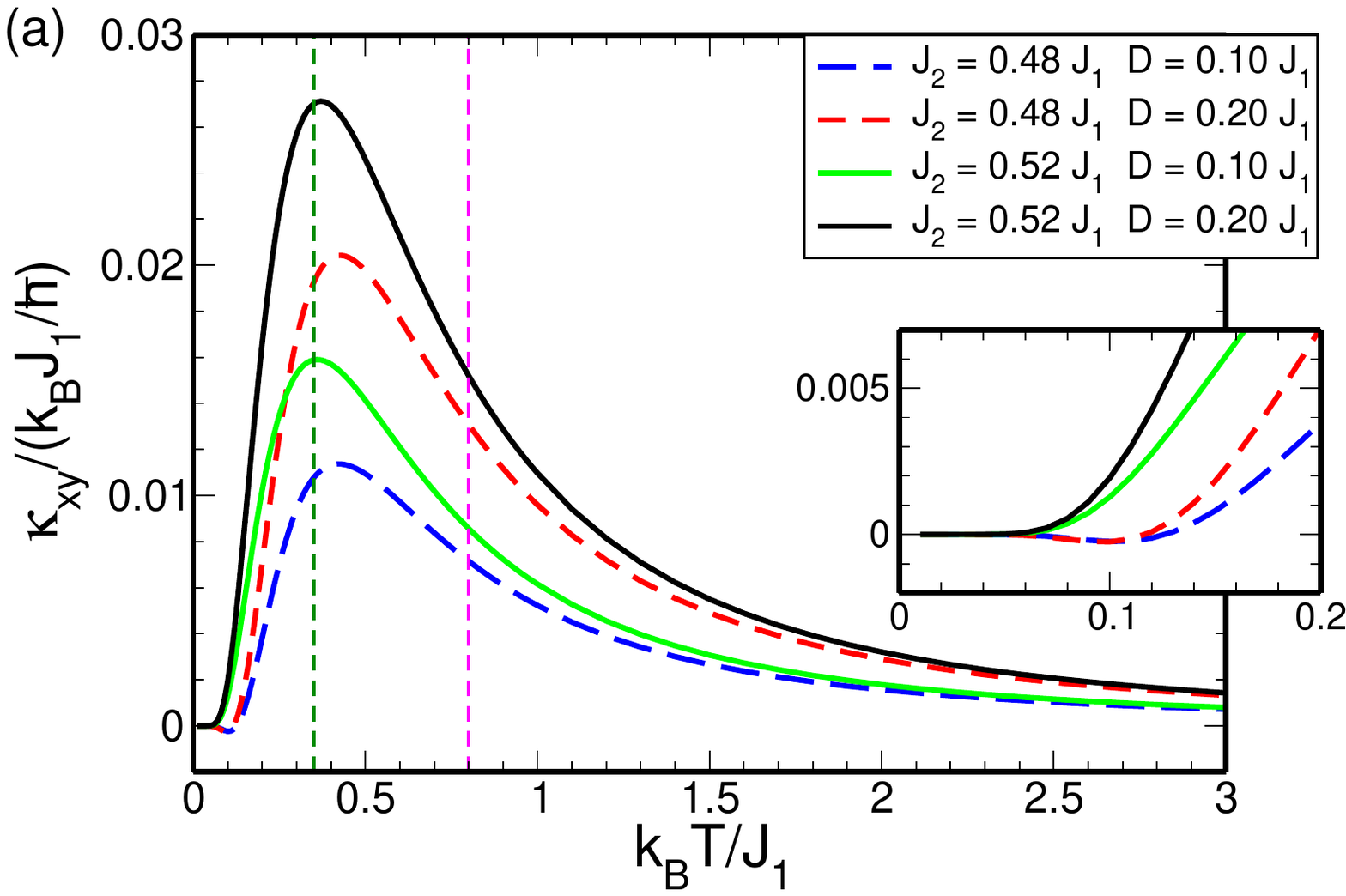}
                   \hskip1.5cm
                   \includegraphics[width=8.0cm]{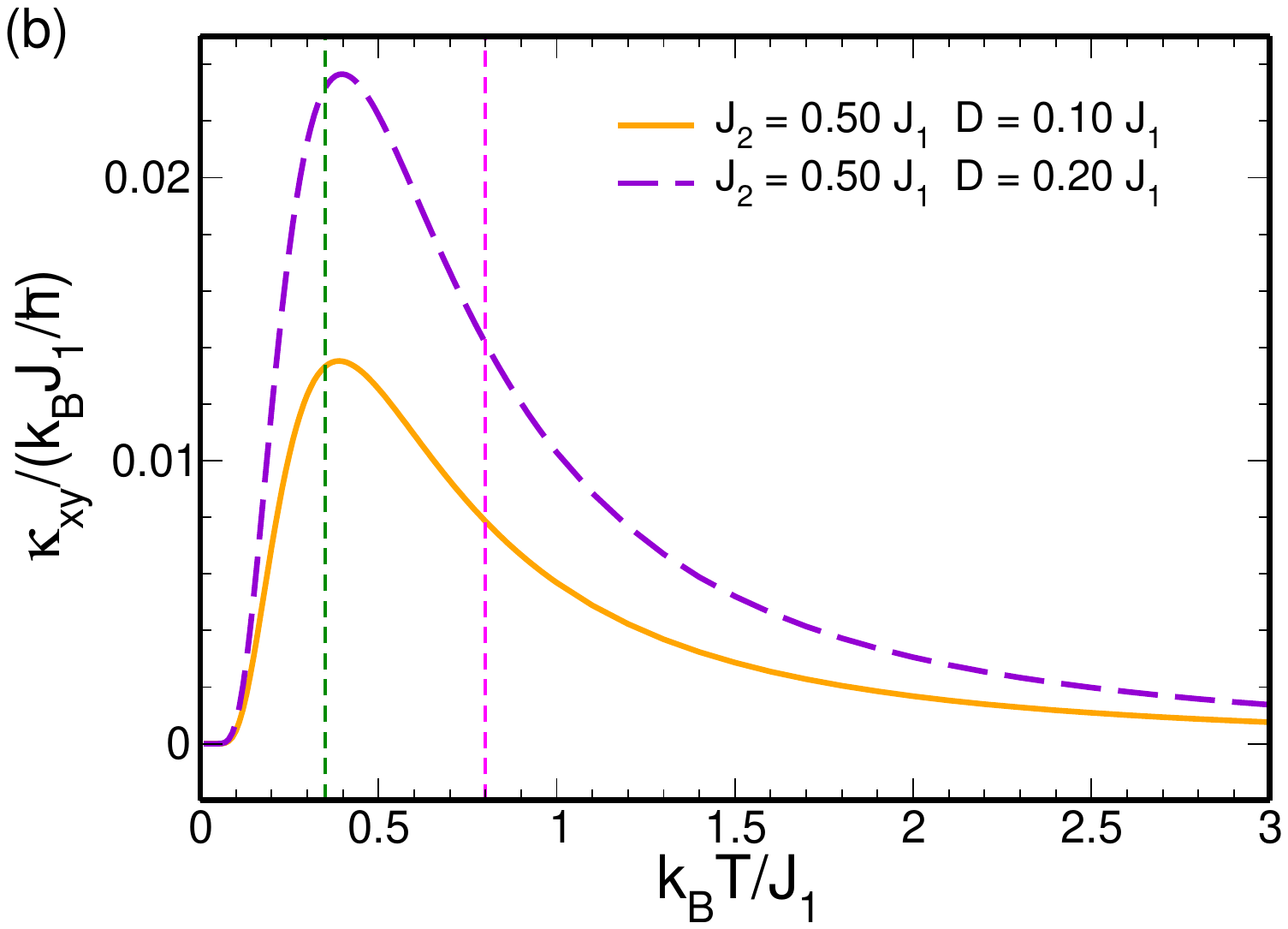}
}
\caption{Thermal Hall conductivity $\kappa_{xy}/(k_B J_1/\hbar)$ as
  a function of the temperature $k_BT/J_1$ for the columnar VBS phase of
  the Heisenberg model \eqref{ham}. Next-nearest-neighbor exchange
  and DM couplings:
  (a) $J_2 = 0.48\, J_1$ and $D = 0.10\, J_1$ (dashed blue line), 
       $J_2 = 0.48\, J_1$ and $D = 0.20\, J_1$ (dashed red line),
       $J_2 = 0.52\, J_1$ and $D = 0.10\, J_1$ (solid green line), and
       $J_2 = 0.52\, J_1$ and $D = 0.20\, J_1$ (solid black line);
  (b) $J_2 = 0.50\, J_1$ and $D = 0.10\, J_1$ (solid orange line) and
       $J_2 = 0.50\, J_1$ and $D = 0.20\, J_1$ (dashed violet line).
  The external magnetic field $h_z = 0.05\, J_1$.
  The vertical dark-green dashed line indicates the temperature above
  which triplon-triplon interactions might be relevant;
  the vertical magenta dashed line is an estimate for the critical
  temperature $T_c$ above which the VBS phase is no longer stable
  (see text for details).
  Inset panel (a): low-temperature behaviour of the thermal Hall conductivity.
}
\label{fig:kappa}
\end{figure*}

\section{Thermal Hall conductivity}
\label{sec:kappa}

We now focus on the transport properties of the columnar
VBS phase of the Heisenberg model \eqref{ham}, in particular, we
determine the thermal Hall conductivity $\kappa_{xy}$ due to triplons. 
We follow the procedure proposed in Refs.~\cite{prl-ryo11,prb-ryo11},
where the thermal Hall conductivity $\kappa_{xy}$ due to the magnon
excitations of a long-range ordered ferromagnet was determined;
such a formalism was also applied to study the triplon thermal Hall
effect in a AFM on a Shastry-Sutherland lattice \cite{penc15,malki17}.

For a system of noniteracting boson excitations, the thermal Hall
conductivity is given by \cite{prl-ryo11,prb-ryo11} 
\begin{equation}
  \kappa_{xy} = - \frac{k_B^2 T}{\hbar V} \sum_{\bk,\alpha}
                     c_2\left[ f_B \left( \omega^\alpha_\bk \right)  \right] \Omega_\alpha(\bk),
\label{kappa}
\end{equation}
where $k_B$ is the Boltzmann constant, $T$ is the temperature,
\begin{equation}
  f_B(\epsilon) = \frac{1}{e^{\epsilon/k_BT} - 1}
\label{bose}
\end{equation}
is the Bose distribution function, and
$\Omega_\alpha(\bk)$ is the Berry curvature \eqref{Berry} of the
(bosonic) triplon excitation band $\omega^\alpha_\bk$, with $\alpha =x,y,z$.
Moreover, the function $c_2(x)$ assumes the form
\begin{align}
  c_2(x) =& \int_{0}^x dt \left(\ln \frac{1+t}{t} \right)^2
\nonumber \\
            =& \left( 1 + x \right) \left(\ln \frac{1+x}{x} \right)^2
               - \left( \ln x \right)^2 - 2 {\rm Li}_2(-x),
\label{c2eq}
\end{align}
with ${\rm Li}_2(x)$ being the polylogarithm, which is defined as
${\rm Li}_n(z) = \sum_{j=1}^\infty z^j/j^n$;
one shows that $c_2(x)$ is a monotonic function with
$c_2(0) = 0$ and $c_2(x\rightarrow \infty) = \pi^2/3$. 
We consider the Berry curvatures calculated in Sec.~\ref{sec:berry}
and determine $\kappa_{xy}$ via a numerical integration of Eq.~\eqref{kappa}.

\begin{figure}[b] 
\centerline{\includegraphics[width=7.0cm]{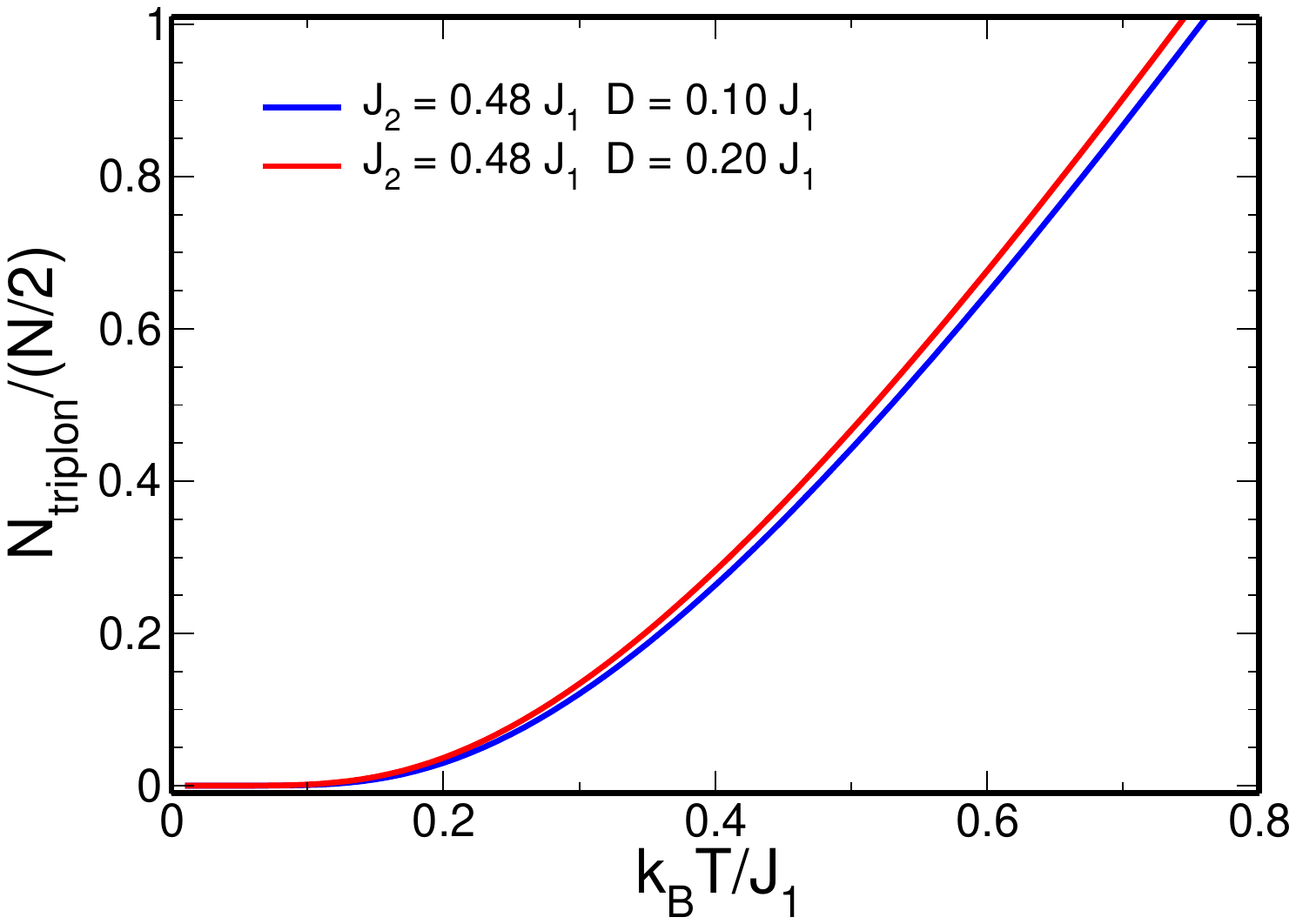}
}
\caption{Number of triplons \eqref{n-triplons} as a function of the temperature
  $k_BT/J_1$ for the columnar VBS phase of the Heisenberg model
  \eqref{ham} with $J_2 = 0.48\, J_1$ and 
  DM couplings $D = 0.10$ and $0.20\, J_1$. 
  The external magnetic field $h_z = 0.05\, J_1$.
  $N/2$ is the number of sites of the dimerized lattice $\mathcal{D}$
  as illustrated in Fig.~\ref{fig:model}(c).
}
\label{fig:nt}
\end{figure}

The behaviour of the thermal Hall conductivity \eqref{kappa} as a
function of the temperature $T$ for $J_2 = 0.48$, $0.50$, and $0.52\, J_1$,
DM interaction $D = 0.10$ and $0.20\, J_1$,  and
external magnetic field $h_z = 0.05\, J_1$
is shown in Fig.~\ref{fig:kappa}. 
One sees that, regardless the values of the exchange coupling $J_2$ and
DM interaction $D$, the thermal Hall conductivities display a peak at 
$T \sim 0.5\, J_1/k_B$, whose height increases with the DM interaction for a
fixed value of the exchange coupling $J_2$.
Moreover, one notices that $\kappa_{xy}$ vanishes
in the low-temperature region, a feature related to the existence of a
finite triplon excitation gap, see Fig.~\ref{fig:nzero}(d). 
Interestingly, only for $J_2 = 0.48\, J_1$, $\kappa_{xy}$ assumes
negative values around $T \sim 0.1\, J_1/k_B$ [see the inset of Fig.~\ref{fig:kappa}(a)]; 
indeed, for $J_2 < 0.50\, J_1$, we find a sign reversal of the thermal Hall conductivity as $T$ varies
in this low-temperature region.
Apart from the sign reversal in the low-temperature region and the
fact that the triplons are topologically trivial, the behaviour of
$\kappa_{xy}$ with the temperature $T$ qualitatively agrees with the 
phenomenological one discussed in Ref.~\cite{yang20}.

A few remarks here about the temperature range below which our results
for the thermal Hall conductivity might be valid are in order:
The number of triplons as a function of the temperature is given by
\begin{equation}
  N_{\rm triplons} =  \sum_{\bk,\alpha} \langle b^\dagger_{\bk,\alpha} b_{\bk,\alpha} \rangle  
                          = \sum_{\bk,\alpha}  f_B \left( \omega^\alpha_\bk \right),
\label{n-triplons}
\end{equation}
with $f_B(\epsilon)$ being the Bose distribution function \eqref{bose}.
Figure~\ref{fig:nt} shows the behaviour of $N_{\rm triplons}$ with the temperature $T$ 
for $J_2 = 0.48\, J_1$, DM interaction $D = 0.10$ and $0.20\, J_1$,  and
external magnetic field $h_z = 0.05\, J_1$;
we find a quite similar quantitative behaviour for $J_2 = 0.50$ and $0.52\, J_1$.
One expects that, for $ T  \gtrsim 0.35 \, k_B/J_1$
(dark-green dashed line in Fig.~\ref{fig:kappa}),
the cubic terms \eqref{h3} and \eqref{hdm3} as well as
the quartic terms \eqref{h4} and \eqref{hdm4}, which are neglected
in our study, might be relevant, and therefore, they could provide
important corrections to the thermal Hall conductivity.
Furthermore, for $T \gtrsim 0.8 \, k_B/J_1$,
one sees that the number of triplons
$N_{\rm triplons}$ is larger than the number of sites $N' = N/2$ of the
dimerized lattice $\mathcal{D}$;  
such a feature indicates that, for $T \sim 0.8 \, k_B/J_1$
(magenta dashed line in Fig.~\ref{fig:kappa}),  
the system might be close to the critical temperature $T_c$ above
which the columnar VBS phase is no longer stable.
We intend to investigate the effects of the triplon-triplon
interaction on the thermal Hall conductivity in a future publication;
such a study will also be important for a proper determination of the
critical temperature $T_c$.

\section{Summary and discussion}
\label{sec:summary}

In spite of the fact that the triplon excitations are topologically trivial,
the columnar VBS phase of the square-lattice AFM Heisenberg model \eqref{ham}
is characterized by a finite thermal Hall conductivity. Such a feature is indeed a
thermal effect: Due to the weight function $c_2(x)$ [Eq.~\eqref{c2eq}],
the integrand of Eq.~\eqref{kappa} vanishes in the vicinity of the
$\mathbf{X}$ point of the dimerized Brillouin zone (not shown here),
implying a finite thermal Hall conductivity;
recall that the Berry curvatures of the triplon bands are finite around the  
$\mathbf{X}$ point of the dimerized Brillouin zone (see Fig.~\ref{fig:berry}),
an important contribution that yields topologically trivial triplon
excitations;
as pointed out in Ref.~\cite{rhine19}, a VBS phase with topologically
nontrivial triplons might be characterized by a thermal Hall
conductivity qualitatively similar to the one found for the columnar
VBS phase of the model \eqref{ham}, but with a higher peak at low
temperatures.

The fact that the topological properties of the triplon bands
enhance the thermal Hall conductivity was indeed observed in
Refs.~\cite{penc15,malki17}, where a Shastry-Sutherland AFM model
relevant for the SrCu$_2$(BO$_3$)$_2$ compound  is considered: 
Here, it is found that, for an external magnetic field $| h_z | < h_{z,c}$, the
Chern numbers of the triplon bands are finite; 
the behaviour of the thermal Hall conductivity as function of the
magnetic field $h_z$ for a fixed temperature $T$ indicates that the
largest values of $\kappa_{xy}$ are within the interval $| h_z | < h_{z,c}$ 
(see, e.g, Fig.~10 from Ref.~\cite{malki17});
interestingly, the thermal Hall conductivity is finite (although
smaller) even for $| h_z | > h_{z,c}$, where the triplon bands are
topologically trivial.
Concerning the AFM Heisenberg model \eqref{ham}, 
a critical magnetic field $h_{z,c}$, where the gaps between the three
triplon bands close and above which the triplon bands are
topologically nontrivial, is not observed.

The fact that a finite thermal Hall conductivity is found for the
columnar VBS phase of the square-lattice AFM Heisenberg model \eqref{ham} 
indicates that the {\sl no-go} condition \cite{lee10}, which was
determined for a long-range ordered magnet and is based on
linear spin-wave theory results, does not apply to the columnar VBS phase.
In fact, as mentioned in the Introduction, exceptions to the {\sl no-go}
condition \cite{lee10} were previously reported in Refs.~\cite{hotta19,rhine19}.  
Again, for the spin-liquid phase on a
square lattice discussed in Ref.~\cite{rhine19},
a finite thermal Hall effect due to spinons was found when
the DM vector pattern associated with the YBCO compound was considered
and, in spite of the fact that $\kappa_{xy}$ is finite, the
spin-liquid phase is characterized by topologically trivial (bosonic)
spinon excitations. 
One sees that the features found here for the columnar VBS phase are
quite similar to the ones observed for the spin-liquid phase \cite{rhine19}. 
Therefore, it seems that the DM interaction corresponding to the YBCO
compound yields square-lattice quantum paramagnetic phases, both
spin-liquid and VBS ones, characterized by a finite thermal Hall
conductivity $\kappa_{xy}$ and topologically trivial elementary
excitations.  
In order to confirm such an issue, it would be interesting
to determine the Chern numbers of the excitation bands and
$\kappa_{xy}$ for the model \eqref{ham}, but now considering both
gapless and gapped spin-liquid phases proposed as 
ground-state candidates for the $J_1$-$J_2$ model 
(see Sec.~\ref{sec:model} for details).
Indeed, it would also be interesting to provide in the near future a discussion
concerning the general conditions for the observation of a thermal
Hall effect in quantum paramagnetic phases.

A sign reversal of the thermal Hall conductivity with the increasing
of the temperature $T$, from negative to positive values as we found for $J_2 < 0.50 \, J_1$  
in the low-temperature region, was also observed for ordered
ferromagnets on the Shastry-Sutherland \cite{malki19} and
honeycomb \cite{wu21} lattices.
In particular, for a ferromagnet on the honeycomb lattice \cite{wu21},
it was pointed out that such a sign change indicates a topological phase
transition, since a gap closure between the upper and lower magnon
bands is also observed as $T$ varies, an effect due to magnon-magnon
interactions.
Such a similar gap closure is not observed in our case.
At the moment, the sign reversal of $\kappa_{xy}$ for the columnar VBS
phase of the model \eqref{ham} with $J_2 < 0.50 \, J_1$ is not
completely understood;
similarly to the phase diagram shown in Fig.~\ref{fig:phasediag}, the
low-temperature behaviour of $\kappa_{xy}$ seems to indicate a distinction between
the parameter regions $J_2 < 0.5\, J_1$ and $J_2 > 0.5 \, J_1$ of the 
$J_1$-$J_2$ model, which qualitatively agrees with
Refs.~\cite{gong14,imada15,wang18,nomura20,ferrari20,liu22};
it is worth mentioning that one of us also find such a distinction in the
study \cite{doretto20} concerning many-triplon states in the columnar
VBS phase of the  $J_1$-$J_2$ model.

In summary, in this paper we studied the spin-$1/2$ $J_1$-$J_2$ AFM
Heisenberg model on a square lattice with an additional DM interaction
between the spins and in the presence of an external magnetic
field. We focused on the columnar VBS phase, which is stable within the
intermediate parameter region of the $J_1$-$J_2$ model.
In particular, we discussed the topological properties of the triplon
excitations of the columnar VBS phase, which is described via an
effective interacting boson model obtained with the aid of a
bond-operator formalism for the spin operators.  
We considered the interacting boson model within a harmonic
approximation and determined the triplon excitation bands. We found
that the DM interaction provides a finite Berry curvature for the
triplon bands, but the corresponding Chern numbers vanish. Although
the triplon excitations are topologically trivial, we found that the
thermal Hall conductivity of the columnar VBS phase is indeed finite.

Finally, we hope that our results may motivate the search for VBS
phases in (dimerized) square-lattice AFMs characterized by a finite
triplon thermal Hall effect.
Indeed, although a thermal Hall effect due to triplons has been predicted for the
Shastry-Sutherland compound SrCu$_2$(BO$_3$)$_2$ \cite{penc15,malki17},
some recently measurements do not observe discernible values for the
thermal Hall conductivity within the experimental resolution \cite{matsuda22}.

\acknowledgments

We thank Eric Andrade and E. Miranda for helpful discussions.
L.S.B. kindly acknowledges the financial support of the
Coordena\c{c}\~ao de Aperfei\c{c}oamento 
de Pessoal de N\'ivel Superior - Brasil (CAPES) - Finance Code 001.

\appendix

\section{Effective boson model in real space}
\label{ap:details-boso}

In this section, we provide the expressions of the Hamiltonians
\eqref{h-boson1}, \eqref{h-boson2}, and \eqref{ham-B-dimer} 
in terms of the singlet $s_i$ and triplet $t_{i \alpha}$ boson operators.

Let us first consider the $J_1$-$J_2$ AFM Heisenberg model \eqref{ham-j1j2}.
After substituting the generalized bond-operator expansion \eqref{spin-bondop}
of the spin operators in terms of the singlet and triplet operators,
one arrives at the Hamiltonian \eqref{h-boson1}, where the
$\mathcal{H}_{J,n}$ terms read
\begin{align}	
    \mathcal{H}_{J,0} =& - \frac{3}{4} J_1 \sum_i s_i^\dagger s_i , 
\nonumber \\
\nonumber \\  
    \mathcal{H}_{J,2} =& \, \frac{J_1}{4} \sum_i t_{i \alpha  }^{\dagger} t_{i \alpha } 
             + \frac{1}{4} \sum_{i, \tau} \zeta_2(\tau) 
                 \left( s_i s_{i+\tau}^\dagger  t_{i \alpha}^ \dagger t_{i+\tau \alpha}  + {\rm H.c.}  \right.
\nonumber \\
                  & \left. + \; s_i^\dagger s_{i+\tau}^\dagger   t_{i\alpha} t_{i+\tau  \alpha} + {\rm H.c.} \right) ,
\nonumber \\
\label{h-boson1ap} \\  
    \mathcal{H}_{J,3} =&  \, \frac{i}{4}    \epsilon_{\alpha \beta \lambda} \sum_{i,\tau }\zeta_3(\tau)\Big[  
                  \left( s_i^\dagger   t_{i \alpha} + t_{i \alpha}^\dagger s_i \right )
                           t_{i + \tau \beta}^\dagger  t_{i + \tau  \lambda}   \Big.
\nonumber \\
                        &\Big.   - \; (i \leftrightarrow i+\tau) \Big] ,
\nonumber \\
\nonumber \\  
   \mathcal{H}_{J,4} =& -\frac{1}{4} \epsilon_{\alpha \beta \lambda} \: \epsilon_{\alpha \mu \nu}
            \sum_{i,\tau} \zeta_4(\tau)t_{i \beta}^\dagger  t_{i+\tau \mu}^\dagger   t_{i \lambda}  t_{i+\tau \nu}.
\nonumber
\end{align}
Here the summation convention over repeated greek indices is implied,
the $\zeta_i(\tau)$ functions are defined as
\begin{eqnarray}
  \zeta_2(\tau) &=& 2(J_1 - J_2)\delta_{\tau, 2} - J_1\delta_{\tau, 1} 
                         - J_2\left( \delta_{\tau, 1+2} + \delta_{\tau,1-2} \right),
\nonumber \\
  \zeta_3(\tau) &=& J_1\delta_{\tau,1} + J_2\left( \delta_{\tau, 1+2} + \delta_{\tau,1-2} \right),
\nonumber \\
  \zeta_4(\tau) &=& 2(J_1 + J_2)\delta_{\tau, 2} + J_1\delta_{\tau, 1} 
                         + J_2\left( \delta_{\tau, 1+2} + \delta_{\tau,1-2} \right),
\nonumber 
\end{eqnarray}
and the index $\tau$ indicates the dimer nearest-neighbor vectors
$\taub_n$ [see Eq.~\eqref{tau-col}]. 
It should be mentioned that, in order to derive the Hamiltonian
\eqref{h-boson1}, it is convenient to consider the identity
\begin{equation}
  \bS^1\cdot\bS^2 = -\frac{3}{4}s^\dagger s  + \frac{1}{4} \sum_\alpha  t^\dagger_\alpha t_\alpha
\label{local-bond-op1}
\end{equation}
for the local term $J_1 ( \bS^1_i\cdot\bS^2_i )$  of the Hamiltonian
\eqref{ham12-dimer} and the expansion \eqref{spin-bondop} for the
nonlocal ones.

Concerning the DM interaction \eqref{ham-dm}, the local term 
of the Hamiltonian \eqref{ham-dm-dimer}, 
$\bD_{i,i}\cdot( \bS_i^1 \times \bS_i^2 )$,  
can be easily treated with the aid of the identity
\begin{equation}
  \bD\cdot\left( \bS^1 \times \bS^2 \right) = \frac{1}{2} i  D_\alpha s^\dagger t_\alpha  + {\rm H.c.},
\label{local-bond-op2}
\end{equation}  
which can be obtained following the same procedure 
that yields the identity \eqref{local-bond-op1}. 
Similarly to the $J_1$-$J_2$ model \eqref{ham-j1j2}, one employs the
expansion \eqref{spin-bondop} for the nonlocal terms of the Hamiltonian 
\eqref{ham-dm-dimer}, and shows that the four terms of the effective
boson model \eqref{h-boson2} assume the form
\begin{align}
 \mathcal{H}_{\rm DM,1} =& \frac{1}{2} i D s_i^\dagger t_{i x}  + {\rm H.c.},                       
\nonumber \\
\nonumber \\
  \mathcal{H}_{\rm DM,2}  =&-\frac{1}{4} D \, \epsilon_{x\beta\gamma} \sum_{i \in \mathcal{D}} 
        \left(        s^\dagger_{i+1} s_i t_{i\beta}^\dagger t_{i+1,\gamma} + {\rm H.c.}  \right.
\nonumber \\
         &+ \left.  s_i^\dagger s_{i+1}^\dagger t_{i\beta} t_{i+1,\gamma} + {\rm H.c.}   \right)  
\nonumber \\
  &-\frac{1}{2} D \, \epsilon_{y\beta\gamma} \sum_{i \in \mathcal{D}} 
        \left(  s^\dagger_{i+2} s_i t_{i\beta}^\dagger t_{i+2,\gamma} + {\rm H.c.} \right.
\nonumber \\
         &+ \left.  s_i^\dagger s_{i+2}^\dagger t_{i\beta} t_{i+2,\gamma} + {\rm H.c.}    \right), 
\nonumber \\
\label{h-boson2ap}\\
  \mathcal{H}_{\rm DM,3} =& \frac{1}{4} i D \, \epsilon_{x\beta\gamma} \sum_{i \in \mathcal{D}}                       
  \epsilon_{\gamma\mu\nu} \left( s_{i}^\dagger t_{i\beta} +  t^\dagger_{i\beta} s_i \right)  t_{i+1,\mu}^\dagger t_{i+1,\nu}
\nonumber \\
    &- \epsilon_{\beta\mu\nu} \, t_{i\mu}^\dagger t_{i\nu} \left( s_{i+1}^\dagger t_{i+1,\gamma} + t^\dagger_{i+1,\gamma}s_{i+1}\right), 
\nonumber \\
\nonumber \\
\mathcal{H}_{\rm DM,4} =& -D \, \epsilon_{\beta\mu\nu} \epsilon_{\gamma\mu'\nu'} \sum_{i \in \mathcal{D}}
     \left( \frac{1}{4} \, \epsilon_{x\beta\gamma}\, t_{i\mu}^\dagger  t_{i\nu} t_{i+1,\mu'}^\dagger t_{i+1,\nu'}  \right. 
\nonumber \\
  &- \left. \frac{1}{2} \, \epsilon_{y\beta\gamma} \, t_{i\mu}^\dagger t_{i\nu} t_{i+2,\mu'}^\dagger t_{i+2,\nu'} \right).  
\nonumber
\end{align}

Finally, one easily shows that, in terms of the singlet and triplet
operators, the Zeeman term \eqref{ham-B-dimer} is given by
\begin{equation}
  \mathcal{H}_{B} = i \epsilon_{\alpha\beta\gamma}\sum_{i \in \mathcal{D}}
                               h_\alpha \, t_{i \beta}^\dagger t_{i \gamma},
\label{h-boson3ap}   
\end{equation}
where $h_\alpha$, with $\alpha = x,y,z$, is the $\alpha$-component of
the external magnetic field $\bh = g\mu_B\bB$.

\section{Linear triplet term of the effective boson model}
\label{ap:linear-term}

In this section, we provide some details of the procedure employed to
treat the linear triplet term $\mathcal{H}_{\rm DM,1}$ 
of the effective boson model [see Eq.~\eqref{h-boson2ap}].
In particular, we adopted the treatment described in
Ref.~\cite{coldea17}. 

As mentioned in Sec.~\ref{sec:boson-model}, the linear term
$\mathcal{H}_{\rm DM,1}$
can be removed via an unitary transformation performed in each site of the
dimerized lattice $\mathcal{D}$. 
Since $\mathcal{H}_{\rm DM,1}$ couples the singlet $s_i$ and the triplet
$t_{i x}$ operators, we consider the Hamiltonian 
\begin{align}
  H_L = &\, \mathcal{H}_{J,0} + \mathcal{H'}_{J,2} + \mathcal{H}_{\rm DM,1}
\nonumber \\
  = & \, J_1\sum_{i} \left( -\frac{3}{4}  s_i^\dagger s_i + \frac{1}{4}  t_{i x}^\dagger t_{i x}
           + i\frac{D}{2J_1} s_{i}^\dagger t_{ix}  + {\rm H.c.} \right),
\label{ham-HL}
\end{align}
which includes the local terms $\mathcal{H}_{J,0}$ and
$\mathcal{H'}_{J,2}$ related to the $J_1$-$J_2$ model [see Eq.~\eqref{h-boson1ap}]
and the local one $\mathcal{H}_{\rm DM,1}$ 
derived from the DM interaction [see Eq.~\eqref{h-boson2ap}].

In matrix form, the Hamiltonian \eqref{ham-HL} reads
\begin{equation}
   H_L    =   J_1 \sum_{i \in \mathcal{D}}
           \begin{pmatrix}
                s_i^\dagger & t_{ix}^\dagger  \end{pmatrix}
           \begin{pmatrix}
               -\frac{3}{4} & i \alpha \\
                 - i \alpha & \frac{1}{4}  \end{pmatrix}
            \begin{pmatrix}
                  s_i \\
                  t_{ix} \end{pmatrix},
\label{ham-HL-matrix}
\end{equation}
with the parameter $\alpha = D/(2J_1)$ defined in terms of the DM
interaction $D$ and the nearest-neighbor exchange coupling $J_1$.
It is quite easy to show that the Hamiltonian \eqref{ham-HL-matrix}
can be diagonalized by an unitary transformation defined by the
unitary matrix
\begin{equation}
    U = \frac{1}{ \sqrt{ 2 \left( b^2 - b \right) }}  \begin{pmatrix}
            2\alpha   &   i(1 - b)   \\
            i (1 - b)   &    2\alpha 
          \end{pmatrix},
\label{u-mat}
\end{equation}
with $b = (1 + 4 \alpha^2)^{1/2} $. 
Assuming that $\alpha \ll 1$, one finds that 
$ b^2 - b \approx 2 \alpha^2 $ and $ 1 - b \approx  -2 \alpha^2 $,
and therefore, up to linear order in the parameter $\alpha$, the
unitary matrix \eqref{u-mat} reads
\begin{equation}
   U  \approx \begin{pmatrix}
                      1 & -i \alpha \\
                      -i \alpha & 1\end{pmatrix}.
\label{u-mat2}
\end{equation}
With the aid of the Eq.~\eqref{u-mat2}, one defines a new set of 
singlet $\tilde{s}_i$ and triplet $\tilde{t}_{i \alpha}$ boson operators, 
\begin{equation}
 \begin{pmatrix}
     \tilde{s}_i \\  \tilde{t}_{i x} \\ \tilde{t}_{i y} \\ \tilde{t}_{i z}  \end{pmatrix}
    = \begin{pmatrix}
                      1 & -i \alpha & 0 & 0\\
                      -i \alpha & 1 & 0& 0 \\
                       0 & 0 & 1 & 0 \\
                       0 & 0 & 0 & 1 
        \end{pmatrix}
        \begin{pmatrix}
                 s_i \\  t_{i x}  \\  t_{i y}   \\  t_{i z} \end{pmatrix},
\label{uni-trans}
\end{equation}
and shows that the Hamiltonian \eqref{ham-HL} assumes a
diagonal form, namely,
\begin{equation}
       H_L = J_1 \sum_{i \in \mathcal{D}} \left( - \frac{3}{4} \tilde{s}_i^\dagger \tilde{s}_i
              + \frac{1}{4} \tilde{t}_{ix}^\dagger \tilde{t}_{ix} \right).
\end{equation}

Substituting the inverse of the transformation \eqref{uni-trans} in
Eqs.~\eqref{h-boson1ap}, \eqref{h-boson2ap}, and \eqref{h-boson3ap},
one finds the expression of the effective boson model in terms of the 
singlet $\tilde{s}_i$ and triplet $\tilde{t}_{i \alpha}$ boson operators.
In particular, it is easy to show that, up to linear order in the
parameter $\alpha$, the quadratic terms $\mathcal{H}_{J,2}$
and $\mathcal{H}_{\rm DM,2}$ assume the forms shown 
in Eqs.~\eqref{h-boson2ap} and \eqref{h-boson3ap},
respectively, with the replacements
$s_i \rightarrow \tilde{s}_i$ and
$t_{i \alpha} \rightarrow \tilde{t}_{i \alpha}$.
Moreover, one also finds that, up to linear order in the parameter $\alpha$, 
the cubic term $\mathcal{H}_{J,3}$ yields an additional quadratic
term, namely,
\begin{equation}
  \mathcal{H} = -i \epsilon_{x\beta\gamma} \sum_\bk  D'_\bk  \left[
                                \tilde{t}^\dagger_{\bk\beta}\tilde{t}_{\bk\gamma}  
   + \frac{1}{2}\left(  \tilde{t}_{-\bk\beta} \tilde{t}_{\bk\gamma}
                             -  \tilde{t}^\dagger_{-\bk\beta} \tilde{t}^\dagger_{\bk\gamma}    \right)   \right],    
 \label{hdm2-L} 
\end{equation}
with
\begin{align}
  D'_\bk  = &\frac{1}{2} D N_0 \sin(2 k_x) + \frac{1}{2} D N_0 \frac{J_2}{J_1} \left[ 
                   \sin\left( 2 k_x + k_y \right) \right.
\nonumber \\
                  & \left. + \sin\left( 2 k_x - k_y \right) \right].
\label{coefs04}
\end{align}
Therefore, the quadratic Hamiltonian \eqref{ham-harm}, considered
within the harmonic approximation, assumes the form defined by
Eqs.~\eqref{ham-harm-mat}-\eqref{AB-matrices}, with the replacements
$t_{i \alpha} \rightarrow \tilde{t}_{i \alpha}$ and
$D_\bk \rightarrow D_\bk + D'_\bk$, see Eq.~\eqref{coefs03}.
In Sec.~\ref{sec:harm}, after performing such replacements, we restore
the notation for the triplet operators, i.e., 
$\tilde{t}_{i \alpha} \rightarrow t_{i \alpha}$.

A few remarks here about the Zeeman term \eqref{h-boson3ap} are in
order: In principle, the Zeeman term should also be included in the
Hamiltonian \eqref{ham-HL}; since we consider a very small external
magnetic field, $h_z = 0.05\, J_1$, we neglect the contribution of
the Zeeman term to the (local) Hamiltonian \eqref{ham-HL};
importantly, the results derived in Sec.~\ref{sec:harm} within the harmonic
approximation for $D < h_z$ may need some further corrections.

\section{Details about the diagonalization of the harmonic Hamiltonian}
\label{ap:diag}

In this appendix, we present the analytical expressions of the triplon
dispersion relations $\omega^\alpha_\bk$, in addition to the matrix
elements of the $\hat{T}_\bk$ matrix \eqref{Tmatrix1}, which relates the
triplet $t$ and triplon $b$ boson operators, see Eq.~\eqref{psi-phi}. 
The analytical procedure employed here is based on Refs.~\cite{colpa78,blaizot}
and it was previously applied by one of us to diagonalize a
$4 \times 4$ and a
$6 \times 6$ boson Hamiltonians,
see Refs.~\cite{doretto12} and \cite{doretto14}, respectively.

As mentioned in Sec.~\ref{sec:harm}, instead of the $\hat{H}_\bk$
matrix \eqref{hk-mat}, one should diagonalize the $\hat{I}_B\hat{H}_\bk$
matrix \eqref{ib-hk-mat}. One finds that the eigenvalues $\omega^\alpha_\bk$ are
the roots of the polynomial
\begin{equation}
  (\omega^\alpha_\bk)^6 + a_{2,\bk}(\omega^\alpha_\bk)^4
  + a_{1,\bk}(\omega^\alpha_\bk)^2 + a_{0,\bk} = 0,
\label{poly}
\end{equation}
where the coefficients $a_{i,\bk}$ are written in terms of
the coefficients $A_\bk$ and $B_\bk$ defined by Eq.~\eqref{coefs01},
the coefficients $C_\bk$ and $D_\bk$ respectively given by
Eqs.~\eqref{coefs02} and \eqref{coefs03}, and the external magnetic
field $h_z$:
\begin{widetext}
\begin{align}
 a_{0,\bk}  =&  \left(B_\bk - A_\bk \right) \left( B_\bk^2 + h_z^2 -  A_\bk^2 \right)
                     \left[  B_\bk \left( B_\bk^2 + h_z^2 - 4D_\bk^2 -  4C_\bk^2 \right)   
                     +  A_\bk \left(  B_\bk^2 + h_z^2 + 4D_\bk^2+ 4C_\bk^2  \right) 
                      - A_\bk^2 \left(  A_\bk  +  B_\bk \right)  \right],
\nonumber \\
\nonumber \\  
  a_{1,\bk}  =& \, h_z^4 + \left( A_\bk - B_\bk \right)^2
                   \left[ 3 \left(A_\bk + B_\bk \right)^2 -4 \left( C_\bk^2 + D_\bk^2 \right) \right],
\\            
\nonumber \\
  a_{2,\bk}  =& -3 \left(  A_\bk^2  - B_\bk^2 \right)  - 2 h_z^2 .
\nonumber 
\end{align}
Due to the form of Eq.~\eqref{poly}, one sees that the eigenvalues $(\omega^\alpha_\bk)^2$
are indeed the roots of a cubic polynomial which can be written as
\begin{align}
  \omega_\bk^{x/y} &= \left[ - \frac{1}{3} a_{2,\bk} - \frac{1}{2} \left( P_{1,\bk} + P_{2,\bk} \right)
                                     \pm i  \frac{\sqrt{3} }{2}  \left( P_{1,\bk} - P_{2,\bk} \right) \right]^{1/2},
\nonumber \\
  \omega_\bk^z &= \left[ - \frac{1}{3} a_{2,\bk} + \left( P_{1,\bk} + P_{2,\bk} \right) \right]^{1/2},
\label{omega}
\end{align}
where,
\begin{align}
  P_{1/2,\bk} =& \left (M_\bk \pm  \sqrt{N_\bk} \right)^{1/3},
  \quad\quad\quad
  N_\bk = Q_\bk^3 +M_\bk^2,
  \quad\quad\quad
  Q_\bk = \frac{1}{9} \left(3 a_{1,\bk} - a_{2,\bk}^2 \right),
\nonumber \\
  M_\bk =& \frac{1}{54} \left(9 a_{2,\bk} a_{1,\bk} - 27 a_{0,\bk} - 2 a_{2,\bk}^3 \right).
\label{aux-omega}
\end{align}

In order to determine the matrix elements of the $\hat{T}_\bk$ matrix \eqref{Tmatrix1},
it is interesting to consider the two eigenvalue problems
\begin{align}
  & \hat{I}_B \hat{H'}_\bk \hat{Z}^i_{1,\bk} - \omega_\bk^i \hat{Z}^i_{1,\bk} = 0, 
\quad\quad\quad {\rm and}  \quad\quad\quad
      \hat{I}_B \hat{H'}_\bk \hat{Z}^i_{2,\bk} + \omega_\bk^i \hat{Z}^i_{2,\bk}  = 0, 
\label{aux01}
\end{align}
with $\omega^i_\bk > 0$ and $i = 1,2,3$.
Here the six-component vectors $\hat{Z}^i_{1/2,\bk}$ are
defined as  
\begin{align}
  \hat{Z}^i_{1,\bk} &= \begin{pmatrix}
                          u^{i1}_\bk & u^{i2}_\bk & u^{i3}_\bk & v^{i1}_\bk & v^{i2}_\bk & v^{i3}_\bk  	
                          \end{pmatrix}^T, 
\quad\quad\quad {\rm and}  \quad\quad\quad
  \hat{Z}^i_{2,\bk}  = \begin{pmatrix}
                            y^{i1}_\bk & y^{i2}_\bk & y^{i3}_\bk & x^{i1}_\bk & x^{i2}_\bk & x^{i3}_\bk  	
                           \end{pmatrix}^T
\nonumber
\end{align}
and one identifies
$\omega_\bk^1 = \omega^x_\bk$,
$\omega_\bk^2 = \omega^y_\bk$, and
$\omega_\bk^3 = \omega^z_\bk$.
Equation \eqref{aux01} allows one to determine
(i) the elements
$u^{i1}_\bk$, $u^{i2}_\bk$, $u^{i3}_\bk$,  $v^{i1}_\bk$, and $v^{i2}_\bk$
in terms of the element $v^{i3}_\bk$  and  	
(ii) the elements
$y^{i1}_\bk$, $y^{i2}_\bk$, $y^{i3}_\bk$,  $x^{i1}_\bk$, and $x^{i2}_\bk$
in terms of the element $x^{i3}_\bk$,
for $i = 1,2,3$.
Moreover, the condition \eqref{conditionT}
allows one to determine the elements $v^{i3}_\bk$ and $x^{i3}_\bk$. 
Indeed, it is useful to write the matrix elements of the $3 \times 3$
matrices $\hat{U}_\bk$, $\hat{V}_\bk$, $\hat{X}_\bk$, and
$\hat{Y}_\bk$ in terms of the (auxiliary) elements
$\bar{u}_{ij,\bk}$, $\bar{v}_{ij,\bk}$, $\bar{x}_{ij,\bk}$, and $\bar{y}_{ij,\bk}$
as
\begin{align}
 \hat{U}_\bk =&
\begin{pmatrix}
 v^{31}_\bk \bar{u}_{11,\bk} &  v^{32}_\bk \bar{u}_{42,\bk} &  v^{33}_\bk \bar{u}_{73,\bk} \\
 v^{31}_\bk \bar{u}_{21,\bk} &  v^{32}_\bk \bar{u}_{52,\bk} &  v^{33}_\bk \bar{u}_{83,\bk} \\
 v^{31}_\bk \bar{u}_{31,\bk} &  v^{32}_\bk \bar{u}_{62,\bk} &  v^{33}_\bk \bar{u}_{93,\bk}
\end{pmatrix},
&\hat{Y}_\bk =
\begin{pmatrix}
x^{31}_\bk \bar{y}_{11,\bk} &  x^{32}_\bk \bar{y}_{42,\bk} &  x^{33}_\bk \bar{y}_{73,\bk}  \\
x^{31}_\bk \bar{y}_{21,\bk} &  x^{32}_\bk \bar{y}_{52,\bk} &  x^{33}_\bk \bar{y}_{83,\bk}  \\
x^{31}_\bk \bar{y}_{31,\bk} &  x^{32}_\bk \bar{y}_{62,\bk} &  x^{33}_\bk \bar{y}_{93,\bk}
\end{pmatrix},
\nonumber \\
\\
 \hat{V}_\bk =& 
\begin{pmatrix}
  v^{31}_\bk \bar{v}_{11,\bk}   &  v^{32}_\bk \bar{v}_{42,\bk}   &  v^{33}_\bk \bar{v}_{73,\bk}   \\
  v^{31}_\bk \bar{v}_{21,\bk}   &  v^{32}_\bk \bar{v}_{52,\bk}   &  v^{33}_\bk \bar{v}_{83,\bk}   \\
  v^{31}_\bk \                       &  v^{32}_\bk                          &  v^{33}_\bk 
\end{pmatrix},
&\hat{X}_\bk =
\begin{pmatrix}
x^{31}_\bk \bar{x}_{11,\bk} &  x^{32}_\bk \bar{x}_{42,\bk} &  x^{33}_\bk \bar{x}_{73,\bk} \\
x^{31}_\bk \bar{x}_{21,\bk} &  x^{32}_\bk \bar{x}_{52,\bk} &  x^{33}_\bk \bar{x}_{83,\bk} \\
x^{31}_\bk                        & x^{32}_\bk                         & x^{33}_\bk
\end{pmatrix}.
\nonumber
\end{align}  
With the aid of the condition \eqref{conditionT}, one then shows that
\begin{align}
 v^{3,j}_\bk = &\left( \left|\bar{u}_{3j-2,1,\bk}\right|^2 - \left|\bar{v}_{3j-2,1,\bk}\right|^2 + \left|\bar{u}_{3j-1,1,\bk}\right|^2 
                              - \left|\bar{v}_{3j-1,1,\bk}\right|^2 - 1 + \left|\bar{u}_{3j,1,\bk}\right|^2 \right)^{-1/2},
\nonumber \\
\label{v3-x3} \\
x^{3,j}_\bk = & \left(\left|\bar{x}_{3j-2,1,\bk}\right|^2 - \left|\bar{y}_{3j-2,1,\bk}\right|^2 + \left|\bar{x}_{3j-1,1,\bk}\right|^2 
                             - \left|\bar{y}_{3j-1,1,\bk}\right|^2 + 1 - \left|\bar{y}_{3j,1,\bk}\right|^2 \right)^{-1/2},
\nonumber
\end{align}
for $j=1,2,3$.
The two eigenvalue problems \eqref{aux01} allow one to determine the elements
$\bar{u}_{ij,\bk}$, $\bar{v}_{ij,\bk}$, $\bar{x}_{ij,\bk}$, and $\bar{y}_{ij,\bk}$.
One then finds that the elements $\bar{u}_{ij,\bk}$ assume the form
\begin{align}
  \bar{u}_{\alpha\beta,\bk} =& \, i \frac{1}{F_{\beta,\bk}}\left( B_\bk + R_{\beta,\bk} \right) \Big[  B_\bk^3 C_\bk
                          + B_\bk^2 \left( L_{4,\bk} - C_\bk S_{\beta,\bk} \right)
                          + \left( L_{4,\bk} + C_\bk R_{\beta,\bk} \right) \left( h_z^2 - S_{\beta,\bk}^2 \right) \Big.
\nonumber \\
  &\Big. + \, B_\bk \left( h_z^2 C_\bk + 2 L_{4,\bk} \omega_\bk^\beta + C_\bk R_{\beta,\bk}  S_{\beta,\bk}\right) \Big],
\quad\quad\quad
          {\rm for} \quad\quad\quad \alpha = 1, 4, 7,
\nonumber \\
\nonumber \\
 \bar{u}_{\alpha\beta,\bk}  =& \frac{1}{F_{\beta,\bk}} \Big[ iL_{1,\bk}L_{6,\bk} \left( L_{5,\bk} - D_\bk L_{6,\bk}  \right)
                          + h_z \left( 2 B_\bk L_{4,\bk} + 3B_\bk^2C_\bk + h_z^2 C_\bk
                          - 2 A_\bk L_{4,\bk}   - 4 A_\bk B_\bk C_\bk + A_\bk^2 C_\bk \right) \omega_\bk^\beta \Big.  
\nonumber \\
       & \Big.   -\, \left( i h_z^2 D_\bk - 2 i A_\bk D_\bk L_{6,\bk}  - h_z C_\bk L_{6,\bk} \right)( \omega_\bk^\beta )^2 
          - h_z C_\bk ( \omega_\bk^\beta )^3
         + i D_\bk  ( \omega_\bk^\beta )^4 \Big], 
\quad\quad
          {\rm for} \quad\quad \alpha = 2, 5, 8,
\nonumber \\
\\
\bar{u}_{\alpha\beta,\bk}  =& \frac{1}{F_{\beta,\bk}} \Big[ 
       B_\bk L_{1,\bk}\left( 2 L_{2,\bk} + A_\bk^2 -h_z^2\right) 
     -L_{1,\bk} \left(  B_\bk^3 + 2 A_\bk L_{2,\bk}  \right)
     - 2 \left( B_\bk^3 + A_\bk L_{2,\bk}   \right) ( \omega_\bk^\beta )^2  \Big.
\nonumber \\
    & \Big. +\, B_\bk \left( h_z^2 + A_\bk^2 +  L_{2,\bk} \right)  ( \omega_\bk^\beta )^2
                - B_\bk ( \omega_\bk^\beta )^4 \Big],
\quad\quad\quad
          {\rm for} \quad\quad\quad \alpha = 3, 6, 9,
\nonumber
\end{align}
the elements $\bar{v}_{ij,\bk}$ are given by
\begin{align}
\bar{v}_{\alpha\beta,\bk}  =& + i \frac{1}{F_{\alpha,\bk}}\left( B_\bk + R_{\alpha,\bk} \right) \Big[B_\bk^3 C_\bk + B_\bk^2
                              \left(  C_\bk R_{\alpha,\bk} - L_{4,\bk}  \right) 
                               - i \left( h_z + R_{\alpha,\bk} \right) \left( h_z +  R_{\alpha,\bk} \right) \left(+ h_z D_\bk + i C_\bk S_{\alpha,\bk} \right) \Big.
\nonumber \\ 
                                       &+ \, B_\bk \left( h_z^2 C_\bk + 2 L_{4,\bk} \omega_\bk^\alpha + C_\bk R_{\alpha,\bk} S_{\alpha,\bk}\right)\Big], 
\quad\quad\quad
       {\rm for} \quad\quad\quad \alpha = 1, 4, 7,
\nonumber \\
\\
\bar{v}_{\alpha\beta,\bk}  =& - i \,\frac{1}{F_{\alpha,\bk}} \left( B_\bk +  R_{\alpha,\bk} \right) \Big[ B_\bk^3 D_\bk 
                                 + B_\bk^2 \left( D_\bk R_{\alpha,\bk} + L_{5,\bk} \right) 
                                 - i \left( h_z + R_{\alpha,\bk} \right) \left( R_{\alpha,\bk} - h_z\right) \left( -h_z C_\bk + i D_\bk S_{\alpha,\bk} \right) \Big. 
\nonumber \\
 &+\, B_\bk \left( h_z^2 D_\bk - 2 L_{5,\bk} \omega_\bk^\alpha + D_\bk R_{\alpha,\bk} S_{\alpha,\bk} \right) \Big],
\quad\quad\quad
          {\rm for} \quad\quad\quad  \alpha = 2, 5, 8,
\nonumber
\end{align}
the elements $\bar{y}_{\alpha\beta,\bk}$ read
\begin{align}
 \bar{y}_{\alpha\beta,\bk}  =& \frac{1}{G_{\beta,\bk}} \Big[ + i B_\bk^4 C_\bk - B_\bk^3\left( 2 i A_\bk C_\bk - h_z D_\bk\right) 
              - \left( h_z + R_{\beta,\bk} \right) \left(  R_{\beta,\bk} - h_z\right) \left( -h_z D_\bk + i C_\bk S_{\beta,\bk} \right) S_{\beta,\bk} \Big.
\nonumber \\
  &\Big. + \, B_\bk \left( h_z^3 D_\bk - 2 i h_z^2 C_\bk S_{\beta,\bk} - 2 i A_\bk C_\bk R_{\beta,\bk}  S_{\beta,\bk} \right)
               - h_z B_\bk  D_\bk \left( A_\bk^2 - 4 A_\bk \omega_\bk^\beta  - (\omega_\bk^\beta)^2 \right)        \Big.
\nonumber \\ 
                  & \Big.  - h_z B_\bk^2 D_\bk \left( A_\bk + 3 \omega_\bk^\beta \right) - h_zB_\bk^2 L_{5,\bk}  \Big],
\quad\quad\quad
          {\rm for} \quad\quad\quad \alpha = 1, 4, 7,
\nonumber \\
\nonumber \\
\bar{y}_{\alpha\beta,\bk}  =& \frac{1}{G_{\beta,\bk}}\Big[- i B_\bk^4D_\bk +  B_\bk^3 \left( 2 i A_\bk D_\bk   + h_z C_\bk\right) 
              + \left( h_z + R_{\beta,\bk} \right) 
                  \left( R_{\beta,\bk} - h_z\right) \left( +h_z C_\bk + i D_\bk S_{\beta,\bk} \right) S_{\beta,\bk} \Big.  
\nonumber \\ 
            &\Big. + \, B_\bk \left( h_z^3 C_\bk  + 2 i h_z^2 D_\bk S_{\beta,\bk} + 2 i A_\bk D_\bk R_{\beta,\bk} S_{\beta,\bk}\right) 
                - h_z B_\bk C_\bk \left( A_\bk^2 - 4 A_\bk \omega_\bk^\beta  - ( \omega_\bk^\beta )^2  \right)    \Big.
\nonumber \\
            & \Big. - \, h_z B_\bk^2 C_\bk \left( A_\bk + 3\omega_\bk^\beta \right) + h_z B^2_\bk L_{4,\bk}   \Big],
\quad\quad\quad
          {\rm for} \quad\quad\quad \alpha = 2, 5, 8,
\nonumber \\
\\
 \bar{y}_{\alpha\beta,\bk}  =&  \frac{1}{G_{\beta,\bk}} \Big[ 
      B_\bk L_{1,\bk} \left( 2 L_{2,\bk} +  A_\bk^2 -h_z^2\right)
    - L_{1,\bk}\left(  B_\bk^3 + 2 A_\bk L_{2,\bk}  \right)  
 \Big. - 2 \left( B_\bk^3 + A_\bk L_{2,\bk}  \right) (\omega_\bk^\beta)^2 
\nonumber \\
  &\Big.  + B_\bk \left( h_z^2+ A_\bk^2 + L_2 \right)  (\omega_\bk^\beta)^2 
      - B_\bk  ( \omega_\bk^\beta )^4  \Big],
\quad\quad
          {\rm for} \quad\quad\quad \alpha = 3, 6, 9,
\nonumber
\end{align}
and the elements $\bar{x}_{i,\bk}$ are given by
\begin{align}
\bar{x}_{\alpha\beta,\bk} =& +i \frac{1}{G_{\beta,\bk}}  \left( B_\bk - S_{\beta,\bk} \right)  
     \Big[ B_\bk^3 C_\bk + \left( h_z - S_{\beta,\bk} \right) 
             \left( h_z + S_{\beta,\bk} \right) \left( C_\bk R_{\beta,\bk} -L_{4,\bk}\right)  \Big. 
\nonumber \\ 
&\Big. - B_\bk^2 \left( L_{4,\bk} + C_\bk S_{\beta,\bk}  \right) 
  + B_\bk \left( h_z^2 C_\bk - 2 L_{4,\bk} \omega_\bk^\beta + C_\bk R_{\beta,\bk} S_{\beta,\bk} \right) \Big],
         && {\rm for} \quad\quad\quad \alpha = 1, 4, 7,
\nonumber \\
\\
 \bar{x}_{\alpha\beta,\bk} =& -i \frac{1}{G_{\beta,\bk}} \left( B_\bk - S_{\beta,\bk} \right) 
         \Big[  B_\bk^3 D_\bk + \left( h_z - S_{\beta,\bk} \right) 
         \left( h_z + S_{\beta,\bk} \right) 
          \left( D_\bk R_{\beta,\bk}  + L_{5,\bk} \right) \Big. 
\nonumber \\ 
  &\Big. + B_\bk^2 \left( L_{5,\bk} - D_\bk S_{\beta,\bk} \right) 
    + B_\bk \left( h_z^2 D_\bk  + 2 L_{5,\bk} \omega_\bk^\beta + D_\bk R_{\beta,\bk} S_{\beta,\bk}  \right)  \Big],
         && {\rm for} \quad\quad\quad \alpha = 2, 5, 8.
\nonumber 
\end{align}
Here, the coefficients $L_{i,\bk}$ are defined as
\begin{align}
  &L_{1,\bk} = B_\bk^2 + h_z^2 - A_\bk^2, 
  &&L_{2,\bk} = D_\bk^2 + C_\bk^2,  
  &&L_{3,\bk}  = B_\bk^2 - h_z^2,    
\nonumber \\
  &L_{4,\bk} = -i h_z D_\bk,  
  &&L_{5,\bk} = -i h_z C_\bk,  
  &&L_{6,\bk} = B_\bk - A_\bk,
\end{align}
the coefficients $O_{i,\bk}$ reads
\begin{align}
  &O_{1,\bk} = A_\bk \left( L_{1,\bk} + A_\bk^2 + 2 L_{2,\bk} \right) 
                      - 2 B_\bk L_{2,\bk}  - A_\bk^3, 
\nonumber \\
  &O_{2,\bk} = A_\bk \left( L_{2,\bk} + L_{3,\bk} \right) - B_\bk L_{2,\bk} - A_\bk^3,
\end{align}
and, finally,
\begin{align}
  &R^\alpha_\bk = \; \omega_\bk^\alpha - A_\bk, \quad\quad\quad\quad
  S^\alpha_\bk = \; \omega_\bk^\alpha + A_\bk,
\label{auxr-auxs}\\        
  &F^\alpha_\bk = \; L_{1,\bk} O_{1,\bk} - L_{1,\bk}^2 \omega_\bk^\alpha + 2O_{2,\bk} \left( \omega_\bk^\alpha \right)^2
     + 2 \left(  A_\bk^2  -L_{3,\bk} \right) \left( \omega_\bk^\alpha \right)^3
            + A_\bk \left( \omega_\bk^\alpha \right)^4 - \left( \omega_\bk^\alpha \right)^5, 
\nonumber \\
  &G^\alpha_\bk  = \; L_{1,\bk} O_{1,\bk} + L_{1,\bk}^2 \omega_\bk^\alpha + 2O_{2,\bk} \left( \omega_\bk^\alpha \right)^2
    + 2 \left( L_{3,\bk} - A_\bk^2 \right) \left( \omega_\bk^\alpha \right)^3
              + A_\bk \left( \omega_\bk^\alpha \right)^4 + \left( \omega_\bk^\alpha \right)^5,
\nonumber
\end{align}
with $\alpha = 1,2,3$,
the coefficients $A_\bk$ and $B_\bk$ given by Eq.~\eqref{coefs01},
the coefficients $C_\bk$ and $D_\bk$ respectively given by
Eqs.~\eqref{coefs02} and \eqref{coefs03}, and
$h_z$ being the external magnetic field.

Finally, it is important to mention that Eqs.~\eqref{v3-x3}-\eqref{auxr-auxs}
are not appropriated for DM interaction $D = 0$: In this case, we find
divergences for the Berry curvatures \eqref{Berry};
for $D = 10^{-3}\, J_1$, well defined results for the Berry curvatures
are found.

\end{widetext}

\bibliography{refs}{}

\begin{thebibliography}{113}%
\makeatletter
\providecommand \@ifxundefined [1]{%
 \@ifx{#1\undefined}
}%
\providecommand \@ifnum [1]{%
 \ifnum #1\expandafter \@firstoftwo
 \else \expandafter \@secondoftwo
 \fi
}%
\providecommand \@ifx [1]{%
 \ifx #1\expandafter \@firstoftwo
 \else \expandafter \@secondoftwo
 \fi
}%
\providecommand \natexlab [1]{#1}%
\providecommand \enquote  [1]{``#1''}%
\providecommand \bibnamefont  [1]{#1}%
\providecommand \bibfnamefont [1]{#1}%
\providecommand \citenamefont [1]{#1}%
\providecommand \href@noop [0]{\@secondoftwo}%
\providecommand \href [0]{\begingroup \@sanitize@url \@href}%
\providecommand \@href[1]{\@@startlink{#1}\@@href}%
\providecommand \@@href[1]{\endgroup#1\@@endlink}%
\providecommand \@sanitize@url [0]{\catcode `\\12\catcode `\$12\catcode
  `\&12\catcode `\#12\catcode `\^12\catcode `\_12\catcode `\%12\relax}%
\providecommand \@@startlink[1]{}%
\providecommand \@@endlink[0]{}%
\providecommand \url  [0]{\begingroup\@sanitize@url \@url }%
\providecommand \@url [1]{\endgroup\@href {#1}{\urlprefix }}%
\providecommand \urlprefix  [0]{URL }%
\providecommand \Eprint [0]{\href }%
\providecommand \doibase [0]{https://doi.org/}%
\providecommand \selectlanguage [0]{\@gobble}%
\providecommand \bibinfo  [0]{\@secondoftwo}%
\providecommand \bibfield  [0]{\@secondoftwo}%
\providecommand \translation [1]{[#1]}%
\providecommand \BibitemOpen [0]{}%
\providecommand \bibitemStop [0]{}%
\providecommand \bibitemNoStop [0]{.\EOS\space}%
\providecommand \EOS [0]{\spacefactor3000\relax}%
\providecommand \BibitemShut  [1]{\csname bibitem#1\endcsname}%
\let\auto@bib@innerbib\@empty
\bibitem [{\citenamefont {Sachdev}(2004)}]{review-sachdev}%
  \BibitemOpen
  \bibfield  {author} {\bibinfo {author} {\bibfnamefont {S.}~\bibnamefont
  {Sachdev}},\ }\bibfield  {title} {\bibinfo {title} {Quantum phases and phase
  transitions of mott insulators},\ }in\ \href@noop {} {\emph {\bibinfo
  {booktitle} {Quantum Magnetism}}},\ \bibinfo {series and number} {Lecture
  Notes in Physics Vol. 645},\ \bibinfo {editor} {edited by\ \bibinfo {editor}
  {\bibfnamefont {U.}~\bibnamefont {Schollwöck}}, \bibinfo {editor}
  {\bibfnamefont {J.}~\bibnamefont {Richter}}, \bibinfo {editor} {\bibfnamefont
  {D.~J.~J.}\ \bibnamefont {Farnell}},\ and\ \bibinfo {editor} {\bibfnamefont
  {R.~A.}\ \bibnamefont {Bishop}}}\ (\bibinfo  {publisher} {Springer},\
  \bibinfo {year} {2004})\BibitemShut {NoStop}%
\bibitem [{\citenamefont {Savary}\ and\ \citenamefont
  {Balents}(2017)}]{review-balents}%
  \BibitemOpen
  \bibfield  {author} {\bibinfo {author} {\bibfnamefont {L.}~\bibnamefont
  {Savary}}\ and\ \bibinfo {author} {\bibfnamefont {L.}~\bibnamefont
  {Balents}},\ }\bibfield  {title} {\bibinfo {title} {{Quantum spin liquids: a
  review}},\ }\href {https://doi.org/10.1088/0034-4885/80/1/016502} {\bibfield
  {journal} {\bibinfo  {journal} {Rep. Prog. Phys.}\ }\textbf {\bibinfo
  {volume} {80}},\ \bibinfo {pages} {016502} (\bibinfo {year}
  {2017})}\BibitemShut {NoStop}%
\bibitem [{\citenamefont {Murakami}\ and\ \citenamefont
  {Okamoto}(2017)}]{murakami17}%
  \BibitemOpen
  \bibfield  {author} {\bibinfo {author} {\bibfnamefont {S.}~\bibnamefont
  {Murakami}}\ and\ \bibinfo {author} {\bibfnamefont {A.}~\bibnamefont
  {Okamoto}},\ }\bibfield  {title} {\bibinfo {title} {Thermal {H}all effect of
  magnons},\ }\href {https://doi.org/10.7566/JPSJ.86.011010} {\bibfield
  {journal} {\bibinfo  {journal} {Journal of the Physical Society of Japan}\
  }\textbf {\bibinfo {volume} {86}},\ \bibinfo {pages} {011010} (\bibinfo
  {year} {2017})}\BibitemShut {NoStop}%
\bibitem [{\citenamefont {Zhang}\ \emph {et~al.}(2023)\citenamefont {Zhang},
  \citenamefont {Gao},\ and\ \citenamefont {Chen}}]{ar-thermal-review}%
  \BibitemOpen
  \bibfield  {author} {\bibinfo {author} {\bibfnamefont {X.-T.}\ \bibnamefont
  {Zhang}}, \bibinfo {author} {\bibfnamefont {Y.~H.}\ \bibnamefont {Gao}},\
  and\ \bibinfo {author} {\bibfnamefont {G.}~\bibnamefont {Chen}},\ }\href@noop
  {} {\bibinfo {title} {Thermal {H}all effects in quantum magnets}} (\bibinfo
  {year} {2023}),\ \Eprint {https://arxiv.org/abs/2305.04830} {arXiv:2305.04830
  [cond-mat.str-el]} \BibitemShut {NoStop}%
\bibitem [{\citenamefont {Owerre}(2016{\natexlab{a}})}]{owerre16-2}%
  \BibitemOpen
  \bibfield  {author} {\bibinfo {author} {\bibfnamefont {S.~A.}\ \bibnamefont
  {Owerre}},\ }\bibfield  {title} {\bibinfo {title} {A first theoretical
  realization of honeycomb topological magnon insulator},\ }\href
  {https://doi.org/10.1088/0953-8984/28/38/386001} {\bibfield  {journal}
  {\bibinfo  {journal} {Journal of Physics: Condensed Matter}\ }\textbf
  {\bibinfo {volume} {28}},\ \bibinfo {pages} {386001} (\bibinfo {year}
  {2016}{\natexlab{a}})}\BibitemShut {NoStop}%
\bibitem [{\citenamefont {Owerre}(2016{\natexlab{b}})}]{owerre16}%
  \BibitemOpen
  \bibfield  {author} {\bibinfo {author} {\bibfnamefont {S.~A.}\ \bibnamefont
  {Owerre}},\ }\bibfield  {title} {\bibinfo {title} {Topological honeycomb
  magnon hall effect: A calculation of thermal {H}all conductivity of magnetic
  spin excitations},\ }\href {https://doi.org/10.1063/1.4959815} {\bibfield
  {journal} {\bibinfo  {journal} {Journal of Applied Physics}\ }\textbf
  {\bibinfo {volume} {120}},\ \bibinfo {pages} {043903} (\bibinfo {year}
  {2016}{\natexlab{b}})}\BibitemShut {NoStop}%
\bibitem [{\citenamefont {Malki}\ and\ \citenamefont {Uhrig}(2019)}]{malki19}%
  \BibitemOpen
  \bibfield  {author} {\bibinfo {author} {\bibfnamefont {M.}~\bibnamefont
  {Malki}}\ and\ \bibinfo {author} {\bibfnamefont {G.~S.}\ \bibnamefont
  {Uhrig}},\ }\bibfield  {title} {\bibinfo {title} {Topological magnon bands
  for magnonics},\ }\href {https://doi.org/10.1103/PhysRevB.99.174412}
  {\bibfield  {journal} {\bibinfo  {journal} {Phys. Rev. B}\ }\textbf {\bibinfo
  {volume} {99}},\ \bibinfo {pages} {174412} (\bibinfo {year}
  {2019})}\BibitemShut {NoStop}%
\bibitem [{\citenamefont {Lu}\ \emph {et~al.}(2021)\citenamefont {Lu},
  \citenamefont {Li},\ and\ \citenamefont {Wu}}]{wu21}%
  \BibitemOpen
  \bibfield  {author} {\bibinfo {author} {\bibfnamefont {Y.-S.}\ \bibnamefont
  {Lu}}, \bibinfo {author} {\bibfnamefont {J.-L.}\ \bibnamefont {Li}},\ and\
  \bibinfo {author} {\bibfnamefont {C.-T.}\ \bibnamefont {Wu}},\ }\bibfield
  {title} {\bibinfo {title} {Topological phase transitions of {D}irac magnons
  in honeycomb ferromagnets},\ }\href
  {https://doi.org/10.1103/PhysRevLett.127.217202} {\bibfield  {journal}
  {\bibinfo  {journal} {Phys. Rev. Lett.}\ }\textbf {\bibinfo {volume} {127}},\
  \bibinfo {pages} {217202} (\bibinfo {year} {2021})}\BibitemShut {NoStop}%
\bibitem [{\citenamefont {Kim}\ and\ \citenamefont {Kim}(2022)}]{kim22}%
  \BibitemOpen
  \bibfield  {author} {\bibinfo {author} {\bibfnamefont {H.}~\bibnamefont
  {Kim}}\ and\ \bibinfo {author} {\bibfnamefont {S.~K.}\ \bibnamefont {Kim}},\
  }\bibfield  {title} {\bibinfo {title} {Topological phase transition in magnon
  bands in a honeycomb ferromagnet driven by sublattice symmetry breaking},\
  }\href {https://doi.org/10.1103/PhysRevB.106.104430} {\bibfield  {journal}
  {\bibinfo  {journal} {Phys. Rev. B}\ }\textbf {\bibinfo {volume} {106}},\
  \bibinfo {pages} {104430} (\bibinfo {year} {2022})}\BibitemShut {NoStop}%
\bibitem [{\citenamefont {Zhang}\ \emph {et~al.}(2013)\citenamefont {Zhang},
  \citenamefont {Ren}, \citenamefont {Wang},\ and\ \citenamefont {Li}}]{li13}%
  \BibitemOpen
  \bibfield  {author} {\bibinfo {author} {\bibfnamefont {L.}~\bibnamefont
  {Zhang}}, \bibinfo {author} {\bibfnamefont {J.}~\bibnamefont {Ren}}, \bibinfo
  {author} {\bibfnamefont {J.-S.}\ \bibnamefont {Wang}},\ and\ \bibinfo
  {author} {\bibfnamefont {B.}~\bibnamefont {Li}},\ }\bibfield  {title}
  {\bibinfo {title} {Topological magnon insulator in insulating ferromagnet},\
  }\href {https://doi.org/10.1103/PhysRevB.87.144101} {\bibfield  {journal}
  {\bibinfo  {journal} {Phys. Rev. B}\ }\textbf {\bibinfo {volume} {87}},\
  \bibinfo {pages} {144101} (\bibinfo {year} {2013})}\BibitemShut {NoStop}%
\bibitem [{\citenamefont {Zyuzin}\ and\ \citenamefont
  {Kovalev}(2016)}]{kovalev16}%
  \BibitemOpen
  \bibfield  {author} {\bibinfo {author} {\bibfnamefont {V.~A.}\ \bibnamefont
  {Zyuzin}}\ and\ \bibinfo {author} {\bibfnamefont {A.~A.}\ \bibnamefont
  {Kovalev}},\ }\bibfield  {title} {\bibinfo {title} {Magnon spin {N}ernst
  effect in antiferromagnets},\ }\href
  {https://doi.org/10.1103/PhysRevLett.117.217203} {\bibfield  {journal}
  {\bibinfo  {journal} {Phys. Rev. Lett.}\ }\textbf {\bibinfo {volume} {117}},\
  \bibinfo {pages} {217203} (\bibinfo {year} {2016})}\BibitemShut {NoStop}%
\bibitem [{\citenamefont {Laurell}\ and\ \citenamefont
  {Fiete}(2018)}]{fiete18}%
  \BibitemOpen
  \bibfield  {author} {\bibinfo {author} {\bibfnamefont {P.}~\bibnamefont
  {Laurell}}\ and\ \bibinfo {author} {\bibfnamefont {G.~A.}\ \bibnamefont
  {Fiete}},\ }\bibfield  {title} {\bibinfo {title} {Magnon thermal {H}all
  effect in kagome antiferromagnets with {D}zyaloshinskii-{M}oriya
  interactions},\ }\href {https://doi.org/10.1103/PhysRevB.98.094419}
  {\bibfield  {journal} {\bibinfo  {journal} {Phys. Rev. B}\ }\textbf {\bibinfo
  {volume} {98}},\ \bibinfo {pages} {094419} (\bibinfo {year}
  {2018})}\BibitemShut {NoStop}%
\bibitem [{\citenamefont {Neumann}\ \emph {et~al.}(2022)\citenamefont
  {Neumann}, \citenamefont {Mook}, \citenamefont {Henk},\ and\ \citenamefont
  {Mertig}}]{mertig22}%
  \BibitemOpen
  \bibfield  {author} {\bibinfo {author} {\bibfnamefont {R.~R.}\ \bibnamefont
  {Neumann}}, \bibinfo {author} {\bibfnamefont {A.}~\bibnamefont {Mook}},
  \bibinfo {author} {\bibfnamefont {J.}~\bibnamefont {Henk}},\ and\ \bibinfo
  {author} {\bibfnamefont {I.}~\bibnamefont {Mertig}},\ }\bibfield  {title}
  {\bibinfo {title} {Thermal {H}all effect of magnons in collinear
  antiferromagnetic insulators: Signatures of magnetic and topological phase
  transitions},\ }\href {https://doi.org/10.1103/PhysRevLett.128.117201}
  {\bibfield  {journal} {\bibinfo  {journal} {Phys. Rev. Lett.}\ }\textbf
  {\bibinfo {volume} {128}},\ \bibinfo {pages} {117201} (\bibinfo {year}
  {2022})}\BibitemShut {NoStop}%
\bibitem [{\citenamefont {Romhányi}\ \emph {et~al.}(2015)\citenamefont
  {Romhányi}, \citenamefont {Penc},\ and\ \citenamefont {Ganesh}}]{penc15}%
  \BibitemOpen
  \bibfield  {author} {\bibinfo {author} {\bibfnamefont {J.}~\bibnamefont
  {Romhányi}}, \bibinfo {author} {\bibfnamefont {K.}~\bibnamefont {Penc}},\
  and\ \bibinfo {author} {\bibfnamefont {R.}~\bibnamefont {Ganesh}},\
  }\bibfield  {title} {\bibinfo {title} {Hall effect of triplons in a dimerized
  quantum magnet},\ }\href {https://doi.org/10.1038/ncomms7805} {\bibfield
  {journal} {\bibinfo  {journal} {Nat. Commun.}\ }\textbf {\bibinfo {volume}
  {6}},\ \bibinfo {pages} {6805} (\bibinfo {year} {2015})}\BibitemShut
  {NoStop}%
\bibitem [{\citenamefont {Malki}\ and\ \citenamefont
  {Schmidt}(2017)}]{malki17}%
  \BibitemOpen
  \bibfield  {author} {\bibinfo {author} {\bibfnamefont {M.}~\bibnamefont
  {Malki}}\ and\ \bibinfo {author} {\bibfnamefont {K.~P.}\ \bibnamefont
  {Schmidt}},\ }\bibfield  {title} {\bibinfo {title} {Magnetic {C}hern bands
  and triplon {H}all effect in an extended {S}hastry-{S}utherland model},\
  }\href {https://doi.org/10.1103/PhysRevB.95.195137} {\bibfield  {journal}
  {\bibinfo  {journal} {Phys. Rev. B}\ }\textbf {\bibinfo {volume} {95}},\
  \bibinfo {pages} {195137} (\bibinfo {year} {2017})}\BibitemShut {NoStop}%
\bibitem [{\citenamefont {McClarty}\ \emph {et~al.}(2017)\citenamefont
  {McClarty}, \citenamefont {Krüger}, \citenamefont {Guidi}, \citenamefont
  {Parker}, \citenamefont {Refson}, \citenamefont {Parker}, \citenamefont
  {Prabhakaran},\ and\ \citenamefont {Coldea}}]{coldea17}%
  \BibitemOpen
  \bibfield  {author} {\bibinfo {author} {\bibfnamefont {P.}~\bibnamefont
  {McClarty}}, \bibinfo {author} {\bibfnamefont {F.}~\bibnamefont {Krüger}},
  \bibinfo {author} {\bibfnamefont {T.}~\bibnamefont {Guidi}}, \bibinfo
  {author} {\bibfnamefont {S.~F.}\ \bibnamefont {Parker}}, \bibinfo {author}
  {\bibfnamefont {K.}~\bibnamefont {Refson}}, \bibinfo {author} {\bibfnamefont
  {A.~W.}\ \bibnamefont {Parker}}, \bibinfo {author} {\bibfnamefont
  {D.}~\bibnamefont {Prabhakaran}},\ and\ \bibinfo {author} {\bibfnamefont
  {R.}~\bibnamefont {Coldea}},\ }\bibfield  {title} {\bibinfo {title}
  {Topological triplon modes and bound states in a {S}hastry–{S}utherland
  magnet},\ }\href {https://doi.org/10.1038/nphys4117} {\bibfield  {journal}
  {\bibinfo  {journal} {Nature Phys.}\ }\textbf {\bibinfo {volume} {13}},\
  \bibinfo {pages} {736} (\bibinfo {year} {2017})}\BibitemShut {NoStop}%
\bibitem [{\citenamefont {Joshi}\ and\ \citenamefont
  {Schnyder}(2017)}]{joshi17}%
  \BibitemOpen
  \bibfield  {author} {\bibinfo {author} {\bibfnamefont {D.~G.}\ \bibnamefont
  {Joshi}}\ and\ \bibinfo {author} {\bibfnamefont {A.~P.}\ \bibnamefont
  {Schnyder}},\ }\bibfield  {title} {\bibinfo {title} {Topological quantum
  paramagnet in a quantum spin ladder},\ }\href
  {https://doi.org/10.1103/PhysRevB.96.220405} {\bibfield  {journal} {\bibinfo
  {journal} {Phys. Rev. B}\ }\textbf {\bibinfo {volume} {96}},\ \bibinfo
  {pages} {220405} (\bibinfo {year} {2017})}\BibitemShut {NoStop}%
\bibitem [{\citenamefont {Haldane}(1988)}]{haldane88}%
  \BibitemOpen
  \bibfield  {author} {\bibinfo {author} {\bibfnamefont {F.~D.~M.}\
  \bibnamefont {Haldane}},\ }\bibfield  {title} {\bibinfo {title} {Model for a
  quantum {H}all effect without {L}andau levels: Condensed-matter realization
  of the "parity anomaly"},\ }\href
  {https://doi.org/10.1103/PhysRevLett.61.2015} {\bibfield  {journal} {\bibinfo
   {journal} {Phys. Rev. Lett.}\ }\textbf {\bibinfo {volume} {61}},\ \bibinfo
  {pages} {2015} (\bibinfo {year} {1988})}\BibitemShut {NoStop}%
\bibitem [{\citenamefont {Kane}(2013)}]{kane13}%
  \BibitemOpen
  \bibfield  {author} {\bibinfo {author} {\bibfnamefont {C.~L.}\ \bibnamefont
  {Kane}},\ }\bibfield  {title} {\bibinfo {title} {Topological insulators},\
  }in\ \href@noop {} {\emph {\bibinfo {booktitle} {Contemporary Concepts of
  Condensed Matter Science}}},\ Vol.~\bibinfo {volume} {6},\ \bibinfo {editor}
  {edited by\ \bibinfo {editor} {\bibfnamefont {M.}~\bibnamefont {Franz}}\ and\
  \bibinfo {editor} {\bibfnamefont {L.}~\bibnamefont {Molenkamp}}}\ (\bibinfo
  {publisher} {Elsevier, Oxford, UK},\ \bibinfo {year} {2013})\BibitemShut
  {NoStop}%
\bibitem [{\citenamefont {Rachel}(2018)}]{rachel18}%
  \BibitemOpen
  \bibfield  {author} {\bibinfo {author} {\bibfnamefont {S.}~\bibnamefont
  {Rachel}},\ }\bibfield  {title} {\bibinfo {title} {Interacting topological
  insulators: a review},\ }\href {https://doi.org/10.1088/1361-6633/aad6a6}
  {\bibfield  {journal} {\bibinfo  {journal} {Reports on Progress in Physics}\
  }\textbf {\bibinfo {volume} {81}},\ \bibinfo {pages} {116501} (\bibinfo
  {year} {2018})}\BibitemShut {NoStop}%
\bibitem [{\citenamefont {Katsura}\ \emph {et~al.}(2010)\citenamefont
  {Katsura}, \citenamefont {Nagaosa},\ and\ \citenamefont {Lee}}]{lee10}%
  \BibitemOpen
  \bibfield  {author} {\bibinfo {author} {\bibfnamefont {H.}~\bibnamefont
  {Katsura}}, \bibinfo {author} {\bibfnamefont {N.}~\bibnamefont {Nagaosa}},\
  and\ \bibinfo {author} {\bibfnamefont {P.~A.}\ \bibnamefont {Lee}},\
  }\bibfield  {title} {\bibinfo {title} {Theory of the thermal {H}all effect in
  quantum magnets},\ }\href {https://doi.org/10.1103/PhysRevLett.104.066403}
  {\bibfield  {journal} {\bibinfo  {journal} {Phys. Rev. Lett.}\ }\textbf
  {\bibinfo {volume} {104}},\ \bibinfo {pages} {066403} (\bibinfo {year}
  {2010})}\BibitemShut {NoStop}%
\bibitem [{\citenamefont {Matsumoto}\ and\ \citenamefont
  {Murakami}(2011{\natexlab{a}})}]{prl-ryo11}%
  \BibitemOpen
  \bibfield  {author} {\bibinfo {author} {\bibfnamefont {R.}~\bibnamefont
  {Matsumoto}}\ and\ \bibinfo {author} {\bibfnamefont {S.}~\bibnamefont
  {Murakami}},\ }\bibfield  {title} {\bibinfo {title} {Theoretical prediction
  of a rotating magnon wave packet in ferromagnets},\ }\href
  {https://doi.org/10.1103/PhysRevLett.106.197202} {\bibfield  {journal}
  {\bibinfo  {journal} {Phys. Rev. Lett.}\ }\textbf {\bibinfo {volume} {106}},\
  \bibinfo {pages} {197202} (\bibinfo {year} {2011}{\natexlab{a}})}\BibitemShut
  {NoStop}%
\bibitem [{\citenamefont {Matsumoto}\ and\ \citenamefont
  {Murakami}(2011{\natexlab{b}})}]{prb-ryo11}%
  \BibitemOpen
  \bibfield  {author} {\bibinfo {author} {\bibfnamefont {R.}~\bibnamefont
  {Matsumoto}}\ and\ \bibinfo {author} {\bibfnamefont {S.}~\bibnamefont
  {Murakami}},\ }\bibfield  {title} {\bibinfo {title} {Rotational motion of
  magnons and the thermal {H}all effect},\ }\href
  {https://doi.org/10.1103/PhysRevB.84.184406} {\bibfield  {journal} {\bibinfo
  {journal} {Phys. Rev. B}\ }\textbf {\bibinfo {volume} {84}},\ \bibinfo
  {pages} {184406} (\bibinfo {year} {2011}{\natexlab{b}})}\BibitemShut
  {NoStop}%
\bibitem [{\citenamefont {Chang}\ \emph {et~al.}(2023)\citenamefont {Chang},
  \citenamefont {Liu},\ and\ \citenamefont {MacDonald}}]{rmp-mc23}%
  \BibitemOpen
  \bibfield  {author} {\bibinfo {author} {\bibfnamefont {C.-Z.}\ \bibnamefont
  {Chang}}, \bibinfo {author} {\bibfnamefont {C.-X.}\ \bibnamefont {Liu}},\
  and\ \bibinfo {author} {\bibfnamefont {A.~H.}\ \bibnamefont {MacDonald}},\
  }\bibfield  {title} {\bibinfo {title} {Colloquium: Quantum anomalous {H}all
  effect},\ }\href {https://doi.org/10.1103/RevModPhys.95.011002} {\bibfield
  {journal} {\bibinfo  {journal} {Rev. Mod. Phys.}\ }\textbf {\bibinfo {volume}
  {95}},\ \bibinfo {pages} {011002} (\bibinfo {year} {2023})}\BibitemShut
  {NoStop}%
\bibitem [{\citenamefont {Ideue}\ \emph {et~al.}(2012)\citenamefont {Ideue},
  \citenamefont {Onose}, \citenamefont {Katsura}, \citenamefont {Shiomi},
  \citenamefont {Ishiwata}, \citenamefont {Nagaosa},\ and\ \citenamefont
  {Tokura}}]{tokura12}%
  \BibitemOpen
  \bibfield  {author} {\bibinfo {author} {\bibfnamefont {T.}~\bibnamefont
  {Ideue}}, \bibinfo {author} {\bibfnamefont {Y.}~\bibnamefont {Onose}},
  \bibinfo {author} {\bibfnamefont {H.}~\bibnamefont {Katsura}}, \bibinfo
  {author} {\bibfnamefont {Y.}~\bibnamefont {Shiomi}}, \bibinfo {author}
  {\bibfnamefont {S.}~\bibnamefont {Ishiwata}}, \bibinfo {author}
  {\bibfnamefont {N.}~\bibnamefont {Nagaosa}},\ and\ \bibinfo {author}
  {\bibfnamefont {Y.}~\bibnamefont {Tokura}},\ }\bibfield  {title} {\bibinfo
  {title} {Effect of lattice geometry on magnon {H}all effect in ferromagnetic
  insulators},\ }\href {https://doi.org/10.1103/PhysRevB.85.134411} {\bibfield
  {journal} {\bibinfo  {journal} {Phys. Rev. B}\ }\textbf {\bibinfo {volume}
  {85}},\ \bibinfo {pages} {134411} (\bibinfo {year} {2012})}\BibitemShut
  {NoStop}%
\bibitem [{\citenamefont {Kawano}\ and\ \citenamefont {Hotta}(2019)}]{hotta19}%
  \BibitemOpen
  \bibfield  {author} {\bibinfo {author} {\bibfnamefont {M.}~\bibnamefont
  {Kawano}}\ and\ \bibinfo {author} {\bibfnamefont {C.}~\bibnamefont {Hotta}},\
  }\bibfield  {title} {\bibinfo {title} {Thermal {H}all effect and topological
  edge states in a square-lattice antiferromagnet},\ }\href
  {https://doi.org/10.1103/PhysRevB.99.054422} {\bibfield  {journal} {\bibinfo
  {journal} {Phys. Rev. B}\ }\textbf {\bibinfo {volume} {99}},\ \bibinfo
  {pages} {054422} (\bibinfo {year} {2019})}\BibitemShut {NoStop}%
\bibitem [{\citenamefont {Cheng}\ \emph {et~al.}(2016)\citenamefont {Cheng},
  \citenamefont {Okamoto},\ and\ \citenamefont {Xiao}}]{xiao16}%
  \BibitemOpen
  \bibfield  {author} {\bibinfo {author} {\bibfnamefont {R.}~\bibnamefont
  {Cheng}}, \bibinfo {author} {\bibfnamefont {S.}~\bibnamefont {Okamoto}},\
  and\ \bibinfo {author} {\bibfnamefont {D.}~\bibnamefont {Xiao}},\ }\bibfield
  {title} {\bibinfo {title} {Spin {N}ernst effect of magnons in collinear
  antiferromagnets},\ }\href {https://doi.org/10.1103/PhysRevLett.117.217202}
  {\bibfield  {journal} {\bibinfo  {journal} {Phys. Rev. Lett.}\ }\textbf
  {\bibinfo {volume} {117}},\ \bibinfo {pages} {217202} (\bibinfo {year}
  {2016})}\BibitemShut {NoStop}%
\bibitem [{\citenamefont {Doki}\ \emph {et~al.}(2018)\citenamefont {Doki},
  \citenamefont {Akazawa}, \citenamefont {Lee}, \citenamefont {Han},
  \citenamefont {Sugii}, \citenamefont {Shimozawa}, \citenamefont {Kawashima},
  \citenamefont {Oda}, \citenamefont {Yoshida},\ and\ \citenamefont
  {Yamashita}}]{doki18}%
  \BibitemOpen
  \bibfield  {author} {\bibinfo {author} {\bibfnamefont {H.}~\bibnamefont
  {Doki}}, \bibinfo {author} {\bibfnamefont {M.}~\bibnamefont {Akazawa}},
  \bibinfo {author} {\bibfnamefont {H.-Y.}\ \bibnamefont {Lee}}, \bibinfo
  {author} {\bibfnamefont {J.~H.}\ \bibnamefont {Han}}, \bibinfo {author}
  {\bibfnamefont {K.}~\bibnamefont {Sugii}}, \bibinfo {author} {\bibfnamefont
  {M.}~\bibnamefont {Shimozawa}}, \bibinfo {author} {\bibfnamefont
  {N.}~\bibnamefont {Kawashima}}, \bibinfo {author} {\bibfnamefont
  {M.}~\bibnamefont {Oda}}, \bibinfo {author} {\bibfnamefont {H.}~\bibnamefont
  {Yoshida}},\ and\ \bibinfo {author} {\bibfnamefont {M.}~\bibnamefont
  {Yamashita}},\ }\bibfield  {title} {\bibinfo {title} {Spin thermal {H}all
  conductivity of a kagome antiferromagnet},\ }\href
  {https://doi.org/10.1103/PhysRevLett.121.097203} {\bibfield  {journal}
  {\bibinfo  {journal} {Phys. Rev. Lett.}\ }\textbf {\bibinfo {volume} {121}},\
  \bibinfo {pages} {097203} (\bibinfo {year} {2018})}\BibitemShut {NoStop}%
\bibitem [{\citenamefont {Samajdar}\ \emph {et~al.}(2019)\citenamefont
  {Samajdar}, \citenamefont {Chatterjee}, \citenamefont {Sachdev},\ and\
  \citenamefont {Scheurer}}]{rhine19}%
  \BibitemOpen
  \bibfield  {author} {\bibinfo {author} {\bibfnamefont {R.}~\bibnamefont
  {Samajdar}}, \bibinfo {author} {\bibfnamefont {S.}~\bibnamefont
  {Chatterjee}}, \bibinfo {author} {\bibfnamefont {S.}~\bibnamefont
  {Sachdev}},\ and\ \bibinfo {author} {\bibfnamefont {M.~S.}\ \bibnamefont
  {Scheurer}},\ }\bibfield  {title} {\bibinfo {title} {Thermal {H}all effect in
  square-lattice spin liquids: A {S}chwinger boson mean-field study},\ }\href
  {https://doi.org/10.1103/PhysRevB.99.165126} {\bibfield  {journal} {\bibinfo
  {journal} {Phys. Rev. B}\ }\textbf {\bibinfo {volume} {99}},\ \bibinfo
  {pages} {165126} (\bibinfo {year} {2019})}\BibitemShut {NoStop}%
\bibitem [{\citenamefont {Esaki}\ \emph {et~al.}(2023)\citenamefont {Esaki},
  \citenamefont {Akagi},\ and\ \citenamefont {Katsura}}]{esaki23}%
  \BibitemOpen
  \bibfield  {author} {\bibinfo {author} {\bibfnamefont {N.}~\bibnamefont
  {Esaki}}, \bibinfo {author} {\bibfnamefont {Y.}~\bibnamefont {Akagi}},\ and\
  \bibinfo {author} {\bibfnamefont {H.}~\bibnamefont {Katsura}},\ }\href@noop
  {} {\bibinfo {title} {Electric field induced thermal {H}all effect of
  triplons in the quantum dimer magnets {$X$}{C}u{C}l$_{3}$ ({$X =$ Tl, K})}}
  (\bibinfo {year} {2023}),\ \Eprint {https://arxiv.org/abs/2309.12812}
  {arXiv:2309.12812 [cond-mat.mes-hall]} \BibitemShut {NoStop}%
\bibitem [{\citenamefont {Chen}\ \emph {et~al.}(2018)\citenamefont {Chen},
  \citenamefont {Chung}, \citenamefont {Gao}, \citenamefont {Chen},
  \citenamefont {Stone}, \citenamefont {Kolesnikov}, \citenamefont {Huang},\
  and\ \citenamefont {Dai}}]{chen18}%
  \BibitemOpen
  \bibfield  {author} {\bibinfo {author} {\bibfnamefont {L.}~\bibnamefont
  {Chen}}, \bibinfo {author} {\bibfnamefont {J.-H.}\ \bibnamefont {Chung}},
  \bibinfo {author} {\bibfnamefont {B.}~\bibnamefont {Gao}}, \bibinfo {author}
  {\bibfnamefont {T.}~\bibnamefont {Chen}}, \bibinfo {author} {\bibfnamefont
  {M.~B.}\ \bibnamefont {Stone}}, \bibinfo {author} {\bibfnamefont {A.~I.}\
  \bibnamefont {Kolesnikov}}, \bibinfo {author} {\bibfnamefont
  {Q.}~\bibnamefont {Huang}},\ and\ \bibinfo {author} {\bibfnamefont
  {P.}~\bibnamefont {Dai}},\ }\bibfield  {title} {\bibinfo {title} {Topological
  spin excitations in honeycomb ferromagnet ${\mathrm{cri}}_{3}$},\ }\href
  {https://doi.org/10.1103/PhysRevX.8.041028} {\bibfield  {journal} {\bibinfo
  {journal} {Phys. Rev. X}\ }\textbf {\bibinfo {volume} {8}},\ \bibinfo {pages}
  {041028} (\bibinfo {year} {2018})}\BibitemShut {NoStop}%
\bibitem [{\citenamefont {Joshi}\ and\ \citenamefont
  {Schnyder}(2019)}]{joshi19}%
  \BibitemOpen
  \bibfield  {author} {\bibinfo {author} {\bibfnamefont {D.~G.}\ \bibnamefont
  {Joshi}}\ and\ \bibinfo {author} {\bibfnamefont {A.~P.}\ \bibnamefont
  {Schnyder}},\ }\bibfield  {title} {\bibinfo {title} {${Z}_{2}$ topological
  quantum paramagnet on a honeycomb bilayer},\ }\href
  {https://doi.org/10.1103/PhysRevB.100.020407} {\bibfield  {journal} {\bibinfo
   {journal} {Phys. Rev. B}\ }\textbf {\bibinfo {volume} {100}},\ \bibinfo
  {pages} {020407} (\bibinfo {year} {2019})}\BibitemShut {NoStop}%
\bibitem [{\citenamefont {Dzyaloshinsky}(1958)}]{dm01}%
  \BibitemOpen
  \bibfield  {author} {\bibinfo {author} {\bibfnamefont {I.}~\bibnamefont
  {Dzyaloshinsky}},\ }\bibfield  {title} {\bibinfo {title} {A thermodynamic
  theory of “weak” ferromagnetism of antiferromagnetics},\ }\href
  {https://doi.org/10.1016/0022-3697(58)90076-3} {\bibfield  {journal}
  {\bibinfo  {journal} {J. Phys. Chem. Solid}\ }\textbf {\bibinfo {volume}
  {4}},\ \bibinfo {pages} {241} (\bibinfo {year} {1958})}\BibitemShut {NoStop}%
\bibitem [{\citenamefont {Moriya}(1960{\natexlab{a}})}]{dm03}%
  \BibitemOpen
  \bibfield  {author} {\bibinfo {author} {\bibfnamefont {T.}~\bibnamefont
  {Moriya}},\ }\bibfield  {title} {\bibinfo {title} {New mechanism of
  anisotropic superexchange interaction},\ }\href
  {https://doi.org/10.1103/PhysRevLett.4.228} {\bibfield  {journal} {\bibinfo
  {journal} {Phys. Rev. Lett.}\ }\textbf {\bibinfo {volume} {4}},\ \bibinfo
  {pages} {228} (\bibinfo {year} {1960}{\natexlab{a}})}\BibitemShut {NoStop}%
\bibitem [{\citenamefont {Moriya}(1960{\natexlab{b}})}]{dm02}%
  \BibitemOpen
  \bibfield  {author} {\bibinfo {author} {\bibfnamefont {T.}~\bibnamefont
  {Moriya}},\ }\bibfield  {title} {\bibinfo {title} {Anisotropic superexchange
  interaction and weak ferromagnetism},\ }\href
  {https://doi.org/10.1103/PhysRev.120.91} {\bibfield  {journal} {\bibinfo
  {journal} {Phys. Rev.}\ }\textbf {\bibinfo {volume} {120}},\ \bibinfo {pages}
  {91} (\bibinfo {year} {1960}{\natexlab{b}})}\BibitemShut {NoStop}%
\bibitem [{\citenamefont {Coffey}\ \emph {et~al.}(1990)\citenamefont {Coffey},
  \citenamefont {Bedell},\ and\ \citenamefont {Trugman}}]{dm-cuprates01}%
  \BibitemOpen
  \bibfield  {author} {\bibinfo {author} {\bibfnamefont {D.}~\bibnamefont
  {Coffey}}, \bibinfo {author} {\bibfnamefont {K.~S.}\ \bibnamefont {Bedell}},\
  and\ \bibinfo {author} {\bibfnamefont {S.~A.}\ \bibnamefont {Trugman}},\
  }\bibfield  {title} {\bibinfo {title} {Effective spin hamiltonian for the
  {C}u{O} planes in ${\mathrm{la}}_{2}$${\mathrm{cuo}}_{4}$ and
  metamagnetism},\ }\href {https://doi.org/10.1103/PhysRevB.42.6509} {\bibfield
   {journal} {\bibinfo  {journal} {Phys. Rev. B}\ }\textbf {\bibinfo {volume}
  {42}},\ \bibinfo {pages} {6509} (\bibinfo {year} {1990})}\BibitemShut
  {NoStop}%
\bibitem [{\citenamefont {Coffey}\ \emph {et~al.}(1991)\citenamefont {Coffey},
  \citenamefont {Rice},\ and\ \citenamefont {Zhang}}]{dm-cuprates02}%
  \BibitemOpen
  \bibfield  {author} {\bibinfo {author} {\bibfnamefont {D.}~\bibnamefont
  {Coffey}}, \bibinfo {author} {\bibfnamefont {T.~M.}\ \bibnamefont {Rice}},\
  and\ \bibinfo {author} {\bibfnamefont {F.~C.}\ \bibnamefont {Zhang}},\
  }\bibfield  {title} {\bibinfo {title} {Dzyaloshinskii-moriya interaction in
  the cuprates},\ }\href {https://doi.org/10.1103/PhysRevB.44.10112} {\bibfield
   {journal} {\bibinfo  {journal} {Phys. Rev. B}\ }\textbf {\bibinfo {volume}
  {44}},\ \bibinfo {pages} {10112} (\bibinfo {year} {1991})}\BibitemShut
  {NoStop}%
\bibitem [{\citenamefont {Koshibae}\ \emph {et~al.}(1994)\citenamefont
  {Koshibae}, \citenamefont {Ohta},\ and\ \citenamefont
  {Maekawa}}]{dm-cuprates03}%
  \BibitemOpen
  \bibfield  {author} {\bibinfo {author} {\bibfnamefont {W.}~\bibnamefont
  {Koshibae}}, \bibinfo {author} {\bibfnamefont {Y.}~\bibnamefont {Ohta}},\
  and\ \bibinfo {author} {\bibfnamefont {S.}~\bibnamefont {Maekawa}},\
  }\bibfield  {title} {\bibinfo {title} {Theory of {D}zyaloshinski-{M}oriya
  antiferromagnetism in distorted ${\mathrm{cuo}}_{2}$ and ${\mathrm{nio}}_{2}$
  planes},\ }\href {https://doi.org/10.1103/PhysRevB.50.3767} {\bibfield
  {journal} {\bibinfo  {journal} {Phys. Rev. B}\ }\textbf {\bibinfo {volume}
  {50}},\ \bibinfo {pages} {3767} (\bibinfo {year} {1994})}\BibitemShut
  {NoStop}%
\bibitem [{\citenamefont {Grissonnanche}\ \emph {et~al.}(2019)\citenamefont
  {Grissonnanche}, \citenamefont {Legros}, \citenamefont {Badoux},
  \citenamefont {Lefrançois}, \citenamefont {Zatko}, \citenamefont {Lizaire},
  \citenamefont {Laliberté}, \citenamefont {Gourgout}, \citenamefont {Zhou},
  \citenamefont {Pyon}, \citenamefont {Takayama}, \citenamefont {Takagi},
  \citenamefont {Ono}, \citenamefont {Doiron-Leyraud},\ and\ \citenamefont
  {Taillefer}}]{kappa-LSCO}%
  \BibitemOpen
  \bibfield  {author} {\bibinfo {author} {\bibfnamefont {G.}~\bibnamefont
  {Grissonnanche}}, \bibinfo {author} {\bibfnamefont {A.}~\bibnamefont
  {Legros}}, \bibinfo {author} {\bibfnamefont {S.}~\bibnamefont {Badoux}},
  \bibinfo {author} {\bibfnamefont {E.}~\bibnamefont {Lefrançois}}, \bibinfo
  {author} {\bibfnamefont {V.}~\bibnamefont {Zatko}}, \bibinfo {author}
  {\bibfnamefont {M.}~\bibnamefont {Lizaire}}, \bibinfo {author} {\bibfnamefont
  {F.}~\bibnamefont {Laliberté}}, \bibinfo {author} {\bibfnamefont
  {A.}~\bibnamefont {Gourgout}}, \bibinfo {author} {\bibfnamefont {J.~S.}\
  \bibnamefont {Zhou}}, \bibinfo {author} {\bibfnamefont {S.}~\bibnamefont
  {Pyon}}, \bibinfo {author} {\bibfnamefont {T.}~\bibnamefont {Takayama}},
  \bibinfo {author} {\bibfnamefont {H.}~\bibnamefont {Takagi}}, \bibinfo
  {author} {\bibfnamefont {S.}~\bibnamefont {Ono}}, \bibinfo {author}
  {\bibfnamefont {N.}~\bibnamefont {Doiron-Leyraud}},\ and\ \bibinfo {author}
  {\bibfnamefont {L.}~\bibnamefont {Taillefer}},\ }\bibfield  {title} {\bibinfo
  {title} {Giant thermal {H}all conductivity in the pseudogap phase of cuprate
  superconductors},\ }\href {https://doi.org/10.1038/s41586-019-1375-0}
  {\bibfield  {journal} {\bibinfo  {journal} {Nature}\ }\textbf {\bibinfo
  {volume} {571}},\ \bibinfo {pages} {376} (\bibinfo {year}
  {2019})}\BibitemShut {NoStop}%
\bibitem [{\citenamefont {Grissonnanche}\ \emph {et~al.}(2020)\citenamefont
  {Grissonnanche}, \citenamefont {Thériault}, \citenamefont {Gourgout},
  \citenamefont {Boulanger}, \citenamefont {Lefrançois}, \citenamefont
  {Ataei}, \citenamefont {Lalibert\'e}, \citenamefont {Dion}, \citenamefont
  {Zhou}, \citenamefont {Pyon}, \citenamefont {Takayama}, \citenamefont
  {Takagi}, \citenamefont {Doiron-Leyraud},\ and\ \citenamefont
  {Taillefer}}]{kappa-LSCO2}%
  \BibitemOpen
  \bibfield  {author} {\bibinfo {author} {\bibfnamefont {G.}~\bibnamefont
  {Grissonnanche}}, \bibinfo {author} {\bibfnamefont {S.}~\bibnamefont
  {Thériault}}, \bibinfo {author} {\bibfnamefont {A.}~\bibnamefont
  {Gourgout}}, \bibinfo {author} {\bibfnamefont {M.~E.}\ \bibnamefont
  {Boulanger}}, \bibinfo {author} {\bibfnamefont {E.}~\bibnamefont
  {Lefrançois}}, \bibinfo {author} {\bibfnamefont {A.}~\bibnamefont {Ataei}},
  \bibinfo {author} {\bibfnamefont {F.}~\bibnamefont {Lalibert\'e}}, \bibinfo
  {author} {\bibfnamefont {M.}~\bibnamefont {Dion}}, \bibinfo {author}
  {\bibfnamefont {J.~S.}\ \bibnamefont {Zhou}}, \bibinfo {author}
  {\bibfnamefont {S.}~\bibnamefont {Pyon}}, \bibinfo {author} {\bibfnamefont
  {T.}~\bibnamefont {Takayama}}, \bibinfo {author} {\bibfnamefont
  {H.}~\bibnamefont {Takagi}}, \bibinfo {author} {\bibfnamefont
  {N.}~\bibnamefont {Doiron-Leyraud}},\ and\ \bibinfo {author} {\bibfnamefont
  {L.}~\bibnamefont {Taillefer}},\ }\bibfield  {title} {\bibinfo {title}
  {Chiral phonons in the pseudogap phase of cuprates},\ }\href
  {https://doi.org/10.1038/s41567-020-0965-y} {\bibfield  {journal} {\bibinfo
  {journal} {Nature Physics}\ }\textbf {\bibinfo {volume} {16}},\ \bibinfo
  {pages} {1108} (\bibinfo {year} {2020})}\BibitemShut {NoStop}%
\bibitem [{\citenamefont {Boulanger}\ \emph {et~al.}(2020)\citenamefont
  {Boulanger}, \citenamefont {Grissonnanche}, \citenamefont {Badoux},
  \citenamefont {Allaire}, \citenamefont {Lefrançois}, \citenamefont {Legros},
  \citenamefont {Gourgout}, \citenamefont {Dion}, \citenamefont {Wang},
  \citenamefont {Chen}, \citenamefont {Liang}, \citenamefont {Hardy},
  \citenamefont {Bonn},\ and\ \citenamefont {Taillefer}}]{kappa-LSCO3}%
  \BibitemOpen
  \bibfield  {author} {\bibinfo {author} {\bibfnamefont {M.-E.}\ \bibnamefont
  {Boulanger}}, \bibinfo {author} {\bibfnamefont {G.}~\bibnamefont
  {Grissonnanche}}, \bibinfo {author} {\bibfnamefont {S.}~\bibnamefont
  {Badoux}}, \bibinfo {author} {\bibfnamefont {A.}~\bibnamefont {Allaire}},
  \bibinfo {author} {\bibfnamefont {E.}~\bibnamefont {Lefrançois}}, \bibinfo
  {author} {\bibfnamefont {A.}~\bibnamefont {Legros}}, \bibinfo {author}
  {\bibfnamefont {A.}~\bibnamefont {Gourgout}}, \bibinfo {author}
  {\bibfnamefont {M.}~\bibnamefont {Dion}}, \bibinfo {author} {\bibfnamefont
  {C.~H.}\ \bibnamefont {Wang}}, \bibinfo {author} {\bibfnamefont {X.~H.}\
  \bibnamefont {Chen}}, \bibinfo {author} {\bibfnamefont {R.}~\bibnamefont
  {Liang}}, \bibinfo {author} {\bibfnamefont {W.~N.}\ \bibnamefont {Hardy}},
  \bibinfo {author} {\bibfnamefont {D.~A.}\ \bibnamefont {Bonn}},\ and\
  \bibinfo {author} {\bibfnamefont {L.}~\bibnamefont {Taillefer}},\ }\bibfield
  {title} {\bibinfo {title} {Thermal {H}all conductivity in the cuprate {M}ott
  insulators {Nd2CuO4} and {Sr2CuO2Cl2}},\ }\href
  {https://doi.org/10.1038/s41467-020-18881-z} {\bibfield  {journal} {\bibinfo
  {journal} {Nature Communications}\ }\textbf {\bibinfo {volume} {11}},\
  \bibinfo {pages} {5325} (\bibinfo {year} {2020})}\BibitemShut {NoStop}%
\bibitem [{\citenamefont {Sachdev}\ and\ \citenamefont
  {Bhatt}(1990)}]{sachdev90}%
  \BibitemOpen
  \bibfield  {author} {\bibinfo {author} {\bibfnamefont {S.}~\bibnamefont
  {Sachdev}}\ and\ \bibinfo {author} {\bibfnamefont {R.~N.}\ \bibnamefont
  {Bhatt}},\ }\bibfield  {title} {\bibinfo {title} {Bond-operator
  representation of quantum spins: Mean-field theory of frustrated quantum
  {H}eisenberg antiferromagnets},\ }\href
  {https://doi.org/10.1103/PhysRevB.41.9323} {\bibfield  {journal} {\bibinfo
  {journal} {Phys. Rev. B}\ }\textbf {\bibinfo {volume} {41}},\ \bibinfo
  {pages} {9323} (\bibinfo {year} {1990})}\BibitemShut {NoStop}%
\bibitem [{\citenamefont {Tabunshchyk}\ and\ \citenamefont
  {Gooding}(2005)}]{gooding05}%
  \BibitemOpen
  \bibfield  {author} {\bibinfo {author} {\bibfnamefont {K.~V.}\ \bibnamefont
  {Tabunshchyk}}\ and\ \bibinfo {author} {\bibfnamefont {R.~J.}\ \bibnamefont
  {Gooding}},\ }\bibfield  {title} {\bibinfo {title} {Magnetic susceptibility
  of a $\mathrm{Cu}{\mathrm{o}}_{2}$ plane in the
  ${\mathrm{la}}_{2}\mathrm{Cu}{\mathrm{o}}_{4}$ system: I. {R}andom-phase
  approximation treatment of the {D}zyaloshinskii-{M}oriya interactions},\
  }\href {https://doi.org/10.1103/PhysRevB.71.214418} {\bibfield  {journal}
  {\bibinfo  {journal} {Phys. Rev. B}\ }\textbf {\bibinfo {volume} {71}},\
  \bibinfo {pages} {214418} (\bibinfo {year} {2005})}\BibitemShut {NoStop}%
\bibitem [{\citenamefont {Silva~Neto}\ \emph {et~al.}(2006)\citenamefont
  {Silva~Neto}, \citenamefont {Benfatto}, \citenamefont {Juricic},\ and\
  \citenamefont {Morais~Smith}}]{marcelo06}%
  \BibitemOpen
  \bibfield  {author} {\bibinfo {author} {\bibfnamefont {M.~B.}\ \bibnamefont
  {Silva~Neto}}, \bibinfo {author} {\bibfnamefont {L.}~\bibnamefont
  {Benfatto}}, \bibinfo {author} {\bibfnamefont {V.}~\bibnamefont {Juricic}},\
  and\ \bibinfo {author} {\bibfnamefont {C.}~\bibnamefont {Morais~Smith}},\
  }\bibfield  {title} {\bibinfo {title} {Magnetic susceptibility anisotropies
  in a two-dimensional quantum {H}eisenberg antiferromagnet with
  {D}zyaloshinskii-{M}oriya interactions},\ }\href
  {https://doi.org/10.1103/PhysRevB.73.045132} {\bibfield  {journal} {\bibinfo
  {journal} {Phys. Rev. B}\ }\textbf {\bibinfo {volume} {73}},\ \bibinfo
  {pages} {045132} (\bibinfo {year} {2006})}\BibitemShut {NoStop}%
\bibitem [{\citenamefont {Benfatto}\ and\ \citenamefont
  {Silva~Neto}(2006)}]{lara06}%
  \BibitemOpen
  \bibfield  {author} {\bibinfo {author} {\bibfnamefont {L.}~\bibnamefont
  {Benfatto}}\ and\ \bibinfo {author} {\bibfnamefont {M.~B.}\ \bibnamefont
  {Silva~Neto}},\ }\bibfield  {title} {\bibinfo {title} {Field dependence of
  the magnetic spectrum in anisotropic and {D}zyaloshinskii-{M}oriya
  antiferromagnets. i. theory},\ }\href
  {https://doi.org/10.1103/PhysRevB.74.024415} {\bibfield  {journal} {\bibinfo
  {journal} {Phys. Rev. B}\ }\textbf {\bibinfo {volume} {74}},\ \bibinfo
  {pages} {024415} (\bibinfo {year} {2006})}\BibitemShut {NoStop}%
\bibitem [{\citenamefont {Chandra}\ and\ \citenamefont
  {Doucot}(1988)}]{docout88}%
  \BibitemOpen
  \bibfield  {author} {\bibinfo {author} {\bibfnamefont {P.}~\bibnamefont
  {Chandra}}\ and\ \bibinfo {author} {\bibfnamefont {B.}~\bibnamefont
  {Doucot}},\ }\bibfield  {title} {\bibinfo {title} {Possible spin-liquid state
  at large ${S}$ for the frustrated square {H}eisenberg lattice},\ }\href
  {https://doi.org/10.1103/PhysRevB.38.9335} {\bibfield  {journal} {\bibinfo
  {journal} {Phys. Rev. B}\ }\textbf {\bibinfo {volume} {38}},\ \bibinfo
  {pages} {9335} (\bibinfo {year} {1988})}\BibitemShut {NoStop}%
\bibitem [{\citenamefont {Gelfand}\ \emph {et~al.}(1989)\citenamefont
  {Gelfand}, \citenamefont {Singh},\ and\ \citenamefont {Huse}}]{huse89}%
  \BibitemOpen
  \bibfield  {author} {\bibinfo {author} {\bibfnamefont {M.~P.}\ \bibnamefont
  {Gelfand}}, \bibinfo {author} {\bibfnamefont {R.~R.~P.}\ \bibnamefont
  {Singh}},\ and\ \bibinfo {author} {\bibfnamefont {D.~A.}\ \bibnamefont
  {Huse}},\ }\bibfield  {title} {\bibinfo {title} {Zero-temperature ordering in
  two-dimensional frustrated quantum {H}eisenberg antiferromagnets},\ }\href
  {https://doi.org/10.1103/PhysRevB.40.10801} {\bibfield  {journal} {\bibinfo
  {journal} {Phys. Rev. B}\ }\textbf {\bibinfo {volume} {40}},\ \bibinfo
  {pages} {10801} (\bibinfo {year} {1989})}\BibitemShut {NoStop}%
\bibitem [{\citenamefont {Dagotto}\ and\ \citenamefont
  {Moreo}(1989)}]{moreo89}%
  \BibitemOpen
  \bibfield  {author} {\bibinfo {author} {\bibfnamefont {E.}~\bibnamefont
  {Dagotto}}\ and\ \bibinfo {author} {\bibfnamefont {A.}~\bibnamefont
  {Moreo}},\ }\bibfield  {title} {\bibinfo {title} {Phase diagram of the
  frustrated spin-1/2 {H}eisenberg antiferromagnet in 2 dimensions},\ }\href
  {https://doi.org/10.1103/PhysRevLett.63.2148} {\bibfield  {journal} {\bibinfo
   {journal} {Phys. Rev. Lett.}\ }\textbf {\bibinfo {volume} {63}},\ \bibinfo
  {pages} {2148} (\bibinfo {year} {1989})}\BibitemShut {NoStop}%
\bibitem [{\citenamefont {Figueirido}\ \emph {et~al.}(1990)\citenamefont
  {Figueirido}, \citenamefont {Karlhede}, \citenamefont {Kivelson},
  \citenamefont {Sondhi}, \citenamefont {Rocek},\ and\ \citenamefont
  {Rokhsar}}]{sondhi90}%
  \BibitemOpen
  \bibfield  {author} {\bibinfo {author} {\bibfnamefont {F.}~\bibnamefont
  {Figueirido}}, \bibinfo {author} {\bibfnamefont {A.}~\bibnamefont
  {Karlhede}}, \bibinfo {author} {\bibfnamefont {S.}~\bibnamefont {Kivelson}},
  \bibinfo {author} {\bibfnamefont {S.}~\bibnamefont {Sondhi}}, \bibinfo
  {author} {\bibfnamefont {M.}~\bibnamefont {Rocek}},\ and\ \bibinfo {author}
  {\bibfnamefont {D.~S.}\ \bibnamefont {Rokhsar}},\ }\bibfield  {title}
  {\bibinfo {title} {Exact diagonalization of finite frustrated spin-1/2
  {H}eisenberg models},\ }\href {https://doi.org/10.1103/PhysRevB.41.4619}
  {\bibfield  {journal} {\bibinfo  {journal} {Phys. Rev. B}\ }\textbf {\bibinfo
  {volume} {41}},\ \bibinfo {pages} {4619} (\bibinfo {year}
  {1990})}\BibitemShut {NoStop}%
\bibitem [{\citenamefont {de~Oliveira}(1991)}]{oliveira91}%
  \BibitemOpen
  \bibfield  {author} {\bibinfo {author} {\bibfnamefont {M.~J.}\ \bibnamefont
  {de~Oliveira}},\ }\bibfield  {title} {\bibinfo {title} {Phase diagram of the
  spin-1/2 {H}eisenberg antiferromagnet on a square lattice with nearest- and
  next-nearest-neighbor couplings},\ }\href
  {https://doi.org/10.1103/PhysRevB.43.6181} {\bibfield  {journal} {\bibinfo
  {journal} {Phys. Rev. B}\ }\textbf {\bibinfo {volume} {43}},\ \bibinfo
  {pages} {6181} (\bibinfo {year} {1991})}\BibitemShut {NoStop}%
\bibitem [{\citenamefont {Chubukov}\ and\ \citenamefont
  {Jolicoeur}(1991)}]{chubukov91}%
  \BibitemOpen
  \bibfield  {author} {\bibinfo {author} {\bibfnamefont {A.~V.}\ \bibnamefont
  {Chubukov}}\ and\ \bibinfo {author} {\bibfnamefont {T.}~\bibnamefont
  {Jolicoeur}},\ }\bibfield  {title} {\bibinfo {title} {Dimer stability region
  in a frustrated quantum {H}eisenberg antiferromagnet},\ }\href
  {https://doi.org/10.1103/PhysRevB.44.12050} {\bibfield  {journal} {\bibinfo
  {journal} {Phys. Rev. B}\ }\textbf {\bibinfo {volume} {44}},\ \bibinfo
  {pages} {12050} (\bibinfo {year} {1991})}\BibitemShut {NoStop}%
\bibitem [{\citenamefont {Read}\ and\ \citenamefont {Sachdev}(1991)}]{read91}%
  \BibitemOpen
  \bibfield  {author} {\bibinfo {author} {\bibfnamefont {N.}~\bibnamefont
  {Read}}\ and\ \bibinfo {author} {\bibfnamefont {S.}~\bibnamefont {Sachdev}},\
  }\bibfield  {title} {\bibinfo {title} {Large-{N} expansion for frustrated
  quantum antiferromagnets},\ }\href
  {https://doi.org/10.1103/PhysRevLett.66.1773} {\bibfield  {journal} {\bibinfo
   {journal} {Phys. Rev. Lett.}\ }\textbf {\bibinfo {volume} {66}},\ \bibinfo
  {pages} {1773} (\bibinfo {year} {1991})}\BibitemShut {NoStop}%
\bibitem [{\citenamefont {Schulz}\ and\ \citenamefont
  {Ziman}(1992)}]{schulz92}%
  \BibitemOpen
  \bibfield  {author} {\bibinfo {author} {\bibfnamefont {H.~J.}\ \bibnamefont
  {Schulz}}\ and\ \bibinfo {author} {\bibfnamefont {T.~A.~L.}\ \bibnamefont
  {Ziman}},\ }\bibfield  {title} {\bibinfo {title} {Finite-size scaling for the
  two-dimensional frustrated quantum {H}eisenberg antiferromagnet},\ }\href
  {https://doi.org/10.1209/0295-5075/18/4/013} {\bibfield  {journal} {\bibinfo
  {journal} {Europhysics Letters}\ }\textbf {\bibinfo {volume} {18}},\ \bibinfo
  {pages} {355} (\bibinfo {year} {1992})}\BibitemShut {NoStop}%
\bibitem [{\citenamefont {Poilblanc}\ \emph {et~al.}(1991)\citenamefont
  {Poilblanc}, \citenamefont {Gagliano}, \citenamefont {Bacci},\ and\
  \citenamefont {Dagotto}}]{dagotto91}%
  \BibitemOpen
  \bibfield  {author} {\bibinfo {author} {\bibfnamefont {D.}~\bibnamefont
  {Poilblanc}}, \bibinfo {author} {\bibfnamefont {E.}~\bibnamefont {Gagliano}},
  \bibinfo {author} {\bibfnamefont {S.}~\bibnamefont {Bacci}},\ and\ \bibinfo
  {author} {\bibfnamefont {E.}~\bibnamefont {Dagotto}},\ }\bibfield  {title}
  {\bibinfo {title} {Static and dynamical correlations in a spin-1/2 frustrated
  antiferromagnet},\ }\href {https://doi.org/10.1103/PhysRevB.43.10970}
  {\bibfield  {journal} {\bibinfo  {journal} {Phys. Rev. B}\ }\textbf {\bibinfo
  {volume} {43}},\ \bibinfo {pages} {10970} (\bibinfo {year}
  {1991})}\BibitemShut {NoStop}%
\bibitem [{\citenamefont {Igarashi}(1993)}]{igarashi93}%
  \BibitemOpen
  \bibfield  {author} {\bibinfo {author} {\bibfnamefont {J.-i.}\ \bibnamefont
  {Igarashi}},\ }\bibfield  {title} {\bibinfo {title} {1/{S} expansion in a
  two-dimensional frustrated {H}eisenberg antiferromagnet},\ }\href
  {https://doi.org/10.1143/JPSJ.62.4449} {\bibfield  {journal} {\bibinfo
  {journal} {Journal of the Physical Society of Japan}\ }\textbf {\bibinfo
  {volume} {62}},\ \bibinfo {pages} {4449} (\bibinfo {year}
  {1993})}\BibitemShut {NoStop}%
\bibitem [{\citenamefont {Einarsson}\ and\ \citenamefont
  {Schulz}(1995)}]{schulz95}%
  \BibitemOpen
  \bibfield  {author} {\bibinfo {author} {\bibfnamefont {T.}~\bibnamefont
  {Einarsson}}\ and\ \bibinfo {author} {\bibfnamefont {H.~J.}\ \bibnamefont
  {Schulz}},\ }\bibfield  {title} {\bibinfo {title} {Direct calculation of the
  spin stiffness in the ${\mathit{j}}_{1}$-${\mathit{j}}_{2}$ {H}eisenberg
  antiferromagnet},\ }\href {https://doi.org/10.1103/PhysRevB.51.6151}
  {\bibfield  {journal} {\bibinfo  {journal} {Phys. Rev. B}\ }\textbf {\bibinfo
  {volume} {51}},\ \bibinfo {pages} {6151} (\bibinfo {year}
  {1995})}\BibitemShut {NoStop}%
\bibitem [{\citenamefont {Oitmaa}\ and\ \citenamefont
  {Weihong}(1996)}]{oitmaa96}%
  \BibitemOpen
  \bibfield  {author} {\bibinfo {author} {\bibfnamefont {J.}~\bibnamefont
  {Oitmaa}}\ and\ \bibinfo {author} {\bibfnamefont {Z.}~\bibnamefont
  {Weihong}},\ }\bibfield  {title} {\bibinfo {title} {Series expansion for the
  ${\mathit{j}}_{1}$-${\mathit{j}}_{2}$ {H}eisenberg antiferromagnet on a
  square lattice},\ }\href {https://doi.org/10.1103/PhysRevB.54.3022}
  {\bibfield  {journal} {\bibinfo  {journal} {Phys. Rev. B}\ }\textbf {\bibinfo
  {volume} {54}},\ \bibinfo {pages} {3022} (\bibinfo {year}
  {1996})}\BibitemShut {NoStop}%
\bibitem [{\citenamefont {Zhitomirsky}\ and\ \citenamefont
  {Ueda}(1996)}]{zhito96}%
  \BibitemOpen
  \bibfield  {author} {\bibinfo {author} {\bibfnamefont {M.~E.}\ \bibnamefont
  {Zhitomirsky}}\ and\ \bibinfo {author} {\bibfnamefont {K.}~\bibnamefont
  {Ueda}},\ }\bibfield  {title} {\bibinfo {title} {Valence-bond crystal phase
  of a frustrated spin-1/2 square-lattice antiferromagnet},\ }\href
  {https://doi.org/10.1103/PhysRevB.54.9007} {\bibfield  {journal} {\bibinfo
  {journal} {Phys. Rev. B}\ }\textbf {\bibinfo {volume} {54}},\ \bibinfo
  {pages} {9007} (\bibinfo {year} {1996})}\BibitemShut {NoStop}%
\bibitem [{\citenamefont {Singh}\ \emph {et~al.}(1999)\citenamefont {Singh},
  \citenamefont {Weihong}, \citenamefont {Hamer},\ and\ \citenamefont
  {Oitmaa}}]{singh99}%
  \BibitemOpen
  \bibfield  {author} {\bibinfo {author} {\bibfnamefont {R.~R.~P.}\
  \bibnamefont {Singh}}, \bibinfo {author} {\bibfnamefont {Z.}~\bibnamefont
  {Weihong}}, \bibinfo {author} {\bibfnamefont {C.~J.}\ \bibnamefont {Hamer}},\
  and\ \bibinfo {author} {\bibfnamefont {J.}~\bibnamefont {Oitmaa}},\
  }\bibfield  {title} {\bibinfo {title} {Dimer order with striped correlations
  in the ${J}_{1}{\ensuremath{-}j}_{2}$ {H}eisenberg model},\ }\href
  {https://doi.org/10.1103/PhysRevB.60.7278} {\bibfield  {journal} {\bibinfo
  {journal} {Phys. Rev. B}\ }\textbf {\bibinfo {volume} {60}},\ \bibinfo
  {pages} {7278} (\bibinfo {year} {1999})}\BibitemShut {NoStop}%
\bibitem [{\citenamefont {Kotov}\ \emph {et~al.}(1999)\citenamefont {Kotov},
  \citenamefont {Oitmaa}, \citenamefont {Sushkov},\ and\ \citenamefont
  {Weihong}}]{kotov99}%
  \BibitemOpen
  \bibfield  {author} {\bibinfo {author} {\bibfnamefont {V.~N.}\ \bibnamefont
  {Kotov}}, \bibinfo {author} {\bibfnamefont {J.}~\bibnamefont {Oitmaa}},
  \bibinfo {author} {\bibfnamefont {O.~P.}\ \bibnamefont {Sushkov}},\ and\
  \bibinfo {author} {\bibfnamefont {Z.}~\bibnamefont {Weihong}},\ }\bibfield
  {title} {\bibinfo {title} {Low-energy singlet and triplet excitations in the
  spin-liquid phase of the two-dimensional ${J}_{1}{\ensuremath{-}j}_{2}$
  model},\ }\href {https://doi.org/10.1103/PhysRevB.60.14613} {\bibfield
  {journal} {\bibinfo  {journal} {Phys. Rev. B}\ }\textbf {\bibinfo {volume}
  {60}},\ \bibinfo {pages} {14613} (\bibinfo {year} {1999})}\BibitemShut
  {NoStop}%
\bibitem [{\citenamefont {Capriotti}\ and\ \citenamefont
  {Sorella}(2000)}]{sorella00}%
  \BibitemOpen
  \bibfield  {author} {\bibinfo {author} {\bibfnamefont {L.}~\bibnamefont
  {Capriotti}}\ and\ \bibinfo {author} {\bibfnamefont {S.}~\bibnamefont
  {Sorella}},\ }\bibfield  {title} {\bibinfo {title} {Spontaneous plaquette
  dimerization in the ${J}_{1}-{J}_{2}$ {H}eisenberg model},\ }\href
  {https://doi.org/10.1103/PhysRevLett.84.3173} {\bibfield  {journal} {\bibinfo
   {journal} {Phys. Rev. Lett.}\ }\textbf {\bibinfo {volume} {84}},\ \bibinfo
  {pages} {3173} (\bibinfo {year} {2000})}\BibitemShut {NoStop}%
\bibitem [{\citenamefont {Capriotti}\ \emph {et~al.}(2001)\citenamefont
  {Capriotti}, \citenamefont {Becca}, \citenamefont {Parola},\ and\
  \citenamefont {Sorella}}]{beca01}%
  \BibitemOpen
  \bibfield  {author} {\bibinfo {author} {\bibfnamefont {L.}~\bibnamefont
  {Capriotti}}, \bibinfo {author} {\bibfnamefont {F.}~\bibnamefont {Becca}},
  \bibinfo {author} {\bibfnamefont {A.}~\bibnamefont {Parola}},\ and\ \bibinfo
  {author} {\bibfnamefont {S.}~\bibnamefont {Sorella}},\ }\bibfield  {title}
  {\bibinfo {title} {Resonating valence bond wave functions for strongly
  frustrated spin systems},\ }\href
  {https://doi.org/10.1103/PhysRevLett.87.097201} {\bibfield  {journal}
  {\bibinfo  {journal} {Phys. Rev. Lett.}\ }\textbf {\bibinfo {volume} {87}},\
  \bibinfo {pages} {097201} (\bibinfo {year} {2001})}\BibitemShut {NoStop}%
\bibitem [{\citenamefont {Takano}\ \emph {et~al.}(2003)\citenamefont {Takano},
  \citenamefont {Kito}, \citenamefont {\ifmmode~\bar{O}\else \={O}\fi{}no},\
  and\ \citenamefont {Sano}}]{takano03}%
  \BibitemOpen
  \bibfield  {author} {\bibinfo {author} {\bibfnamefont {K.}~\bibnamefont
  {Takano}}, \bibinfo {author} {\bibfnamefont {Y.}~\bibnamefont {Kito}},
  \bibinfo {author} {\bibfnamefont {Y.}~\bibnamefont {\ifmmode~\bar{O}\else
  \={O}\fi{}no}},\ and\ \bibinfo {author} {\bibfnamefont {K.}~\bibnamefont
  {Sano}},\ }\bibfield  {title} {\bibinfo {title} {Nonlinear
  $\ensuremath{\sigma}$ model method for the
  ${J}_{1}\mathrm{\text{\ensuremath{-}}}{J}_{2}$ {H}eisenberg model: Disordered
  ground state with plaquette symmetry},\ }\href
  {https://doi.org/10.1103/PhysRevLett.91.197202} {\bibfield  {journal}
  {\bibinfo  {journal} {Phys. Rev. Lett.}\ }\textbf {\bibinfo {volume} {91}},\
  \bibinfo {pages} {197202} (\bibinfo {year} {2003})}\BibitemShut {NoStop}%
\bibitem [{\citenamefont {Zhang}\ \emph {et~al.}(2003)\citenamefont {Zhang},
  \citenamefont {Hu},\ and\ \citenamefont {Yu}}]{zhang03}%
  \BibitemOpen
  \bibfield  {author} {\bibinfo {author} {\bibfnamefont {G.-M.}\ \bibnamefont
  {Zhang}}, \bibinfo {author} {\bibfnamefont {H.}~\bibnamefont {Hu}},\ and\
  \bibinfo {author} {\bibfnamefont {L.}~\bibnamefont {Yu}},\ }\bibfield
  {title} {\bibinfo {title} {Valence-bond spin-liquid state in two-dimensional
  frustrated spin-$1/2$ {H}eisenberg antiferromagnets},\ }\href
  {https://doi.org/10.1103/PhysRevLett.91.067201} {\bibfield  {journal}
  {\bibinfo  {journal} {Phys. Rev. Lett.}\ }\textbf {\bibinfo {volume} {91}},\
  \bibinfo {pages} {067201} (\bibinfo {year} {2003})}\BibitemShut {NoStop}%
\bibitem [{\citenamefont {Mambrini}\ \emph {et~al.}(2006)\citenamefont
  {Mambrini}, \citenamefont {L\"auchli}, \citenamefont {Poilblanc},\ and\
  \citenamefont {Mila}}]{mila06}%
  \BibitemOpen
  \bibfield  {author} {\bibinfo {author} {\bibfnamefont {M.}~\bibnamefont
  {Mambrini}}, \bibinfo {author} {\bibfnamefont {A.}~\bibnamefont {L\"auchli}},
  \bibinfo {author} {\bibfnamefont {D.}~\bibnamefont {Poilblanc}},\ and\
  \bibinfo {author} {\bibfnamefont {F.}~\bibnamefont {Mila}},\ }\bibfield
  {title} {\bibinfo {title} {Plaquette valence-bond crystal in the frustrated
  {H}eisenberg quantum antiferromagnet on the square lattice},\ }\href
  {https://doi.org/10.1103/PhysRevB.74.144422} {\bibfield  {journal} {\bibinfo
  {journal} {Phys. Rev. B}\ }\textbf {\bibinfo {volume} {74}},\ \bibinfo
  {pages} {144422} (\bibinfo {year} {2006})}\BibitemShut {NoStop}%
\bibitem [{\citenamefont {Sirker}\ \emph {et~al.}(2006)\citenamefont {Sirker},
  \citenamefont {Weihong}, \citenamefont {Sushkov},\ and\ \citenamefont
  {Oitmaa}}]{sirker06}%
  \BibitemOpen
  \bibfield  {author} {\bibinfo {author} {\bibfnamefont {J.}~\bibnamefont
  {Sirker}}, \bibinfo {author} {\bibfnamefont {Z.}~\bibnamefont {Weihong}},
  \bibinfo {author} {\bibfnamefont {O.~P.}\ \bibnamefont {Sushkov}},\ and\
  \bibinfo {author} {\bibfnamefont {J.}~\bibnamefont {Oitmaa}},\ }\bibfield
  {title} {\bibinfo {title} {${J}_1-{J}_2$ model: First-order phase transition
  versus deconfinement of spinons},\ }\href
  {https://doi.org/10.1103/PhysRevB.73.184420} {\bibfield  {journal} {\bibinfo
  {journal} {Phys. Rev. B}\ }\textbf {\bibinfo {volume} {73}},\ \bibinfo
  {pages} {184420} (\bibinfo {year} {2006})}\BibitemShut {NoStop}%
\bibitem [{\citenamefont {Darradi}\ \emph {et~al.}(2008)\citenamefont
  {Darradi}, \citenamefont {Derzhko}, \citenamefont {Zinke}, \citenamefont
  {Schulenburg}, \citenamefont {Kr\"uger},\ and\ \citenamefont
  {Richter}}]{darradi08}%
  \BibitemOpen
  \bibfield  {author} {\bibinfo {author} {\bibfnamefont {R.}~\bibnamefont
  {Darradi}}, \bibinfo {author} {\bibfnamefont {O.}~\bibnamefont {Derzhko}},
  \bibinfo {author} {\bibfnamefont {R.}~\bibnamefont {Zinke}}, \bibinfo
  {author} {\bibfnamefont {J.}~\bibnamefont {Schulenburg}}, \bibinfo {author}
  {\bibfnamefont {S.~E.}\ \bibnamefont {Kr\"uger}},\ and\ \bibinfo {author}
  {\bibfnamefont {J.}~\bibnamefont {Richter}},\ }\bibfield  {title} {\bibinfo
  {title} {Ground state phases of the spin-1/2 ${J}_{1}-{J}_{2}$ {H}eisenberg
  antiferromagnet on the square lattice: A high-order coupled cluster
  treatment},\ }\href {https://doi.org/10.1103/PhysRevB.78.214415} {\bibfield
  {journal} {\bibinfo  {journal} {Phys. Rev. B}\ }\textbf {\bibinfo {volume}
  {78}},\ \bibinfo {pages} {214415} (\bibinfo {year} {2008})}\BibitemShut
  {NoStop}%
\bibitem [{\citenamefont {Arlego}\ and\ \citenamefont
  {Brenig}(2008)}]{arlego08}%
  \BibitemOpen
  \bibfield  {author} {\bibinfo {author} {\bibfnamefont {M.}~\bibnamefont
  {Arlego}}\ and\ \bibinfo {author} {\bibfnamefont {W.}~\bibnamefont
  {Brenig}},\ }\bibfield  {title} {\bibinfo {title} {Plaquette order in the
  ${J}_{1}\text{\ensuremath{-}}{J}_{2}\text{\ensuremath{-}}{J}_{3}$ model:
  Series expansion analysis},\ }\href
  {https://doi.org/10.1103/PhysRevB.78.224415} {\bibfield  {journal} {\bibinfo
  {journal} {Phys. Rev. B}\ }\textbf {\bibinfo {volume} {78}},\ \bibinfo
  {pages} {224415} (\bibinfo {year} {2008})}\BibitemShut {NoStop}%
\bibitem [{\citenamefont {Murg}\ \emph {et~al.}(2009)\citenamefont {Murg},
  \citenamefont {Verstraete},\ and\ \citenamefont {Cirac}}]{cirac09}%
  \BibitemOpen
  \bibfield  {author} {\bibinfo {author} {\bibfnamefont {V.}~\bibnamefont
  {Murg}}, \bibinfo {author} {\bibfnamefont {F.}~\bibnamefont {Verstraete}},\
  and\ \bibinfo {author} {\bibfnamefont {J.~I.}\ \bibnamefont {Cirac}},\
  }\bibfield  {title} {\bibinfo {title} {Exploring frustrated spin systems
  using projected entangled pair states},\ }\href
  {https://doi.org/10.1103/PhysRevB.79.195119} {\bibfield  {journal} {\bibinfo
  {journal} {Phys. Rev. B}\ }\textbf {\bibinfo {volume} {79}},\ \bibinfo
  {pages} {195119} (\bibinfo {year} {2009})}\BibitemShut {NoStop}%
\bibitem [{\citenamefont {Beach}(2009)}]{beach09}%
  \BibitemOpen
  \bibfield  {author} {\bibinfo {author} {\bibfnamefont {K.~S.~D.}\
  \bibnamefont {Beach}},\ }\bibfield  {title} {\bibinfo {title} {Master
  equation approach to computing {R}{V}{B} bond amplitudes},\ }\href
  {https://doi.org/10.1103/PhysRevB.79.224431} {\bibfield  {journal} {\bibinfo
  {journal} {Phys. Rev. B}\ }\textbf {\bibinfo {volume} {79}},\ \bibinfo
  {pages} {224431} (\bibinfo {year} {2009})}\BibitemShut {NoStop}%
\bibitem [{\citenamefont {Ralko}\ \emph {et~al.}(2009)\citenamefont {Ralko},
  \citenamefont {Mambrini},\ and\ \citenamefont {Poilblanc}}]{ralko09}%
  \BibitemOpen
  \bibfield  {author} {\bibinfo {author} {\bibfnamefont {A.}~\bibnamefont
  {Ralko}}, \bibinfo {author} {\bibfnamefont {M.}~\bibnamefont {Mambrini}},\
  and\ \bibinfo {author} {\bibfnamefont {D.}~\bibnamefont {Poilblanc}},\
  }\bibfield  {title} {\bibinfo {title} {Generalized quantum dimer model
  applied to the frustrated {H}eisenberg model on the square lattice: Emergence
  of a mixed columnar-plaquette phase},\ }\href
  {https://doi.org/10.1103/PhysRevB.80.184427} {\bibfield  {journal} {\bibinfo
  {journal} {Phys. Rev. B}\ }\textbf {\bibinfo {volume} {80}},\ \bibinfo
  {pages} {184427} (\bibinfo {year} {2009})}\BibitemShut {NoStop}%
\bibitem [{\citenamefont {Isaev}\ \emph {et~al.}(2009)\citenamefont {Isaev},
  \citenamefont {Ortiz},\ and\ \citenamefont {Dukelsky}}]{ortiz09}%
  \BibitemOpen
  \bibfield  {author} {\bibinfo {author} {\bibfnamefont {L.}~\bibnamefont
  {Isaev}}, \bibinfo {author} {\bibfnamefont {G.}~\bibnamefont {Ortiz}},\ and\
  \bibinfo {author} {\bibfnamefont {J.}~\bibnamefont {Dukelsky}},\ }\bibfield
  {title} {\bibinfo {title} {Hierarchical mean-field approach to the
  ${J}_{1}\text{\ensuremath{-}}{J}_{2}$ {H}eisenberg model on a square
  lattice},\ }\href {https://doi.org/10.1103/PhysRevB.79.024409} {\bibfield
  {journal} {\bibinfo  {journal} {Phys. Rev. B}\ }\textbf {\bibinfo {volume}
  {79}},\ \bibinfo {pages} {024409} (\bibinfo {year} {2009})}\BibitemShut
  {NoStop}%
\bibitem [{\citenamefont {Reuther}\ \emph {et~al.}(2011)\citenamefont
  {Reuther}, \citenamefont {W\"olfle}, \citenamefont {Darradi}, \citenamefont
  {Brenig}, \citenamefont {Arlego},\ and\ \citenamefont {Richter}}]{arlego11}%
  \BibitemOpen
  \bibfield  {author} {\bibinfo {author} {\bibfnamefont {J.}~\bibnamefont
  {Reuther}}, \bibinfo {author} {\bibfnamefont {P.}~\bibnamefont {W\"olfle}},
  \bibinfo {author} {\bibfnamefont {R.}~\bibnamefont {Darradi}}, \bibinfo
  {author} {\bibfnamefont {W.}~\bibnamefont {Brenig}}, \bibinfo {author}
  {\bibfnamefont {M.}~\bibnamefont {Arlego}},\ and\ \bibinfo {author}
  {\bibfnamefont {J.}~\bibnamefont {Richter}},\ }\bibfield  {title} {\bibinfo
  {title} {Quantum phases of the planar antiferromagnetic
  ${J}_{1}\ensuremath{-}{J}_{2}\ensuremath{-}{J}_{3}$ {H}eisenberg model},\
  }\href {https://doi.org/10.1103/PhysRevB.83.064416} {\bibfield  {journal}
  {\bibinfo  {journal} {Phys. Rev. B}\ }\textbf {\bibinfo {volume} {83}},\
  \bibinfo {pages} {064416} (\bibinfo {year} {2011})}\BibitemShut {NoStop}%
\bibitem [{\citenamefont {G\"otze}\ \emph {et~al.}(2012)\citenamefont
  {G\"otze}, \citenamefont {Kr\"uger}, \citenamefont {Fleck}, \citenamefont
  {Schulenburg},\ and\ \citenamefont {Richter}}]{richter12}%
  \BibitemOpen
  \bibfield  {author} {\bibinfo {author} {\bibfnamefont {O.}~\bibnamefont
  {G\"otze}}, \bibinfo {author} {\bibfnamefont {S.~E.}\ \bibnamefont
  {Kr\"uger}}, \bibinfo {author} {\bibfnamefont {F.}~\bibnamefont {Fleck}},
  \bibinfo {author} {\bibfnamefont {J.}~\bibnamefont {Schulenburg}},\ and\
  \bibinfo {author} {\bibfnamefont {J.}~\bibnamefont {Richter}},\ }\bibfield
  {title} {\bibinfo {title} {Ground-state phase diagram of the
  spin-$\frac{1}{2}$ square-lattice ${J}_{1}$-${J}_{2}$ model with plaquette
  structure},\ }\href {https://doi.org/10.1103/PhysRevB.85.224424} {\bibfield
  {journal} {\bibinfo  {journal} {Phys. Rev. B}\ }\textbf {\bibinfo {volume}
  {85}},\ \bibinfo {pages} {224424} (\bibinfo {year} {2012})}\BibitemShut
  {NoStop}%
\bibitem [{\citenamefont {Yu}\ and\ \citenamefont {Kao}(2012)}]{yu12}%
  \BibitemOpen
  \bibfield  {author} {\bibinfo {author} {\bibfnamefont {J.-F.}\ \bibnamefont
  {Yu}}\ and\ \bibinfo {author} {\bibfnamefont {Y.-J.}\ \bibnamefont {Kao}},\
  }\bibfield  {title} {\bibinfo {title} {Spin-$\frac{1}{2}$ ${J}_{1}$-${J}_{2}$
  {H}eisenberg antiferromagnet on a square lattice: A plaquette renormalized
  tensor network study},\ }\href {https://doi.org/10.1103/PhysRevB.85.094407}
  {\bibfield  {journal} {\bibinfo  {journal} {Phys. Rev. B}\ }\textbf {\bibinfo
  {volume} {85}},\ \bibinfo {pages} {094407} (\bibinfo {year}
  {2012})}\BibitemShut {NoStop}%
\bibitem [{\citenamefont {Jiang}\ \emph {et~al.}(2012)\citenamefont {Jiang},
  \citenamefont {Yao},\ and\ \citenamefont {Balents}}]{jiang12}%
  \BibitemOpen
  \bibfield  {author} {\bibinfo {author} {\bibfnamefont {H.-C.}\ \bibnamefont
  {Jiang}}, \bibinfo {author} {\bibfnamefont {H.}~\bibnamefont {Yao}},\ and\
  \bibinfo {author} {\bibfnamefont {L.}~\bibnamefont {Balents}},\ }\bibfield
  {title} {\bibinfo {title} {Spin liquid ground state of the spin-$\frac{1}{2}$
  square ${J}_{1}$-${J}_{2}$ {H}eisenberg model},\ }\href
  {https://doi.org/10.1103/PhysRevB.86.024424} {\bibfield  {journal} {\bibinfo
  {journal} {Phys. Rev. B}\ }\textbf {\bibinfo {volume} {86}},\ \bibinfo
  {pages} {024424} (\bibinfo {year} {2012})}\BibitemShut {NoStop}%
\bibitem [{\citenamefont {Mezzacapo}(2012)}]{mezzacapo12}%
  \BibitemOpen
  \bibfield  {author} {\bibinfo {author} {\bibfnamefont {F.}~\bibnamefont
  {Mezzacapo}},\ }\bibfield  {title} {\bibinfo {title} {Ground-state phase
  diagram of the quantum ${J}_{1}\ensuremath{-}{J}_{2}$ model on the square
  lattice},\ }\href {https://doi.org/10.1103/PhysRevB.86.045115} {\bibfield
  {journal} {\bibinfo  {journal} {Phys. Rev. B}\ }\textbf {\bibinfo {volume}
  {86}},\ \bibinfo {pages} {045115} (\bibinfo {year} {2012})}\BibitemShut
  {NoStop}%
\bibitem [{\citenamefont {Li}\ \emph {et~al.}(2012)\citenamefont {Li},
  \citenamefont {Becca}, \citenamefont {Hu},\ and\ \citenamefont
  {Sorella}}]{li12}%
  \BibitemOpen
  \bibfield  {author} {\bibinfo {author} {\bibfnamefont {T.}~\bibnamefont
  {Li}}, \bibinfo {author} {\bibfnamefont {F.}~\bibnamefont {Becca}}, \bibinfo
  {author} {\bibfnamefont {W.}~\bibnamefont {Hu}},\ and\ \bibinfo {author}
  {\bibfnamefont {S.}~\bibnamefont {Sorella}},\ }\bibfield  {title} {\bibinfo
  {title} {Gapped spin-liquid phase in the ${J}_{1}\ensuremath{-}{J}_{2}$
  {H}eisenberg model by a bosonic resonating valence-bond ansatz},\ }\href
  {https://doi.org/10.1103/PhysRevB.86.075111} {\bibfield  {journal} {\bibinfo
  {journal} {Phys. Rev. B}\ }\textbf {\bibinfo {volume} {86}},\ \bibinfo
  {pages} {075111} (\bibinfo {year} {2012})}\BibitemShut {NoStop}%
\bibitem [{\citenamefont {Wang}\ \emph {et~al.}(2013)\citenamefont {Wang},
  \citenamefont {Poilblanc}, \citenamefont {Gu}, \citenamefont {Wen},\ and\
  \citenamefont {Verstraete}}]{wang13}%
  \BibitemOpen
  \bibfield  {author} {\bibinfo {author} {\bibfnamefont {L.}~\bibnamefont
  {Wang}}, \bibinfo {author} {\bibfnamefont {D.}~\bibnamefont {Poilblanc}},
  \bibinfo {author} {\bibfnamefont {Z.-C.}\ \bibnamefont {Gu}}, \bibinfo
  {author} {\bibfnamefont {X.-G.}\ \bibnamefont {Wen}},\ and\ \bibinfo {author}
  {\bibfnamefont {F.}~\bibnamefont {Verstraete}},\ }\bibfield  {title}
  {\bibinfo {title} {Constructing a gapless spin-liquid state for the
  spin-$1/2$ ${J}_{1}\ensuremath{-}{J}_{2}$ {H}eisenberg model on a square
  lattice},\ }\href {https://doi.org/10.1103/PhysRevLett.111.037202} {\bibfield
   {journal} {\bibinfo  {journal} {Phys. Rev. Lett.}\ }\textbf {\bibinfo
  {volume} {111}},\ \bibinfo {pages} {037202} (\bibinfo {year}
  {2013})}\BibitemShut {NoStop}%
\bibitem [{\citenamefont {Hu}\ \emph {et~al.}(2013)\citenamefont {Hu},
  \citenamefont {Becca}, \citenamefont {Parola},\ and\ \citenamefont
  {Sorella}}]{hu13}%
  \BibitemOpen
  \bibfield  {author} {\bibinfo {author} {\bibfnamefont {W.-J.}\ \bibnamefont
  {Hu}}, \bibinfo {author} {\bibfnamefont {F.}~\bibnamefont {Becca}}, \bibinfo
  {author} {\bibfnamefont {A.}~\bibnamefont {Parola}},\ and\ \bibinfo {author}
  {\bibfnamefont {S.}~\bibnamefont {Sorella}},\ }\bibfield  {title} {\bibinfo
  {title} {Direct evidence for a gapless ${Z}_{2}$ spin liquid by frustrating
  {N}\'eel antiferromagnetism},\ }\href
  {https://doi.org/10.1103/PhysRevB.88.060402} {\bibfield  {journal} {\bibinfo
  {journal} {Phys. Rev. B}\ }\textbf {\bibinfo {volume} {88}},\ \bibinfo
  {pages} {060402} (\bibinfo {year} {2013})}\BibitemShut {NoStop}%
\bibitem [{\citenamefont {Qi}\ and\ \citenamefont {Gu}(2014)}]{gu14}%
  \BibitemOpen
  \bibfield  {author} {\bibinfo {author} {\bibfnamefont {Y.}~\bibnamefont
  {Qi}}\ and\ \bibinfo {author} {\bibfnamefont {Z.-C.}\ \bibnamefont {Gu}},\
  }\bibfield  {title} {\bibinfo {title} {Continuous phase transition from
  {N}\'eel state to ${Z}_{2}$ spin-liquid state on a square lattice},\ }\href
  {https://doi.org/10.1103/PhysRevB.89.235122} {\bibfield  {journal} {\bibinfo
  {journal} {Phys. Rev. B}\ }\textbf {\bibinfo {volume} {89}},\ \bibinfo
  {pages} {235122} (\bibinfo {year} {2014})}\BibitemShut {NoStop}%
\bibitem [{\citenamefont {Chou}\ and\ \citenamefont {Chen}(2014)}]{chou14}%
  \BibitemOpen
  \bibfield  {author} {\bibinfo {author} {\bibfnamefont {C.-P.}\ \bibnamefont
  {Chou}}\ and\ \bibinfo {author} {\bibfnamefont {H.-Y.}\ \bibnamefont
  {Chen}},\ }\bibfield  {title} {\bibinfo {title} {Simulating a two-dimensional
  frustrated spin system with fermionic resonating-valence-bond states},\
  }\href {https://doi.org/10.1103/PhysRevB.90.041106} {\bibfield  {journal}
  {\bibinfo  {journal} {Phys. Rev. B}\ }\textbf {\bibinfo {volume} {90}},\
  \bibinfo {pages} {041106} (\bibinfo {year} {2014})}\BibitemShut {NoStop}%
\bibitem [{\citenamefont {Metavitsiadis}\ \emph {et~al.}(2014)\citenamefont
  {Metavitsiadis}, \citenamefont {Sellmann},\ and\ \citenamefont
  {Eggert}}]{eggert14}%
  \BibitemOpen
  \bibfield  {author} {\bibinfo {author} {\bibfnamefont {A.}~\bibnamefont
  {Metavitsiadis}}, \bibinfo {author} {\bibfnamefont {D.}~\bibnamefont
  {Sellmann}},\ and\ \bibinfo {author} {\bibfnamefont {S.}~\bibnamefont
  {Eggert}},\ }\bibfield  {title} {\bibinfo {title} {Spin-liquid versus dimer
  phases in an anisotropic ${J}_{1}$-${J}_{2}$ frustrated square
  antiferromagnet},\ }\href {https://doi.org/10.1103/PhysRevB.89.241104}
  {\bibfield  {journal} {\bibinfo  {journal} {Phys. Rev. B}\ }\textbf {\bibinfo
  {volume} {89}},\ \bibinfo {pages} {241104} (\bibinfo {year}
  {2014})}\BibitemShut {NoStop}%
\bibitem [{\citenamefont {Doretto}(2014)}]{doretto14}%
  \BibitemOpen
  \bibfield  {author} {\bibinfo {author} {\bibfnamefont {R.~L.}\ \bibnamefont
  {Doretto}},\ }\bibfield  {title} {\bibinfo {title} {Plaquette valence-bond
  solid in the square-lattice ${J}_{1}$-${J}_{2}$ antiferromagnet {H}eisenberg
  model: A bond operator approach},\ }\href
  {https://doi.org/10.1103/PhysRevB.89.104415} {\bibfield  {journal} {\bibinfo
  {journal} {Phys. Rev. B}\ }\textbf {\bibinfo {volume} {89}},\ \bibinfo
  {pages} {104415} (\bibinfo {year} {2014})}\BibitemShut {NoStop}%
\bibitem [{\citenamefont {Gong}\ \emph {et~al.}(2014)\citenamefont {Gong},
  \citenamefont {Zhu}, \citenamefont {Sheng}, \citenamefont {Motrunich},\ and\
  \citenamefont {Fisher}}]{gong14}%
  \BibitemOpen
  \bibfield  {author} {\bibinfo {author} {\bibfnamefont {S.-S.}\ \bibnamefont
  {Gong}}, \bibinfo {author} {\bibfnamefont {W.}~\bibnamefont {Zhu}}, \bibinfo
  {author} {\bibfnamefont {D.~N.}\ \bibnamefont {Sheng}}, \bibinfo {author}
  {\bibfnamefont {O.~I.}\ \bibnamefont {Motrunich}},\ and\ \bibinfo {author}
  {\bibfnamefont {M.~P.~A.}\ \bibnamefont {Fisher}},\ }\bibfield  {title}
  {\bibinfo {title} {Plaquette ordered phase and quantum phase diagram in the
  spin-$\frac{1}{2}$ ${J}_{1}\text{\ensuremath{-}}{J}_{2}$ square {H}eisenberg
  model},\ }\href {https://doi.org/10.1103/PhysRevLett.113.027201} {\bibfield
  {journal} {\bibinfo  {journal} {Phys. Rev. Lett.}\ }\textbf {\bibinfo
  {volume} {113}},\ \bibinfo {pages} {027201} (\bibinfo {year}
  {2014})}\BibitemShut {NoStop}%
\bibitem [{\citenamefont {Richter}\ \emph {et~al.}(2015)\citenamefont
  {Richter}, \citenamefont {Zinke},\ and\ \citenamefont {Farnell}}]{richter15}%
  \BibitemOpen
  \bibfield  {author} {\bibinfo {author} {\bibfnamefont {J.}~\bibnamefont
  {Richter}}, \bibinfo {author} {\bibfnamefont {R.}~\bibnamefont {Zinke}},\
  and\ \bibinfo {author} {\bibfnamefont {D.~J.~J.}\ \bibnamefont {Farnell}},\
  }\bibfield  {title} {\bibinfo {title} {The spin-$1/2$ square-lattice
  ${J}_{1}-{J}_{2}$ model: The spin-gap issue},\ }\href
  {https://doi.org/10.1140/epjb/e2014-50589-x} {\bibfield  {journal} {\bibinfo
  {journal} {Eur. Phys. J. B}\ }\textbf {\bibinfo {volume} {88}},\ \bibinfo
  {pages} {2} (\bibinfo {year} {2015})}\BibitemShut {NoStop}%
\bibitem [{\citenamefont {Morita}\ \emph {et~al.}(2015)\citenamefont {Morita},
  \citenamefont {Kaneko},\ and\ \citenamefont {Imada}}]{imada15}%
  \BibitemOpen
  \bibfield  {author} {\bibinfo {author} {\bibfnamefont {S.}~\bibnamefont
  {Morita}}, \bibinfo {author} {\bibfnamefont {R.}~\bibnamefont {Kaneko}},\
  and\ \bibinfo {author} {\bibfnamefont {M.}~\bibnamefont {Imada}},\ }\bibfield
   {title} {\bibinfo {title} {Quantum spin liquid in spin 1/2 j1–j2
  {H}eisenberg model on square lattice: Many-variable variational {M}onte
  {C}arlo study combined with quantum-number projections},\ }\href
  {https://doi.org/10.7566/JPSJ.84.024720} {\bibfield  {journal} {\bibinfo
  {journal} {Journal of the Physical Society of Japan}\ }\textbf {\bibinfo
  {volume} {84}},\ \bibinfo {pages} {024720} (\bibinfo {year}
  {2015})}\BibitemShut {NoStop}%
\bibitem [{\citenamefont {Wang}\ \emph {et~al.}(2016)\citenamefont {Wang},
  \citenamefont {Gu}, \citenamefont {Verstraete},\ and\ \citenamefont
  {Wen}}]{wang16}%
  \BibitemOpen
  \bibfield  {author} {\bibinfo {author} {\bibfnamefont {L.}~\bibnamefont
  {Wang}}, \bibinfo {author} {\bibfnamefont {Z.-C.}\ \bibnamefont {Gu}},
  \bibinfo {author} {\bibfnamefont {F.}~\bibnamefont {Verstraete}},\ and\
  \bibinfo {author} {\bibfnamefont {X.-G.}\ \bibnamefont {Wen}},\ }\bibfield
  {title} {\bibinfo {title} {Tensor-product state approach to
  spin-$\frac{1}{2}$ square ${J}_{1}\text{\ensuremath{-}}{J}_{2}$
  antiferromagnetic {H}eisenberg model: Evidence for deconfined quantum
  criticality},\ }\href {https://doi.org/10.1103/PhysRevB.94.075143} {\bibfield
   {journal} {\bibinfo  {journal} {Phys. Rev. B}\ }\textbf {\bibinfo {volume}
  {94}},\ \bibinfo {pages} {075143} (\bibinfo {year} {2016})}\BibitemShut
  {NoStop}%
\bibitem [{\citenamefont {Yang}\ and\ \citenamefont {Wang}(2016)}]{yang16}%
  \BibitemOpen
  \bibfield  {author} {\bibinfo {author} {\bibfnamefont {X.}~\bibnamefont
  {Yang}}\ and\ \bibinfo {author} {\bibfnamefont {F.}~\bibnamefont {Wang}},\
  }\bibfield  {title} {\bibinfo {title} {Schwinger boson spin-liquid states on
  square lattice},\ }\href {https://doi.org/10.1103/PhysRevB.94.035160}
  {\bibfield  {journal} {\bibinfo  {journal} {Phys. Rev. B}\ }\textbf {\bibinfo
  {volume} {94}},\ \bibinfo {pages} {035160} (\bibinfo {year}
  {2016})}\BibitemShut {NoStop}%
\bibitem [{\citenamefont {Haghshenas}\ and\ \citenamefont
  {Sheng}(2018)}]{sheng18}%
  \BibitemOpen
  \bibfield  {author} {\bibinfo {author} {\bibfnamefont {R.}~\bibnamefont
  {Haghshenas}}\ and\ \bibinfo {author} {\bibfnamefont {D.~N.}\ \bibnamefont
  {Sheng}},\ }\bibfield  {title} {\bibinfo {title} {$u(1)$-symmetric infinite
  projected entangled-pair states study of the spin-1/2 square
  ${J}_{1}\text{\ensuremath{-}}{J}_{2}$ {H}eisenberg model},\ }\href
  {https://doi.org/10.1103/PhysRevB.97.174408} {\bibfield  {journal} {\bibinfo
  {journal} {Phys. Rev. B}\ }\textbf {\bibinfo {volume} {97}},\ \bibinfo
  {pages} {174408} (\bibinfo {year} {2018})}\BibitemShut {NoStop}%
\bibitem [{\citenamefont {Ferrari}\ and\ \citenamefont
  {Becca}(2018)}]{becca18}%
  \BibitemOpen
  \bibfield  {author} {\bibinfo {author} {\bibfnamefont {F.}~\bibnamefont
  {Ferrari}}\ and\ \bibinfo {author} {\bibfnamefont {F.}~\bibnamefont
  {Becca}},\ }\bibfield  {title} {\bibinfo {title} {Spectral signatures of
  fractionalization in the frustrated {H}eisenberg model on the square
  lattice},\ }\href {https://doi.org/10.1103/PhysRevB.98.100405} {\bibfield
  {journal} {\bibinfo  {journal} {Phys. Rev. B}\ }\textbf {\bibinfo {volume}
  {98}},\ \bibinfo {pages} {100405} (\bibinfo {year} {2018})}\BibitemShut
  {NoStop}%
\bibitem [{\citenamefont {Liu}\ \emph {et~al.}(2018)\citenamefont {Liu},
  \citenamefont {Dong}, \citenamefont {Wang}, \citenamefont {Han},
  \citenamefont {An}, \citenamefont {Guo},\ and\ \citenamefont {He}}]{dong18}%
  \BibitemOpen
  \bibfield  {author} {\bibinfo {author} {\bibfnamefont {W.-Y.}\ \bibnamefont
  {Liu}}, \bibinfo {author} {\bibfnamefont {S.}~\bibnamefont {Dong}}, \bibinfo
  {author} {\bibfnamefont {C.}~\bibnamefont {Wang}}, \bibinfo {author}
  {\bibfnamefont {Y.}~\bibnamefont {Han}}, \bibinfo {author} {\bibfnamefont
  {H.}~\bibnamefont {An}}, \bibinfo {author} {\bibfnamefont {G.-C.}\
  \bibnamefont {Guo}},\ and\ \bibinfo {author} {\bibfnamefont {L.}~\bibnamefont
  {He}},\ }\bibfield  {title} {\bibinfo {title} {Gapless spin liquid ground
  state of the spin-$\frac{1}{2}$ ${J}_{1}\ensuremath{-}{J}_{2}$ {H}eisenberg
  model on square lattices},\ }\href
  {https://doi.org/10.1103/PhysRevB.98.241109} {\bibfield  {journal} {\bibinfo
  {journal} {Phys. Rev. B}\ }\textbf {\bibinfo {volume} {98}},\ \bibinfo
  {pages} {241109} (\bibinfo {year} {2018})}\BibitemShut {NoStop}%
\bibitem [{\citenamefont {Wang}\ and\ \citenamefont {Sandvik}(2018)}]{wang18}%
  \BibitemOpen
  \bibfield  {author} {\bibinfo {author} {\bibfnamefont {L.}~\bibnamefont
  {Wang}}\ and\ \bibinfo {author} {\bibfnamefont {A.~W.}\ \bibnamefont
  {Sandvik}},\ }\bibfield  {title} {\bibinfo {title} {Critical level crossings
  and gapless spin liquid in the square-lattice spin-$1/2$
  ${J}_{1}\ensuremath{-}{J}_{2}$ {H}eisenberg antiferromagnet},\ }\href
  {https://doi.org/10.1103/PhysRevLett.121.107202} {\bibfield  {journal}
  {\bibinfo  {journal} {Phys. Rev. Lett.}\ }\textbf {\bibinfo {volume} {121}},\
  \bibinfo {pages} {107202} (\bibinfo {year} {2018})}\BibitemShut {NoStop}%
\bibitem [{\citenamefont {Poilblanc}\ and\ \citenamefont
  {Mambrini}(2017)}]{didier17}%
  \BibitemOpen
  \bibfield  {author} {\bibinfo {author} {\bibfnamefont {D.}~\bibnamefont
  {Poilblanc}}\ and\ \bibinfo {author} {\bibfnamefont {M.}~\bibnamefont
  {Mambrini}},\ }\bibfield  {title} {\bibinfo {title} {Quantum critical phase
  with infinite projected entangled paired states},\ }\href
  {https://doi.org/10.1103/PhysRevB.96.014414} {\bibfield  {journal} {\bibinfo
  {journal} {Phys. Rev. B}\ }\textbf {\bibinfo {volume} {96}},\ \bibinfo
  {pages} {014414} (\bibinfo {year} {2017})}\BibitemShut {NoStop}%
\bibitem [{\citenamefont {Poilblanc}\ \emph {et~al.}(2019)\citenamefont
  {Poilblanc}, \citenamefont {Mambrini},\ and\ \citenamefont
  {Capponi}}]{capponi19}%
  \BibitemOpen
  \bibfield  {author} {\bibinfo {author} {\bibfnamefont {D.}~\bibnamefont
  {Poilblanc}}, \bibinfo {author} {\bibfnamefont {M.}~\bibnamefont
  {Mambrini}},\ and\ \bibinfo {author} {\bibfnamefont {S.}~\bibnamefont
  {Capponi}},\ }\bibfield  {title} {\bibinfo {title} {{Critical
  colored-{R}{V}{B} states in the frustrated quantum {H}eisenberg model on the
  square lattice}},\ }\href {https://doi.org/10.21468/SciPostPhys.7.4.041}
  {\bibfield  {journal} {\bibinfo  {journal} {SciPost Phys.}\ }\textbf
  {\bibinfo {volume} {7}},\ \bibinfo {pages} {041} (\bibinfo {year}
  {2019})}\BibitemShut {NoStop}%
\bibitem [{\citenamefont {Syromyatnikov}\ and\ \citenamefont
  {Aktersky}(2019)}]{yu19}%
  \BibitemOpen
  \bibfield  {author} {\bibinfo {author} {\bibfnamefont {A.~V.}\ \bibnamefont
  {Syromyatnikov}}\ and\ \bibinfo {author} {\bibfnamefont {A.~Y.}\ \bibnamefont
  {Aktersky}},\ }\bibfield  {title} {\bibinfo {title} {Elementary excitations
  in the ordered phase of spin-$\frac{1}{2}$ ${J}_{1}$--${J}_{2}$ model on
  square lattice},\ }\href {https://doi.org/10.1103/PhysRevB.99.224402}
  {\bibfield  {journal} {\bibinfo  {journal} {Phys. Rev. B}\ }\textbf {\bibinfo
  {volume} {99}},\ \bibinfo {pages} {224402} (\bibinfo {year}
  {2019})}\BibitemShut {NoStop}%
\bibitem [{\citenamefont {Doretto}(2020)}]{doretto20}%
  \BibitemOpen
  \bibfield  {author} {\bibinfo {author} {\bibfnamefont {R.~L.}\ \bibnamefont
  {Doretto}},\ }\bibfield  {title} {\bibinfo {title} {Mean-field theory of
  interacting triplons in a two-dimensional valence-bond solid: Stability and
  properties of many-triplon states},\ }\href
  {https://doi.org/10.1103/PhysRevB.102.014415} {\bibfield  {journal} {\bibinfo
   {journal} {Phys. Rev. B}\ }\textbf {\bibinfo {volume} {102}},\ \bibinfo
  {pages} {014415} (\bibinfo {year} {2020})}\BibitemShut {NoStop}%
\bibitem [{\citenamefont {Nomura}\ and\ \citenamefont
  {Imada}(2021)}]{nomura20}%
  \BibitemOpen
  \bibfield  {author} {\bibinfo {author} {\bibfnamefont {Y.}~\bibnamefont
  {Nomura}}\ and\ \bibinfo {author} {\bibfnamefont {M.}~\bibnamefont {Imada}},\
  }\bibfield  {title} {\bibinfo {title} {Dirac-type nodal spin liquid revealed
  by refined quantum many-body solver using neural-network wave function,
  correlation ratio, and level spectroscopy},\ }\href
  {https://doi.org/10.1103/PhysRevX.11.031034} {\bibfield  {journal} {\bibinfo
  {journal} {Phys. Rev. X}\ }\textbf {\bibinfo {volume} {11}},\ \bibinfo
  {pages} {031034} (\bibinfo {year} {2021})}\BibitemShut {NoStop}%
\bibitem [{\citenamefont {Ferrari}\ and\ \citenamefont
  {Becca}(2020)}]{ferrari20}%
  \BibitemOpen
  \bibfield  {author} {\bibinfo {author} {\bibfnamefont {F.}~\bibnamefont
  {Ferrari}}\ and\ \bibinfo {author} {\bibfnamefont {F.}~\bibnamefont
  {Becca}},\ }\bibfield  {title} {\bibinfo {title} {Gapless spin liquid and
  valence-bond solid in the ${J}_{1}$-${J}_{2}$ {H}eisenberg model on the
  square lattice: Insights from singlet and triplet excitations},\ }\href
  {https://doi.org/10.1103/PhysRevB.102.014417} {\bibfield  {journal} {\bibinfo
   {journal} {Phys. Rev. B}\ }\textbf {\bibinfo {volume} {102}},\ \bibinfo
  {pages} {014417} (\bibinfo {year} {2020})}\BibitemShut {NoStop}%
\bibitem [{\citenamefont {Hasik}\ \emph {et~al.}(2021)\citenamefont {Hasik},
  \citenamefont {Poilblanc},\ and\ \citenamefont {Becca}}]{hasik21}%
  \BibitemOpen
  \bibfield  {author} {\bibinfo {author} {\bibfnamefont {J.}~\bibnamefont
  {Hasik}}, \bibinfo {author} {\bibfnamefont {D.}~\bibnamefont {Poilblanc}},\
  and\ \bibinfo {author} {\bibfnamefont {F.}~\bibnamefont {Becca}},\ }\bibfield
   {title} {\bibinfo {title} {{Investigation of the {N}éel phase of the
  frustrated Heisenberg antiferromagnet by differentiable symmetric tensor
  networks}},\ }\href {https://doi.org/10.21468/SciPostPhys.10.1.012}
  {\bibfield  {journal} {\bibinfo  {journal} {SciPost Phys.}\ }\textbf
  {\bibinfo {volume} {10}},\ \bibinfo {pages} {012} (\bibinfo {year}
  {2021})}\BibitemShut {NoStop}%
\bibitem [{\citenamefont {Liu}\ \emph {et~al.}(2022)\citenamefont {Liu},
  \citenamefont {Gong}, \citenamefont {Li}, \citenamefont {Poilblanc},
  \citenamefont {Chen},\ and\ \citenamefont {Gu}}]{liu22}%
  \BibitemOpen
  \bibfield  {author} {\bibinfo {author} {\bibfnamefont {W.-Y.}\ \bibnamefont
  {Liu}}, \bibinfo {author} {\bibfnamefont {S.-S.}\ \bibnamefont {Gong}},
  \bibinfo {author} {\bibfnamefont {Y.-B.}\ \bibnamefont {Li}}, \bibinfo
  {author} {\bibfnamefont {D.}~\bibnamefont {Poilblanc}}, \bibinfo {author}
  {\bibfnamefont {W.-Q.}\ \bibnamefont {Chen}},\ and\ \bibinfo {author}
  {\bibfnamefont {Z.-C.}\ \bibnamefont {Gu}},\ }\bibfield  {title} {\bibinfo
  {title} {Gapless quantum spin liquid and global phase diagram of the spin-1/2
  j1-j2 square antiferromagnetic {H}eisenberg model},\ }\href
  {https://doi.org/https://doi.org/10.1016/j.scib.2022.03.010} {\bibfield
  {journal} {\bibinfo  {journal} {Science Bulletin}\ }\textbf {\bibinfo
  {volume} {67}},\ \bibinfo {pages} {1034} (\bibinfo {year}
  {2022})}\BibitemShut {NoStop}%
\bibitem [{\citenamefont {Liu}\ \emph {et~al.}(2023)\citenamefont {Liu},
  \citenamefont {Poilblanc}, \citenamefont {Gong}, \citenamefont {Chen},\ and\
  \citenamefont {Gu}}]{liu23}%
  \BibitemOpen
  \bibfield  {author} {\bibinfo {author} {\bibfnamefont {W.-Y.}\ \bibnamefont
  {Liu}}, \bibinfo {author} {\bibfnamefont {D.}~\bibnamefont {Poilblanc}},
  \bibinfo {author} {\bibfnamefont {S.-S.}\ \bibnamefont {Gong}}, \bibinfo
  {author} {\bibfnamefont {W.-Q.}\ \bibnamefont {Chen}},\ and\ \bibinfo
  {author} {\bibfnamefont {Z.-C.}\ \bibnamefont {Gu}},\ }\href@noop {}
  {\bibinfo {title} {Tensor network study of the spin-1/2 square-lattice
  ${J}_1$-${J}_2$-${J}_3$ model: incommensurate spiral order, mixed
  valence-bond solids, and multicritical points}} (\bibinfo {year} {2023}),\
  \Eprint {https://arxiv.org/abs/2309.13301} {arXiv:2309.13301
  [cond-mat.str-el]} \BibitemShut {NoStop}%
\bibitem [{\citenamefont {Yang}\ \emph {et~al.}(2023)\citenamefont {Yang},
  \citenamefont {Chen}, \citenamefont {Cheng},\ and\ \citenamefont
  {Luo}}]{luo23}%
  \BibitemOpen
  \bibfield  {author} {\bibinfo {author} {\bibfnamefont {Y.-T.}\ \bibnamefont
  {Yang}}, \bibinfo {author} {\bibfnamefont {F.-Z.}\ \bibnamefont {Chen}},
  \bibinfo {author} {\bibfnamefont {C.}~\bibnamefont {Cheng}},\ and\ \bibinfo
  {author} {\bibfnamefont {H.-G.}\ \bibnamefont {Luo}},\ }\href@noop {}
  {\bibinfo {title} {An explicit evolution from {N}\'eel to striped
  antiferromagnetic states in the spin-1/2 ${J}_{1}$-${J}_{2}$ {H}eisenberg
  model on the square lattice}} (\bibinfo {year} {2023}),\ \Eprint
  {https://arxiv.org/abs/2310.09174} {arXiv:2310.09174 [cond-mat.str-el]}
  \BibitemShut {NoStop}%
\bibitem [{\citenamefont {Voigt}\ and\ \citenamefont
  {Richter}(1996)}]{voigt96}%
  \BibitemOpen
  \bibfield  {author} {\bibinfo {author} {\bibfnamefont {A.}~\bibnamefont
  {Voigt}}\ and\ \bibinfo {author} {\bibfnamefont {J.}~\bibnamefont
  {Richter}},\ }\bibfield  {title} {\bibinfo {title} {The ${J}_1$-${J}_2$
  antiferromagnet on the square lattice with {D}zyaloshinskii-{M}oriya
  interaction: an exact diagonalization study},\ }\href
  {https://doi.org/10.1088/0953-8984/8/27/015} {\bibfield  {journal} {\bibinfo
  {journal} {Journal of Physics: Condensed Matter}\ }\textbf {\bibinfo {volume}
  {8}},\ \bibinfo {pages} {5059} (\bibinfo {year} {1996})}\BibitemShut
  {NoStop}%
\bibitem [{\citenamefont {Merino}\ and\ \citenamefont
  {Ralko}(2022)}]{merino22}%
  \BibitemOpen
  \bibfield  {author} {\bibinfo {author} {\bibfnamefont {J.}~\bibnamefont
  {Merino}}\ and\ \bibinfo {author} {\bibfnamefont {A.}~\bibnamefont {Ralko}},\
  }\bibfield  {title} {\bibinfo {title} {Majorana chiral spin liquid in a model
  for {M}ott insulating cuprates},\ }\href
  {https://doi.org/10.1103/PhysRevResearch.4.023122} {\bibfield  {journal}
  {\bibinfo  {journal} {Phys. Rev. Res.}\ }\textbf {\bibinfo {volume} {4}},\
  \bibinfo {pages} {023122} (\bibinfo {year} {2022})}\BibitemShut {NoStop}%
\bibitem [{\citenamefont {Leite}\ and\ \citenamefont
  {Doretto}(2019)}]{leite19}%
  \BibitemOpen
  \bibfield  {author} {\bibinfo {author} {\bibfnamefont {L.~S.~G.}\
  \bibnamefont {Leite}}\ and\ \bibinfo {author} {\bibfnamefont {R.~L.}\
  \bibnamefont {Doretto}},\ }\bibfield  {title} {\bibinfo {title} {Entanglement
  entropy for the valence bond solid phases of two-dimensional dimerized
  {H}eisenberg antiferromagnets},\ }\href
  {https://doi.org/10.1103/PhysRevB.100.045113} {\bibfield  {journal} {\bibinfo
   {journal} {Phys. Rev. B}\ }\textbf {\bibinfo {volume} {100}},\ \bibinfo
  {pages} {045113} (\bibinfo {year} {2019})}\BibitemShut {NoStop}%
\bibitem [{\citenamefont {Colpa}(1978)}]{colpa78}%
  \BibitemOpen
  \bibfield  {author} {\bibinfo {author} {\bibfnamefont {J.}~\bibnamefont
  {Colpa}},\ }\bibfield  {title} {\bibinfo {title} {Diagonalization of the
  quadratic boson hamiltonian},\ }\href
  {https://doi.org/https://doi.org/10.1016/0378-4371(78)90160-7} {\bibfield
  {journal} {\bibinfo  {journal} {Physica A: Statistical Mechanics and its
  Applications}\ }\textbf {\bibinfo {volume} {93}},\ \bibinfo {pages} {327}
  (\bibinfo {year} {1978})}\BibitemShut {NoStop}%
\bibitem [{\citenamefont {Blaizot}\ and\ \citenamefont
  {Ripka}(1986)}]{blaizot}%
  \BibitemOpen
  \bibfield  {author} {\bibinfo {author} {\bibfnamefont {J.~P.}\ \bibnamefont
  {Blaizot}}\ and\ \bibinfo {author} {\bibfnamefont {G.}~\bibnamefont
  {Ripka}},\ }\href@noop {} {\emph {\bibinfo {title} {Quantum {T}heory of
  {F}inite {S}ystems}}}\ (\bibinfo  {publisher} {MIT, Cambridge, MA},\ \bibinfo
  {year} {1986})\BibitemShut {NoStop}%
\bibitem [{\citenamefont {Shindou}\ \emph {et~al.}(2013)\citenamefont
  {Shindou}, \citenamefont {Matsumoto}, \citenamefont {Murakami},\ and\
  \citenamefont {Ohe}}]{ryo13}%
  \BibitemOpen
  \bibfield  {author} {\bibinfo {author} {\bibfnamefont {R.}~\bibnamefont
  {Shindou}}, \bibinfo {author} {\bibfnamefont {R.}~\bibnamefont {Matsumoto}},
  \bibinfo {author} {\bibfnamefont {S.}~\bibnamefont {Murakami}},\ and\
  \bibinfo {author} {\bibfnamefont {J.-i.}\ \bibnamefont {Ohe}},\ }\bibfield
  {title} {\bibinfo {title} {Topological chiral magnonic edge mode in a
  magnonic crystal},\ }\href {https://doi.org/10.1103/PhysRevB.87.174427}
  {\bibfield  {journal} {\bibinfo  {journal} {Phys. Rev. B}\ }\textbf {\bibinfo
  {volume} {87}},\ \bibinfo {pages} {174427} (\bibinfo {year}
  {2013})}\BibitemShut {NoStop}%
\bibitem [{\citenamefont {Matsumoto}\ \emph {et~al.}(2014)\citenamefont
  {Matsumoto}, \citenamefont {Shindou},\ and\ \citenamefont
  {Murakami}}]{ryo14}%
  \BibitemOpen
  \bibfield  {author} {\bibinfo {author} {\bibfnamefont {R.}~\bibnamefont
  {Matsumoto}}, \bibinfo {author} {\bibfnamefont {R.}~\bibnamefont {Shindou}},\
  and\ \bibinfo {author} {\bibfnamefont {S.}~\bibnamefont {Murakami}},\
  }\bibfield  {title} {\bibinfo {title} {Thermal {H}all effect of magnons in
  magnets with dipolar interaction},\ }\href
  {https://doi.org/10.1103/PhysRevB.89.054420} {\bibfield  {journal} {\bibinfo
  {journal} {Phys. Rev. B}\ }\textbf {\bibinfo {volume} {89}},\ \bibinfo
  {pages} {054420} (\bibinfo {year} {2014})}\BibitemShut {NoStop}%
\bibitem [{\citenamefont {Yang}\ \emph {et~al.}(2020)\citenamefont {Yang},
  \citenamefont {Zhang},\ and\ \citenamefont {Zhang}}]{yang20}%
  \BibitemOpen
  \bibfield  {author} {\bibinfo {author} {\bibfnamefont {Y.-f.}\ \bibnamefont
  {Yang}}, \bibinfo {author} {\bibfnamefont {G.-M.}\ \bibnamefont {Zhang}},\
  and\ \bibinfo {author} {\bibfnamefont {F.-C.}\ \bibnamefont {Zhang}},\
  }\bibfield  {title} {\bibinfo {title} {Universal behavior of the thermal
  {H}all conductivity},\ }\href
  {https://doi.org/10.1103/PhysRevLett.124.186602} {\bibfield  {journal}
  {\bibinfo  {journal} {Phys. Rev. Lett.}\ }\textbf {\bibinfo {volume} {124}},\
  \bibinfo {pages} {186602} (\bibinfo {year} {2020})}\BibitemShut {NoStop}%
\bibitem [{\citenamefont {Suetsugu}\ \emph {et~al.}(2022)\citenamefont
  {Suetsugu}, \citenamefont {Yokoi}, \citenamefont {Totsuka}, \citenamefont
  {Ono}, \citenamefont {Tanaka}, \citenamefont {Kasahara}, \citenamefont
  {Kasahara}, \citenamefont {Chengchao}, \citenamefont {Kageyama},\ and\
  \citenamefont {Matsuda}}]{matsuda22}%
  \BibitemOpen
  \bibfield  {author} {\bibinfo {author} {\bibfnamefont {S.}~\bibnamefont
  {Suetsugu}}, \bibinfo {author} {\bibfnamefont {T.}~\bibnamefont {Yokoi}},
  \bibinfo {author} {\bibfnamefont {K.}~\bibnamefont {Totsuka}}, \bibinfo
  {author} {\bibfnamefont {T.}~\bibnamefont {Ono}}, \bibinfo {author}
  {\bibfnamefont {I.}~\bibnamefont {Tanaka}}, \bibinfo {author} {\bibfnamefont
  {S.}~\bibnamefont {Kasahara}}, \bibinfo {author} {\bibfnamefont
  {Y.}~\bibnamefont {Kasahara}}, \bibinfo {author} {\bibfnamefont
  {Z.}~\bibnamefont {Chengchao}}, \bibinfo {author} {\bibfnamefont
  {H.}~\bibnamefont {Kageyama}},\ and\ \bibinfo {author} {\bibfnamefont
  {Y.}~\bibnamefont {Matsuda}},\ }\bibfield  {title} {\bibinfo {title}
  {Intrinsic suppression of the topological thermal {H}all effect in an exactly
  solvable quantum magnet},\ }\href
  {https://doi.org/10.1103/PhysRevB.105.024415} {\bibfield  {journal} {\bibinfo
   {journal} {Phys. Rev. B}\ }\textbf {\bibinfo {volume} {105}},\ \bibinfo
  {pages} {024415} (\bibinfo {year} {2022})}\BibitemShut {NoStop}%
\bibitem [{\citenamefont {Doretto}\ \emph {et~al.}(2012)\citenamefont
  {Doretto}, \citenamefont {Morais~Smith},\ and\ \citenamefont
  {Caldeira}}]{doretto12}%
  \BibitemOpen
  \bibfield  {author} {\bibinfo {author} {\bibfnamefont {R.~L.}\ \bibnamefont
  {Doretto}}, \bibinfo {author} {\bibfnamefont {C.}~\bibnamefont
  {Morais~Smith}},\ and\ \bibinfo {author} {\bibfnamefont {A.~O.}\ \bibnamefont
  {Caldeira}},\ }\bibfield  {title} {\bibinfo {title} {Finite-momentum
  condensate of magnetic excitons in a bilayer quantum {H}all system},\ }\href
  {https://doi.org/10.1103/PhysRevB.86.035326} {\bibfield  {journal} {\bibinfo
  {journal} {Phys. Rev. B}\ }\textbf {\bibinfo {volume} {86}},\ \bibinfo
  {pages} {035326} (\bibinfo {year} {2012})}\BibitemShut {NoStop}%
\end{thebibliography}%

\end{document}